\documentclass[11pt]{article}
\usepackage{amsmath,amssymb}
\usepackage{amsthm}
\usepackage{float}
\usepackage[dvipdfmx]{graphicx} 
\usepackage{graphicx}
\usepackage{subfigure}
\usepackage[usenames]{color}
\usepackage[all]{xy}
\usepackage{comment}
\usepackage[vcentermath]{youngtab}

\usepackage[bookmarksnumbered=true,colorlinks=true,filecolor=blue,linkcolor=blue,urlcolor=blue,citecolor=blue,linktocpage=true]{hyperref}

\definecolor{refkey}{rgb}{1,1,1}

\numberwithin{equation}{section}

\newcommand{\e}{\epsilon} %epsilon

\newcommand{\G}{\Gamma}
\newcommand{\la}{\lambda} %lambda

\newcommand{\Om}{\Omega}

\newcommand{\Ncal}{\mathcal{N}}
\newcommand{\Ocal}{\mathcal{O}}

\newcommand{\Xcal}{\mathcal{X}}

\newcommand{\Zcal}{\mathcal{Z}}

\newcommand{\Cbb}{\mathbb{C}}

\newcommand{\Rbb}{\mathbb{R}}

\newcommand{\Zbb}{\mathbb{Z}}

\newcommand{\qfrak}{\mathfrak{q}}
\newcommand{\rfrak}{\mathfrak{r}}
\newcommand{\sfrak}{\mathfrak{s}}

\newcommand{\lp}{\left(}
\newcommand{\rp}{\right)}
\newcommand{\lc}{\left\{}
\newcommand{\rc}{\right\}}

\newcommand{\Tr}{\operatorname{Tr}}
\newcommand{\rem}[1]{\textcolor{blue}{\bf[#1]}}
\newcommand{\Res}[1]{\underset{#1}{\operatorname{Res}}}

\newcommand{\Psf}{\mathsf{P}}
\newcommand{\Ssf}{\mathsf{S}}
\newcommand{\Tsf}{\mathsf{T}}
\newcommand{\Ysf}{\mathsf{Y}}

\begin{document}

%\definecolor{labelkey}{rgb}{1,1,1}
%\renewcommand{\include}[1]{}
%\renewcommand\documentclass[2][]{}

\renewcommand{\thefootnote}{\fnsymbol{footnote}}
\setcounter{page}{0}
%%%%%%%%%%%%%%%%% Title page %%%%%%%%%%%%%%%%%%%%%%%%%%%%
\thispagestyle{empty}
\begin{flushright} YITP-17-47 \\ OU-HET 933 \end{flushright} %paper number

\vskip3cm
\begin{center}
{\LARGE {\bf Refined geometric transition and $qq$-characters}}
 \vskip1.5cm
{\large 
 {%\bf
 \sc Taro Kimura%
 \footnote{\href{mailto:taro.kimura@keio.jp}{\tt taro.kimura@keio.jp}}%
,
 %\bf
 \sc Hironori Mori%
 \footnote{\href{mailto:hironori.mori@yukawa.kyoto-u.ac.jp}{\tt hironori.mori@yukawa.kyoto-u.ac.jp}},
and
 %\bf
 \sc Yuji Sugimoto%
 \footnote{\href{mailto:sugimoto@het.phys.sci.osaka-u.ac.jp}{\tt sugimoto@het.phys.sci.osaka-u.ac.jp}}}
%name address

\vskip1.5cm
\it $^{*}$Department of Physics, %and Research and Education Center for Natural Sciences,
 Keio University, % Hiyoshi 4-1-1,
 Kanagawa 223-8521, Japan\\
 Fields, Gravity \& Strings, CTPU, Institute for Basic Science, Daejeon 34047, Korea
%affiliation

\vskip.5cm
\it $^{\dagger}$Yukawa Institute for Theoretical Physics, Kyoto University, Kyoto 606-8502, Japan %affiliation

\vskip.5cm
\it $^{\ddagger}$Department of Physics, Graduate School of Science, Osaka University, \\ Toyonaka, Osaka 560-0043, Japan} %affiliation
\end{center}

\vskip1cm
\begin{abstract} %abstract
We show the refinement of the prescription for the geometric transition in the refined topological string theory and, as its application, discuss a possibility to describe $qq$-characters from the string theory point of view. Though the suggested way to operate the refined geometric transition has passed through several checks, it is additionally found in this paper that the presence of the preferred direction brings a nontrivial effect. We provide the modified formula involving this point. We then apply our prescription of the refined geometric transition to proposing the stringy description of doubly quantized Seiberg--Witten curves called $qq$-characters in certain cases.
\end{abstract}

%%%%%%%%%%%%%%%%%%%%%%%%%%%%%%%%%%%%%%%%%%%%%%%%%%
\renewcommand{\thefootnote}{\arabic{footnote}}
\setcounter{footnote}{0}

\vfill\eject

\tableofcontents

%%%%%%%%%%%%%%%%%% section 1 %%%%%%%%%%%%%%%%%%%%%%%%%%%
\section{Introduction}
We have encountered the great developments of exact methods and a variety of their applications in quantum field theory, for instance, the Seiberg--Witten theory \cite{Seiberg:1994rs, Seiberg:1994aj} and the Nekrasov partition function for instanton counting problem \cite{Nekrasov:2002qd, Nekrasov:2003rj} as prominent landmarks, which are part of subjects in this paper. Correspondingly, the string theory and M-theory realization of these ingredients have been established and passed through a pile of checks in literatures. Specifically, (the 5d uplift of) the Nekrasov partition function can be systematically obtained by using the topological vertex \cite{Aganagic:2003db} that is a powerful ingredient to calculate the amplitude in the topological string theory \cite{Witten1988, Iqbal:2003ix, Iqbal:2003zz, Eguchi:2003sj} for a given Calabi--Yau threefold as the target space. The free energy of the topological string amplitude is expanded standardly with respect to the genus and the string coupling constant. The latter is translated into the $\Om$-backgrounds $( \e_1, \e_2 )$ in a special limit $\e_1 + \e_2 = 0$ which is called the unrefined (self-dual) limit. Since the Nekrasov partition function could be actually formulated for a general value of $( \e_1, \e_2 )$, the refined version of the topological vertex to include two parameters given by
\begin{align} %
q_1 = e^{2 \pi \mathrm{i} \e_1}, \hspace{2em} q_2 = e^{2 \pi \mathrm{i} \e_2}
\label{ombk}
\end{align}
has been suggested by \cite{Awata:2005fa, Iqbal:2007ii}, which was named the refined topological vertex\footnote{The convention here is translated into $( q, t )$ for the $\Om$-backgrounds used in \cite{Iqbal:2007ii} as $( q, t^{- 1} ) \to ( q_1, q_2 )$.}. Their definition could successfully reproduce the Nekrasov partition function with general $\Om$-background in many circumstances and bring us to meaningful outcomes from string theory to supersymmetric gauge theories (basically with eight supercharges).

It has been shown in \cite{Gopakumar:1998ii, Gopakumar:1998ki, Gopakumar:1998jq} that the open string and closed string sector in the usual (i.e. unrefined) topological sting theory is just linked by the geometric transition (open/closed duality). However, underlying physics for the geometric transition in the refined topological string theory that we would refer to as the refined geometric transition is not yet well understood mainly because there is no known world-sheet interpretation of it. Recently, great quantitative support for the refined geometric transition was reported by \cite{Kameyama:2017ryw}.

The prescription for geometric transition in terms of the refined topological vertex has been proposed \cite{Dimofte:2010tz} and basically checked in the context of the AGT correspondence \cite{Alday:2009aq, Alday:2009fs}, but it is not complete due to the possible choice of the so-called the preferred direction on the refined topological vertex. The topological vertex is graphically a trivalent vertex, and the proper point of the refined vertex different from the unrefined one is the existence of the preferred direction that is a special direction out of three edges of the vertex. This does result from the inclusion of $( q_1, q_2 )$ into the topological vertex. It is labeled by a Young diagram assigned on each edge, and as well, we pick up two of three edges to put $( q_1, q_2 )$ on. This means that the the preferred direction as the last edge has a special role on the computation of the refined topological string amplitude. In the first half of the paper, it will be argued that the refined geometric transition has to be sensitive to the choice of the preferred direction, and we will provide another prescription to implement the refined geometric transition on the web diagram constructed by vertices with the preferred direction that differs from the conventional one mentioned above.

In order to check the consistency of our prescription, we explore double quantization of the Seiberg--Witten geometry, which is called the $qq$-character, by utilizing the refined geometric transition.
The $qq$-character has been recently introduced by Nekrasov in the context of the BPS/CFT correspondence~\cite{Nekrasov:2015wsu,Nekrasov:2016qym,Nekrasov:2016ydq}.
It is a natural gauge theoretical generalization of the $q$-character of quantum affine algebra~\cite{Frenkel:1998}, corresponding to the Nekrasov--Shatashvili limit $(q_1,q_2) \to (e^\hbar, 0)$~\cite{Nekrasov:2009rc,Nekrasov:2013xda}, because the $qq$-character is obtained with generic $\Omega$-background parameter $(q_1,q_2)$.
There are a lot of interesting connections with, for example, quiver gauge theory construction of W-algebra (quiver W-algebra)~\cite{Kimura:2015rgi,Kimura:2016dys}%
\footnote{%
See also an overview article \cite{Kimura:2016ebq}.}, double affine Hecke algebra (DAHA) and Ding--Iohara--Miki (DIM) algebra~\cite{Bourgine:2015szm,Bourgine:2016vsq,Bourgine:2017jsi,Mironov:2016yue,Awata:2016riz}, and so on.

The $qq$-character plays a role as a generating function of the chiral ring operator, and is realized as a defect operator.
For example, it becomes a line operator in 5d gauge theory, which is a codimension-4 defect~\cite{Kim:2016qqs}.
In this paper we propose how to realize the $qq$-character in refined topological string by the brane insertion, analyzed using the refined geometric transition.
In particular, the codimension-2 defect operator, corresponding to the surface operator in gauge theory, is obtained by inserting a defect brane to the Lagrangian submanifold of the Calabi--Yau threefold~\cite{Kozcaz:2010af, Dimofte:2010tz, Taki:2010bj, Awata:2010bz, Bonelli:2011fq, Chen:2013dda}.
We show that the $\Ysf$-operator, which is a codimension-4 building block of the $qq$-character, can be constructed by inserting two codimension-2 defect operators.
Although the $\Ysf$-operator itself has a pole singularity, we obtain the $qq$-character, having no singularity, as a proper combination of $\Ysf$-operators.%
\footnote{%
The (log of) $\Ysf$-operator plays essentially the same role as the resolvent in matrix model, which is a generating function of the gauge invariant single-trace operator.}
The regularity of the $qq$-character is a nontrivial check of our prescription for refined geometric transition.

The remaining part of this paper is organized as follows:
In Sec.~\ref{sec:prescription} we propose a new prescription for geometric transition in refined topological string.
In order to obtain a proper contribution of the Lagrange submanifold, we have to consider the shift of parameters, which is not realized as a shift of the K\"ahler parameter, when the defect brane is inserted to the inner brane.
In Sec.~\ref{sec:qq-ch_geom} we apply the prescription of the refined geometric transition to the $qq$-character, which is a generating function of the chiral ring operator.
We examine several examples, especially $A_1$ and $A_2$ quivers, and obtain a consistent result with quiver gauge theory.
This shows a nontrivial check of our prescription of refined transition.
We conclude with summary and discussions in Sec.~\ref{sec:summary}.

%%%%%%%%%%%%%%%%%% section 2 %%%%%%%%%%%%%%%%%%%%%%%%%%%
\section{Geometric transition in the refined topological string}
\label{sec:prescription}
\begin{figure}[t] % web diagram without brane
	\begin{center}
	\includegraphics[width=5cm,bb=240 135 600 460,clip]{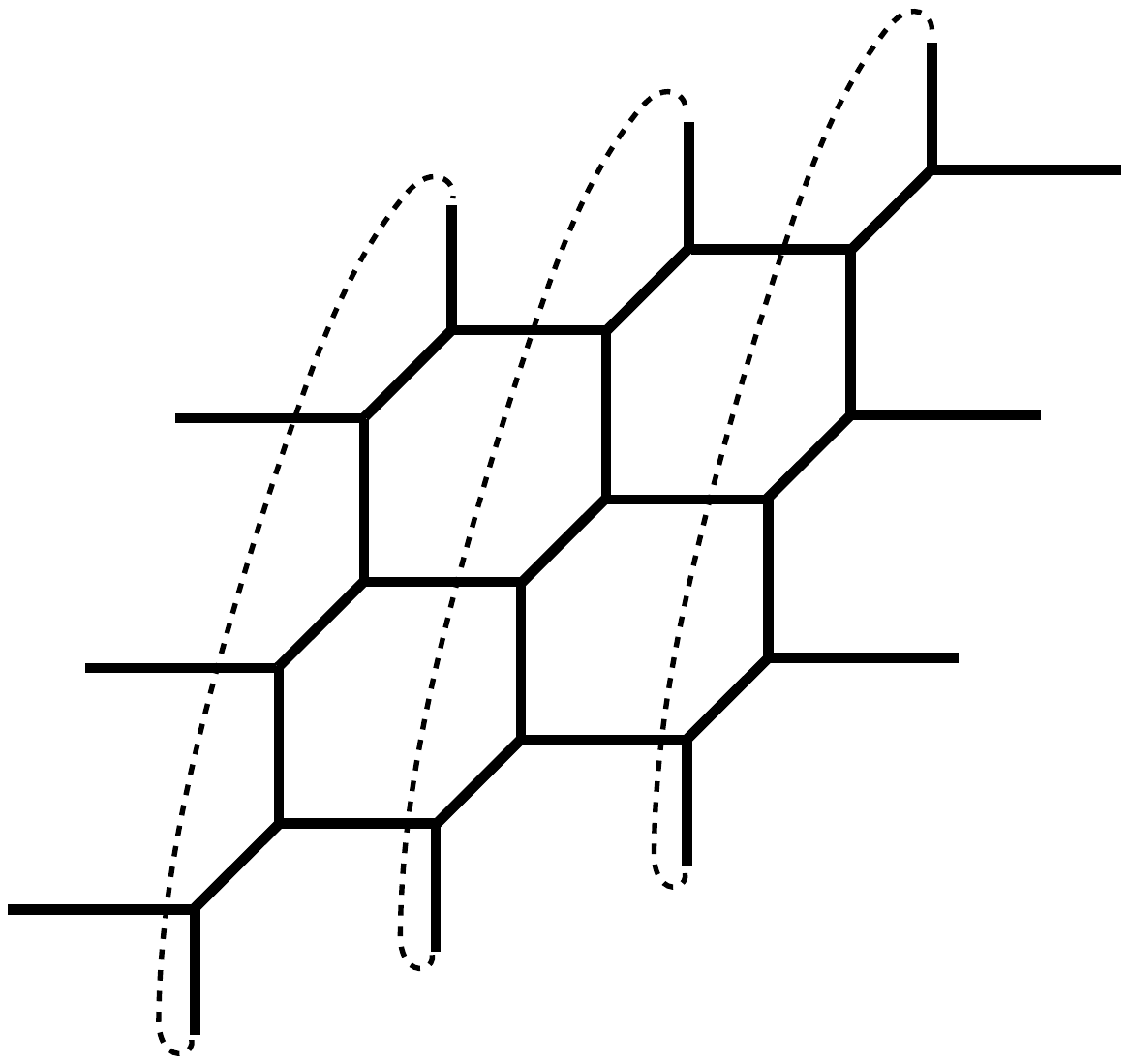}
	\caption{A compactified web diagram which we are considering in the paper.}
	\label{web0}
	\end{center}
\end{figure}

We would upgrade the operation of the geometric transition in the refined topological string theory where the partition function can be in principle evaluated by the refined topological vertex \cite{Awata:2005fa, Iqbal:2007ii} for a given Calabi--Yau geometry (see Appendix \ref{Rtv} for our convention). As there is a much wide variety of Calabi--Yau geometries, for simplicity and a purpose of the application to $qq$-characters, we restrict our argument to a simple class of the geometries visualized by a web diagram in Fig. \ref{web0}.
The thin dotted line connecting the upper and lower end of the diagram represents a compactified direction in the geometry. Note that this type is essentially equipped with the structure of the resolved conifold. It is known that this geometrical data can be dualized to type IIB string theory with D5-branes, NS5-branes, and $( 1, 1 )$-fivebranes.

One of crucial ingredients in calculating the refined topological string amplitude is the preferred direction on the refined topological vertex. It is an artificial technique for formalism, and final results with different choices of the preferred direction have to coincide (at least without any normalization). However, we claim in this paper that the refined geometric transition should be sensitive to where the preferred direction is set. To explain this point, at first we give a brief review of the prescription for the refined geometric transition that has been used in the literatures \cite{Kozcaz:2010af, Dimofte:2010tz, Taki:2010bj, Awata:2010bz, Bonelli:2011fq, Chen:2013dda} in Section \ref{Conv}, and then Section \ref{New} contains our proposal that actually clarifies the effect of the different selection of the preferred direction. The quantitative argument which we rely on is shown in Section \ref{Deri}.

%In this section, we consider the geometric transition in the refined topological string. First, we calculate the partition function of the refined topological string. After that, we consider the geometric transition by setting the K\"ahler parameters which is defined in the toric Calabi--Yau manifold. Then, we will find that we can observe the geometric transition by $t^{-1}q$-shift in some factors. 

%%%%%%%%%%%%%%%%%% section 2.1 %%%%%%%%%%%%%%%%%%%%%%%%%%
\subsection{Conventional prescription} \label{Conv}
Since there is no established world--sheet description of the refined topological string theory so far, one need to fix a guiding principle for the refined geometric transition from another context. One of frameworks to provide such a principle is the AGT correspondence \cite{Alday:2009aq} and its 5d uplift \cite{Awata:2009ur, Awata:2010yy}. This duality can be encoded into type IIB string theory presented by the $( p, q )$-fivebrane web diagram like Fig.\,\ref{web0}. The dictionary between the $( p, q )$-web and the geometry allows us to compute the partition function of the corresponding gauge theory by utilizing the refined topological vertex \cite{Awata:2005fa, Iqbal:2007ii}, which in the 4d limit turns out to be consistent with the correlation function on the 2d conformal field theory (CFT) side in some cases. Soon after finding the AGT relation, its statement has been extended to include the correspondence between a surface operator in the 4d $\Ncal = 2$ SU$(2)$ gauge theory and a degenerate field in the Liouville CFT \cite{Alday:2009fs}. This circumstance can also be realized in the framework of the $( p, q )$-web. The surface operator is engineered by inserting a D3-brane into the $( p, q )$-web, which is further mapped to a Lagrangian brane\footnote{This is often called a toric brane, however, we do not use this term in the paper since the concerned diagrams here are non-toric.} representing a Lagrangian submanifold in the corresponding Calabi--Yau. The computation of the topological string partition function must be incorporated with contributions from open strings when the target space is a Calabi--Yau with specified Lagrangian submanifolds. Although there is no established formula of the refined version of the open topological vertex, this can be evaluated by implementing the geometric transition. In the 4d limit, the result obtained in this way is actually compatible with the correlation function in the presence of a degenerate field in the Liouville CFT.

\begin{figure}[t] %
\begin{center}
\begin{tabular}{ccc}
\hspace{-2.5em}
	\begin{minipage}[b]{.4\hsize}
		\begin{center}
		\includegraphics[width=5.5cm,bb=220 100 580 470,clip]{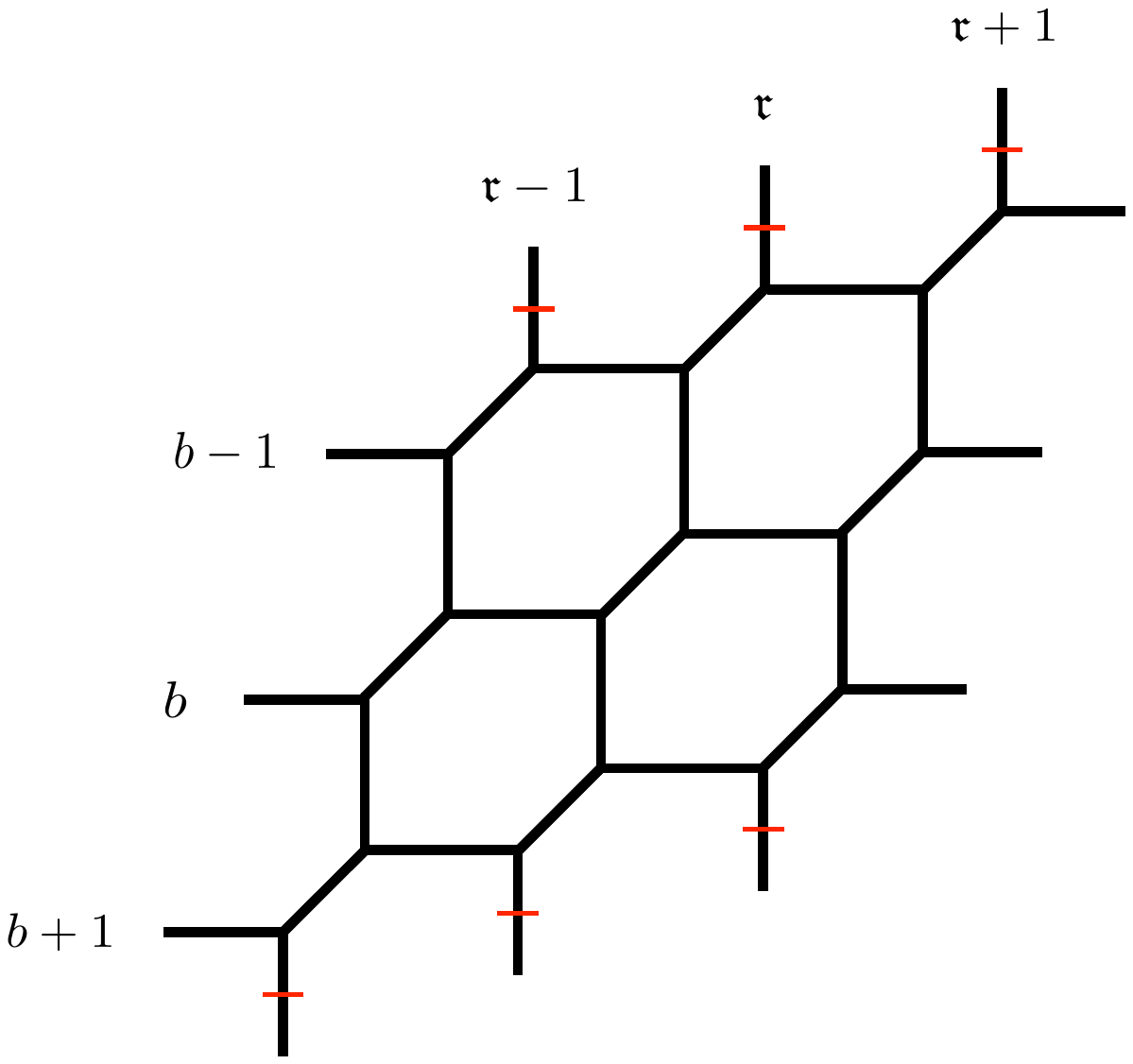}
		\end{center}
	\end{minipage}
	&
	\raisebox{7em}{$\longleftrightarrow$}
	&
	\hspace{-1em}
	\begin{minipage}[b]{.4\hsize}
		\begin{center}
		\includegraphics[width=5.5cm,bb=220 100 580 470,clip]{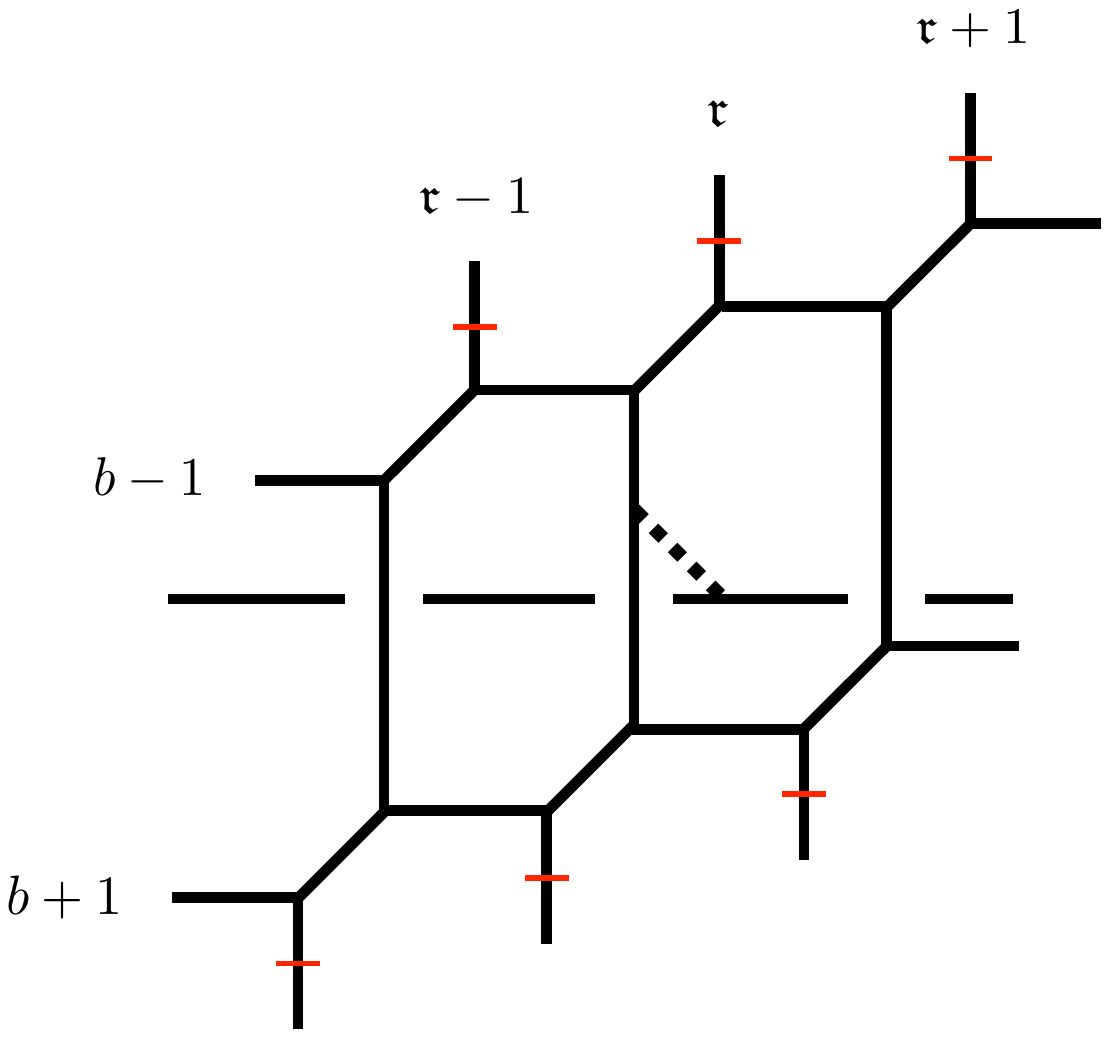}
		\end{center}
	\end{minipage}
\end{tabular}
\vspace{-1.5em}
\caption{\label{geoconv1}The geometric transition operated on the unpreferred direction. The horizontal axis is compactified and the dots indicate the preferred direction.}
\end{center}
\end{figure}

We would sketch concretely the rule of the refined geometric transition that has been lead from the AGT story. On the web diagram as shown in Fig.\,\ref{web0}, each internal line implies the topology of $\Cbb\mathbf{P}^1$ and is equipped with a K\"ahler modulus. Let $Q_{a}^{( \sfrak )}$ be a K\"ahler modulus for the $\sfrak$-th diagonal internal segment from the left in the $a$-th horizontal (uncompactified) line from the top (see Fig.\,\ref{web1} for our convention). The point of calculations along the AGT story with this web diagram is that the preferred direction is chosen on the vertical (compactified) direction, which is depicted as black dots in Fig.\,\ref{geoconv1} (throughout the paper, the vertical axis is always the compactified direction and the horizontal one is uncompactified). The geometric transition can be implemented with the horizontal (uncompactified) line: with appropriately tuning K\"ahler moduli for diagonal lines attached to the $b$-th horizontal line, this line is detached from the vertical lines and moved away. The geometric transition for the web diagram of our interest is essentially the same as that of the conifold, passing through from the resolved conifold to the deformed conifold and vice versa. If one would like to suspend a Lagrangian brane on the $\rfrak$-th vertical line in the process shown in Fig.\,\ref{geoconv1}, the K\"ahler moduli are specialized as\footnote{Note that the combination of $q_1$ and $q_2$ depends on ones convention.}
\begin{xalignat}{2} %
Q_{b}^{( \rfrak )} &= \frac{q_1^{m} q_2^{n}}{\sqrt{q_1 q_2}}, &
Q_{b}^{( \sfrak )} &= \frac{1}{\sqrt{q_1 q_2}} \hspace{.5em} \text{ for } \sfrak \neq \rfrak,
\label{gtunpre1}
\end{xalignat}
%\rem{The convention should be $(e^{\epsilon_1},e^{\epsilon_2}) = (q_1, q_2) = (t^{-1},q)$, and thus $qt^{-1} = q_1 q_2$. We also use $\beta = - \epsilon_1/\epsilon_2$ and $t=q^\beta$. The unrefined limit is given by $\epsilon_1 + \epsilon_2 = 0$; $q = 1$; $\beta=1$; $q = t$. -$>$ fixed so that $( q_1, q_2 ) = ( t, q^{-1} )$, compared with the previous paper \cite{Mori:2016qof}}
with $m, n \in \Zbb$. This prescription can nicely produce the AGT relation with the surface operator. Consequently, the refined geometric transition associated with the {\it unpreferred} direction is operated by \eqref{gtunpre1}.

We are closing the review with commenting on the integers $m, n$ in \eqref{gtunpre1}. It has been argued in \cite{Dimofte:2010tz} that, in the 4d limit, the adjustment such that $m, n > 0$ might produce general (non--elementary) surface operators supported on the surface described by
\begin{align} %
z_1^m z_2^n = 0,
\label{gtunpre3}
\end{align}
where $( z_1, z_2 )$ are complex coordinates on two-dimensional planes respecting the rotation by the $\Om$-background parameters $( \e_1, \e_2 )$, respectively. This discussion seems to work for such physical surface operators, however, for the present we do not have a requirement to restrict the range of $m, n$ to be non-negative from the refined topological string point of view. This is why we take $m, n$ to run for all integers. Although the refined geometric transition with $m, n < 0$ would engineer the unphysical surface operators in the sense that these do not follow the standard discussion \eqref{gtunpre3}, such branes at least in the unrefined ($q_1 q_2 = 1$) context are referred to as anti-branes \cite{Vafa:2001qf}. We would return to this point in Section \ref{sec:summary}.

%We comment on the generalization of the operation \eqref{gtunpre1}. This tuning of the K\"ahler moduli generates the Lagrangian brane that turns to be the simplest (or an elementary in \cite{Dimofte:2010tz}) surface operator in the 4d gauge theory supported on the plane respecting the rotation by one of the $\Om$-background parameters. It is naturally expected that, through the refined geometric transition, more general surface operators might be engineered by adjusting them as
%\begin{align} %
%Q_{a}^{( s )} = q_1^{\alpha} q_2^{\beta} ( q_1 q_2 )^{1 \over 2} \hspace{1em} \text{ with} \hspace{1em} \alpha, \beta \in \Zbb_{\geq 0}.
%\label{gtunpre2}
%\end{align}
%Consequently, the surface operator would have integer labels $( \alpha, \beta )$. This is still highly speculation, and we need more precise verification for this generalization.

%%%%%%%%%%%%%%%%%% section 2.2 %%%%%%%%%%%%%%%%%%%%%
\subsection{New prescription} \label{New}
\begin{figure}[t] %
\begin{center}
\begin{tabular}{ccc}
\hspace{-2.5em}
	\begin{minipage}[b]{.4\hsize}
		\begin{center}
		\includegraphics[width=5.5cm,bb=220 100 580 470,clip]{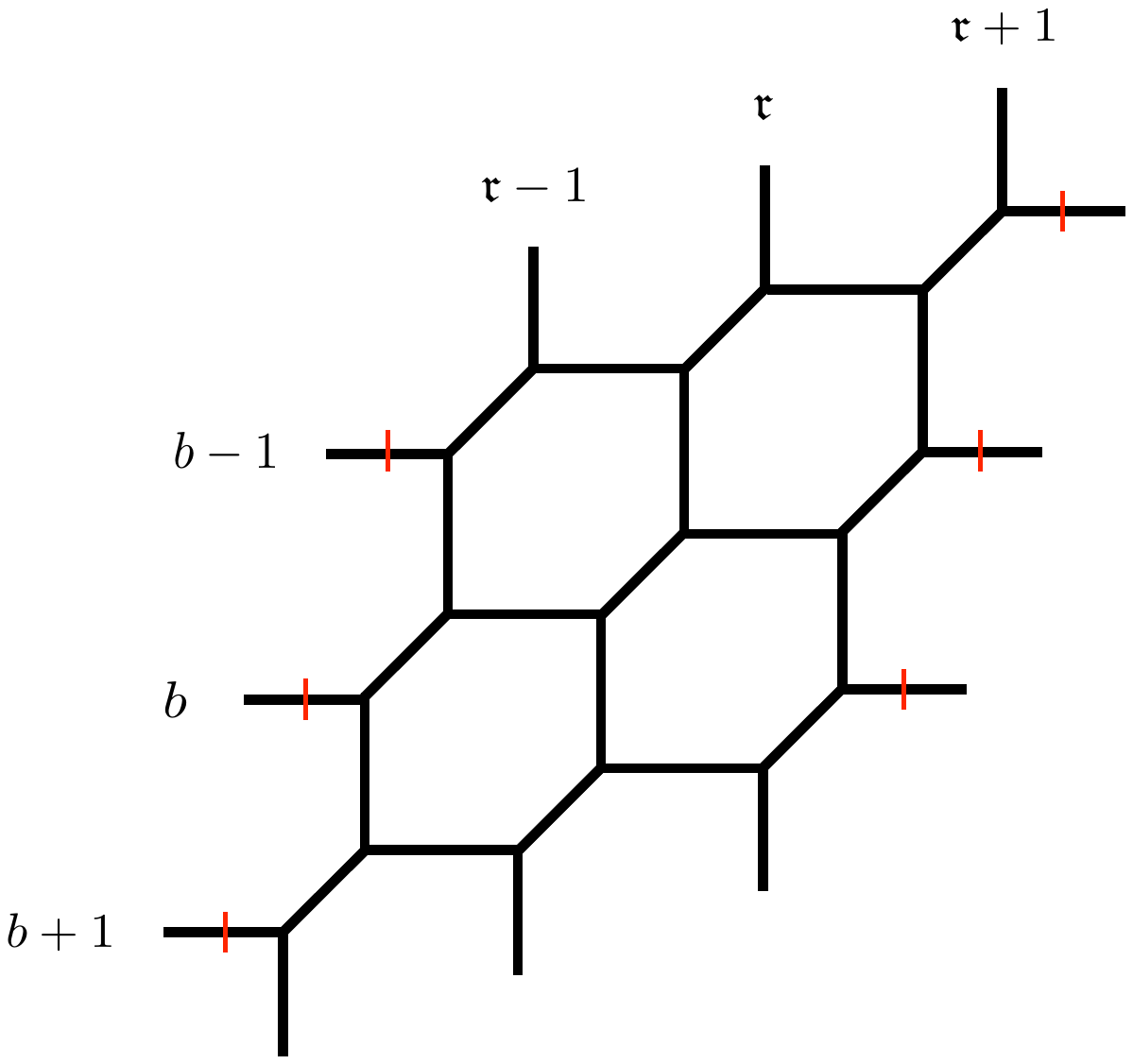}
		\end{center}
	\end{minipage}
	&
	\raisebox{7em}{$\longleftrightarrow$}
	&
	\hspace{-1em}
	\begin{minipage}[b]{.4\hsize}
		\begin{center}
		\includegraphics[width=5.5cm,bb=220 100 580 470,clip]{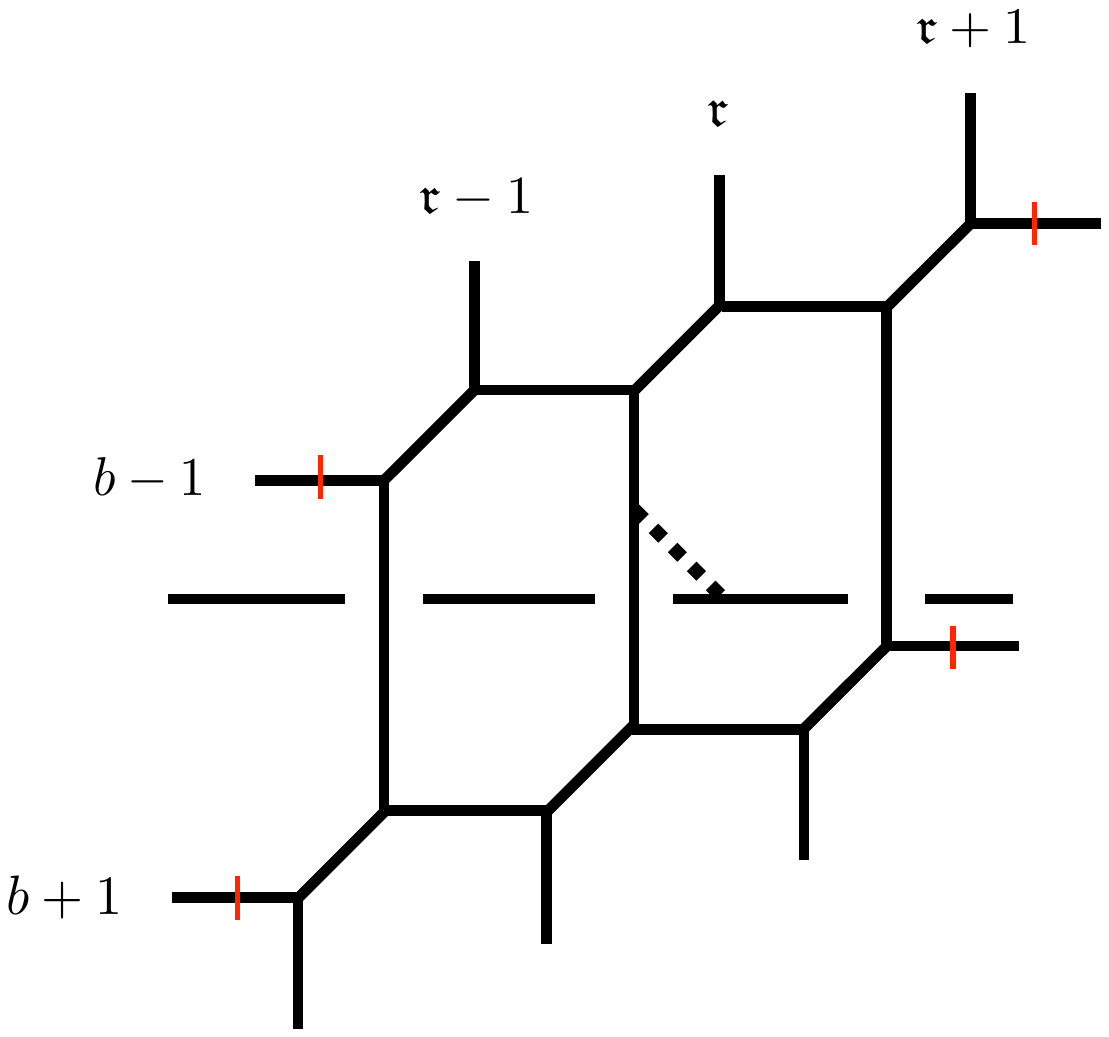}
		\end{center}
	\end{minipage}
\end{tabular}
\vspace{-1.5em}
\caption{\label{geonew1}The geometric transition operated on the preferred direction. The horizontal axis is compactified and the dots indicate the preferred direction.}
\end{center}
\end{figure}

We turn to giving our new prescription for the refined geometric transition that takes the issues of the preferred direction into account. On computing the refined topological string amplitude for the web diagram of our main interest, the preferred direction is chosen along the horizontal, i.e., uncompactified direction, marked by dots in Fig.\,\ref{geonew1}. The difference of the preferred direction from the previous situation requires us to introduce small modification for the refined geometric transition. In this subsection, we write down the process to implement the refined geometric transition for the current choice of the preferred direction.
%Let us turn to giving our new prescription for the refined geometric transition that takes the issues of the preferred direction into account. On computing the refined topological string amplitude for the web diagram of our main interest, the preferred direction is set along the horizontal, i.e., uncompactified direction, that is again marked by red fragments in Fig.\,\ref{geonew1}. This difference from the previous setting requires us to become careful and come up with the refined treatment for the refined geometric transition. In this subsection, we only describe formulas how to implement the refined geometric transition for the current choice of the preferred direction.

A point which we should stress is to put the {\it preferred} direction on the uncompactified (horizontal) direction where the geometric transition can be carried out. In addition, for consistency, it is required that the contributions from the Lagrangian brane is not produced if the web with $( M, N )$ lines simply reduces to the one with $( M, N - 1 )$ lines without the Lagrangian brane after the geometric transition, where $M$ and $N$ stand for the number of compactified (vertical) and uncompactified (horizontal) lines, respectively (see Fig.\,\ref{web1a}).
%A point which we should stress is to put the {\it preferred} direction on the uncompactified direction where the geometric transition can be carried out. In addition, it is required for consistency that the contributions from the Lagrangian brane do not arise if the geometric transition is not accompanied with it, in other words, the partition function for the web with $( M, N )$ lines simply reduces to the one for the web with $( M, N - 1 )$ lines after the geometric transition without the Lagrangian brane, where $M$ and $N$ stand for the number of compactified (vertical) and uncompactified (horizontal) lines, respectively (see Fig.\,\ref{web1a}).

Let us consider the geometric transition that is executed on the $b$-th horizontal line with the Lagrangian brane emerging on the $\rfrak$-th vertical line (Fig.\,\ref{geonew1}). Our proposal for the refined geometric transition under the above requirement is comprised of the following three steps:
%Consider the geometric transition executed on the $b$-th horizontal (uncompactified) line with the Lagrangian brane emerging on the $\rfrak$-th vertical (compactified) line (Fig.\,\ref{geonew1}). Our proposal for the refined geometric transition under the above requirement is comprised of the following three steps:
\begin{enumerate} % prescription
\setcounter{enumi}{-1}
\item As a supposition the preferred direction is taken to be the uncompactified axis (horizontal here), and then one computes the refined closed topological string amplitude as done in \cite{Haghighat:2013gba, Haghighat:2013tka}.
%The preferred direction is taken to be the horizontal (uncompactified) axis as a supposition, and then one computes the refined closed topological string amplitude as done in \cite{Haghighat:2013gba, Haghighat:2013tka}.

\item For $\sfrak \geq \rfrak$, variables $w_{a b}^{( \sfrak )} ( i, j )$ and $u_{a b}^{( \sfrak )} ( i, j )$ defined in \eqref{deri9}, which appear in the generic partition function \eqref{TSpart.fn.}, are shifted by using the inversion \eqref{inversiont1} and the difference equation \eqref{difft1} of the theta function,
%For $\sfrak \geq \rfrak$, variables $w_{a b}^{( \sfrak )} ( i, j )$ and $u_{a b}^{( \sfrak )} ( i, j )$ in the partition function given by \eqref{TSpart.fn.} are shifted by using the inversion \eqref{inversiont1} and the difference equation \eqref{difft1} of the theta function,
\begin{align} %
    \begin{aligned} %
    \theta_1 \lp e^{2 \pi \mathrm{i} w_{a b}^{( \sfrak )} ( i, j )} \rp
    &=
    Q_\tau^{1/2} e^{2 \pi \mathrm{i} w_{a b}^{( \sfrak )} ( i, j )}
    \times
    \theta_1 \lp Q_\tau^{- 1} e^{- 2 \pi \mathrm{i} w_{a b}^{( \sfrak )} ( i, j )} \rp, \\ %1
    \theta_1 \lp e^{2 \pi \mathrm{i} u_{a b}^{( \sfrak )} ( i, j )} \rp
    &=
    Q_\tau^{1/2} e^{2 \pi \mathrm{i} u_{a b}^{( \sfrak )} ( i, j )}
    \times
    \theta_1 \lp Q_\tau^{- 1} e^{- 2 \pi \mathrm{i} u_{a b}^{( \sfrak )} ( i, j )} \rp. %2
    \end{aligned}
\label{gtpre3}
\end{align}

\item Then, tuning the K\"ahler moduli as
\begin{xalignat}{3} %
Q_{b}^{( \sfrak )} &= \frac{1}{\sqrt{q_1 q_2}} \hspace{1.5em} ( \sfrak < \rfrak ), &
Q_{b}^{( \rfrak )} &= \frac{q_1^m q_2^n}{\sqrt{q_1 q_2}}, &
Q_{b}^{( \sfrak )} &= \sqrt{q_1 q_2} \hspace{1.5em} ( \sfrak > \rfrak ),
\label{gtpre1}
\end{xalignat}
with $m, n \in \Zbb$.

\item Finally, shifting variables $w_{a b}^{( \rfrak )} ( i, j )$ for all $a$ by hand,
\begin{align} %
w_{a b}^{( \rfrak )} ( i, j )
\to
w_{a b}^{( \rfrak )} ( i, j ) - \e_1 - \e_2,
\label{gtpre2}
\end{align}
while others in \eqref{deri9} are kept unchanged.
%or equivalently,
%\begin{align} %
%\lp Q_{b a}^{( \rfrak )} \rp^{- 1}
%\to
%\lp Q_{b a}^{( \rfrak )} q_1 q_2 \rp^{- 1},
%\end{align}
%where $Q_{b a}^{( \rfrak )}$ is a certain product of K\"ahler moduli assigned on the $\rfrak$-th line, which is defined by \eqref{deri9}.

\end{enumerate}
We should make a comment on the shift of step 3 in our prescription. The shift \eqref{gtpre2} has nothing to do with the K\"ahler parameters: any K\"ahler parameter is not shifted together with this operation, but rather, with viewing $w_{a b}^{( \rfrak )} ( i, j )$ as a single variable, it is just to add $- \e_1 - \e_2$ to it. This is purely a technical thing which is originated from the difference of the specialization \eqref{gtpre1} of the K\"ahler moduli $Q_b^{( \sfrak )}$ for $\sfrak < \rfrak$ and $\sfrak > \rfrak$. The reason why we need this shift is to satisfy the requirement for consistency that the refined geometric transition without generating a Lagrangian brane reproduces the closed topological string amplitude (see below for numerical details). The step 2 and 3 reflect the dependence of the refined geometric transition on the preferred direction. Indeed, it is expected that, even though the closed topological string amplitude should be independent of the preferred direction, the open one really depends on whether or not the Lagrangian brane is attached to the preferred direction (see, e.g., \cite{Gaiotto:2014ina, Zenkevich:2016xqu}). This is basically because the Lagrangian brane can end on the $( p, q )$-fivebrane with general $( p, q )$, therefore the geometric transition should be characterized by $( p, q )$ in addition to $( m, n )$. This implies that the position of the preferred direction put on the $( p, q )$-fivebrane leads to the inequivalent result of the open topological string amplitude. Both procedures of the refined geometric transition can reproduce correctly the identical result in the unrefined limit $q_1 q_2 = 1$ as expected. Our prescription seems compatible with this suggestion.

A Lagrangian brane appears on only one vertical line upon a single sequence of the above geometric transition. If one desires to generate several Lagrangian branes on different vertical lines for a web diagram, it is necessary to consider a bigger web and repeat the procedure \eqref{gtpre3}-\eqref{gtpre2} many times (as demonstrated in Section \ref{sec:qq-ch_geom}).

We will devote the next subsection to showing quantitative clarification how this process works and produces the refined topological string amplitude incorporating the contribution of the Lagrangian brane. In Section \ref{sec:qq-ch_geom}, it will be discussed that the refined geometric transition initiated by our prescription gives possibly how to realize the $qq$-character from string theory.

\subsection{Derivation} \label{Deri}
Our prescription given above seems a bit intricate rather than \eqref{gtunpre1}, and we would explain here why this works when the uncompactified line assigned with the preferred direction is removed upon the geometric transition.

%%%%%%%%%%%%%%%%%% section 2.3.1 %%%%%%%%%%%%%%%%%%%%
\subsubsection{General formula for the partition function}
\begin{figure}[t] %web diagram
\begin{center}
\begin{tabular}{cc}
	\hspace{-5em}
	\begin{minipage}[b]{.5\hsize}
		\begin{center}
		\subfigure[Entire web]{\label{web1a}\includegraphics[width=10cm,bb=100 70 710 550,clip]{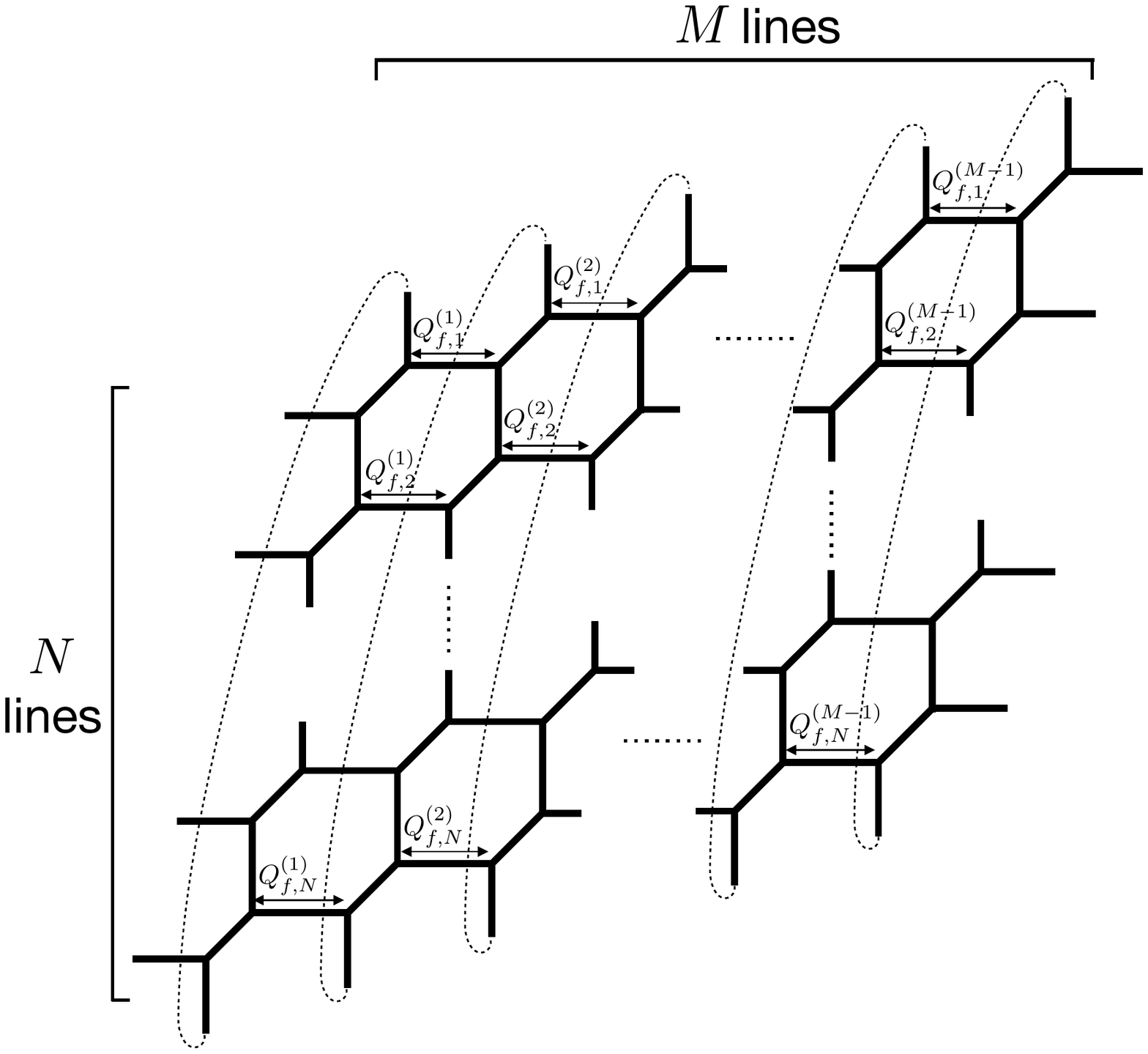}}
		\end{center}
	\end{minipage}
	&
	\hspace{4.5em}
	\begin{minipage}[b]{.5\hsize}
		\begin{center}
		\subfigure[Internal hexagons]{\label{web1b}\includegraphics[width=7.5cm,bb=260 100 590 470,clip]{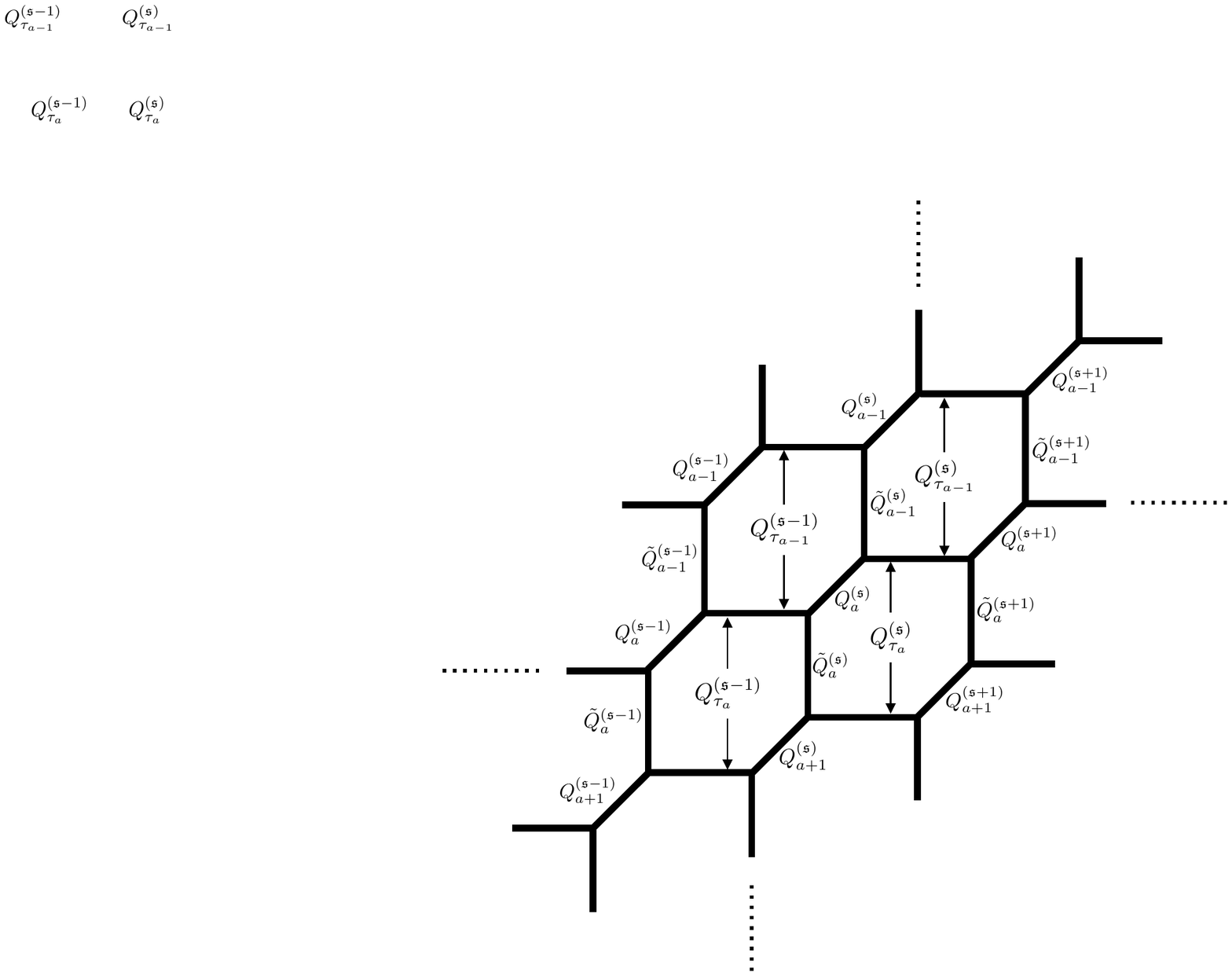}}
		\end{center}
	\end{minipage}
\end{tabular}
%\vspace{-1.5em}
\caption{\label{web1}The web diagram with $M$ vertical and $N$ horizontal lines.}
\end{center}
\end{figure}

We are now concentrating on the compactified web shown as Fig.\,\ref{web1} with general $( M, N )$ lines. On the technique of the refined topological vertex \eqref{rtv}, the partition function $\Zcal_{M, N}$ for this web diagram has been derived as \cite{Haghighat:2013tka}
\begin{align} %
\mathcal{Z}_{M, N}
=
\sum_{\{ \mu_{a}^{( \sfrak )} \}_{a = 1, \cdots, N}^{\sfrak = 1, \cdots, M - 1}}
\prod_{\sfrak = 1}^{M - 1}
\prod_{a = 1}^{N}
\lp \bar{Q}_{f, a}^{( \sfrak )} \rp^{|\mu^{( \sfrak )}_{a}|}
\prod_{( i, j ) \in \mu^{( \sfrak )}_{a}}
\prod_{b = 1}^{N}
{
\theta_{1} ( e^{2 \pi \mathrm{i} z_{a b}^{( \sfrak )} ( i, j )} )
\theta_{1} ( e^{2 \pi \mathrm{i} w_{a b}^{( \sfrak )} ( i, j )} )
\over
\theta_{1} ( e^{2 \pi \mathrm{i} u_{a b}^{( \sfrak )} ( i, j )} )
\theta_{1} ( e^{2 \pi \mathrm{i} v_{a b}^{( \sfrak )} ( i, j )} )
},
\label{TSpart.fn.}
\end{align}
where
\begin{align} %
    \begin{aligned}
    e^{2 \pi \mathrm{i} z_{a b}^{( \sfrak )} ( i, j )}
    &=
    \lp Q_{a b}^{( \sfrak + 1 )} \rp^{- 1}
    q_1^{- \mu_{b, j}^{( \sfrak + 1 ), t} + i - 1/2}
    q_2^{\mu_{a, i}^{( \sfrak )} - j + 1/2}, \\ %1
    e^{2 \pi \mathrm{i} w_{a b}^{( \sfrak )} ( i, j )}
    &=
    \lp Q_{b a}^{( \sfrak )} \rp^{- 1}
    q_1^{\mu_{b, j}^{( \sfrak - 1 ), t} - i + 1/2}
    q_2^{- \mu_{a, i}^{( \sfrak )} + j - 1/2}, \\ %2
    e^{2 \pi \mathrm{i} u_{a b}^{( \sfrak )} ( i, j )}
    &=
    \lp \hat{Q}_{b a}^{( \sfrak )} \rp^{- 1}
    q_1^{\mu_{b, j}^{( \sfrak ), t} - i + 1}
    q_2^{- \mu_{a, i}^{( \sfrak )} + j}, \\ %3
    e^{2 \pi \mathrm{i} v_{a b}^{( \sfrak )} ( i, j )}
    &=
    \lp \hat{Q}_{a b}^{( \sfrak )} \rp^{- 1}
    q_1^{- \mu_{b, j}^{( \sfrak ), t} + i}
    q_2^{\mu_{a, i}^{( \sfrak )} - j + 1}, %4
    \end{aligned}
    \label{deri9}
\end{align}
with $t$ representing the transpose of the Young diagram (Fig.\,\ref{yd3}). We collect the definitions and notations in Appendix \ref{secA}. Note that, for simplicity, we omit a complex parameter $Q_\tau := e^{2 \pi \mathrm{i} \tau}$ in the theta function as $\theta_1 ( x )$ unless otherwise stated. We denote the K\"ahler moduli for diagonal, vertical, and horizontal internal segments by $Q_{a}^{( \sfrak )}$, $\tilde{Q}_{a}^{( \sfrak )}$, and $Q_{f, a}^{( \sfrak )}$, respectively, which are visualized in Fig.\,\ref{web1a} and \ref{web1b}. The weights in the partition function, corresponding to instanton counting parameters in the Nekrasov partition function, are given by
\begin{align} %
\bar{Q}_{f, a}^{( \sfrak )}
=
\lp q_1 q_2 \rp^{N - 1 \over 2} Q_{f, a}^{( \sfrak )}
\prod_{b = 1}^{N} Q_{b}^{( \sfrak )}.
\end{align}
Also, we use the simplified symbols for the products of the K\"ahler moduli in the variables \eqref{deri9}, defined as
\begin{align} %
Q_{a b}^{( \sfrak )}
&=
\lc \begin{aligned}
    & Q_{a}^{( \sfrak )} \prod_{i = b}^{N} Q_{\tau_{i}}^{( \sfrak )}
    && (\text{mod } Q_{\tau})
    && \text{for } a = 1, \\ %1
    & Q_{a}^{( \sfrak )} \prod_{i = 1}^{a - 1} Q_{\tau_{i}}^{( \sfrak )} \prod_{j = b}^{N} Q_{\tau_{j}}^{( \sfrak )}
    && (\text{mod } Q_{\tau})
    && \text{for } a \neq 1, %2
    \end{aligned} \right.
\label{def}
\end{align}
for the numerator, and
\begin{align} %
\hat{Q}_{a b}^{( \sfrak )}
&=
\lc \begin{aligned}
    & \prod_{i = b}^{a - 1} Q_{\tau_{i}}^{( \sfrak )}
    && \text{for } a > b, \\ %1
    & 1
    && \text{for } a = b, \\ %2
    & Q_{\tau} \left/ \prod_{i = a}^{b - 1} Q^{( \sfrak )}_{\tau_{i}} \right.
    && \text{for } a < b, %3
    \end{aligned} \right.
%\notag \\ %1
%\tilde{Q}_{a b}^{( \ssf )}
%&=
%\lc \begin{aligned}
%    & \prod_{i = b}^{a - 1} Q_{\tau_{i}}^{( \ssf )}
%    && \text{for } a > b, \\ %
%    %& Q_{\tau}
%    %&& \text{for } a = b, \\ %
%    & Q_{\tau} \left/ \prod_{i = a}^{b - 1} Q^{( \ssf )}_{\tau_{i}} \right.
%    && \text{for } a < b. %
%    \end{aligned} \right. %2
\end{align}
for the denominator, where
\begin{align} %
Q^{( \sfrak )}_{\tau_{i}} := Q_{i}^{( \sfrak )} \tilde{Q}_{i}^{( \sfrak )} = \tilde{Q}_{i}^{( \sfrak + 1 )} Q_{i + 1}^{( \sfrak + 1 )}.
\label{deri4}
\end{align}
The second equality follows from the consistency condition to form internal hexagons (Fig.\,\ref{web1b}). It has been revealed that this geometry actually realizes an elliptically fibered Calabi--Yau with the complex modulus $Q_\tau$ identified as
\begin{align} %
Q_\tau = \prod_{a = 1}^{N} Q_{\tau_a}^{( \sfrak )} \hspace{2em} \text{for } \forall \sfrak.
\label{deri8}
\end{align}

We comment on the M-theory uplift of this picture. It is well known that type IIB string theory compactified on $S^1$ is dual to M-theory on the torus $T^2$. The web as in Fig.\,\ref{web1} is rendered to the M-theory brane configuration where the stacks of M2-branes are suspended between separated $M$ M5-branes on an asymptotically locally Euclidean (ALE) space equipped with $\Zbb_N$ orbifolding. This duality supports the fact that the low energy theory on the present $( p, q )$-web are described by the tensor branch of the corresponding 6d $\Ncal = ( 1 ,0 )$ theory. It has been argued in \cite{Haghighat:2013gba, Haghighat:2013tka} that the partition function \eqref{TSpart.fn.} captures the spectra of self-dual strings, called M-strings, wrapping $T^2$ in the 6d theory, and the Young diagrams $\mu_{a}^{( \sfrak )}$ label the ground states of M-strings.

%%%%%%%%%%%%%%%%%% section 2.3.2 %%%%%%%%%%%%%%%%%%%%
\subsubsection{Actual process of the geometric transition}
When we perform the geometric transition on the $b$-th horizontal line such that the Lagrangian brane appears on the $\rfrak$-th vertical line, the main contribution that should be carefully treated is
\begin{align} %
\prod_{( i, j ) \in \mu_{a}^{( \sfrak )}}
{
\theta_1 \lp e^{2 \pi \mathrm{i} z_{a b}^{( \sfrak )} ( i, j )} \rp
\theta_1 \lp e^{2 \pi \mathrm{i} w_{a b}^{( \sfrak )} ( i, j )} \rp
\over
\theta_1 \lp e^{2 \pi \mathrm{i} u_{a b}^{( \sfrak )} ( i, j )} \rp
\theta_1 \lp e^{2 \pi \mathrm{i} v_{a b}^{( \sfrak )} ( i, j )} \rp
}
\label{deri5}
\end{align}
for all $a$. We would divide the calculation process for this into two parts with $\sfrak < \rfrak$ and $\sfrak \geq \rfrak$. Remark that we sometimes implicitly use the relation \eqref{deri8} to change the variables.

\subsubsection*{For $\sfrak < \rfrak$} %%%%%%%%%%%%%%%%%%%%%%%%%%%
We firstly focus on the sector for $\sfrak < \rfrak$ where the geometric transition \eqref{gtpre1} can straightforwardly work. One can easily see that \eqref{deri5} does not produce the nontrivial contribution unless
\begin{align} %
\mu_{b}^{( \sfrak )} = \emptyset \text{ for } \forall \sfrak.
\label{deri7}
\end{align}
Accordingly, this condition is necessary in order to obtain the appropriate result for the partition function obtained via the geometric transition. Then, variables $z_{a b}^{( \sfrak )} ( i, j )$ and $w_{a b}^{( \sfrak )} ( i, j )$ can be rewritten as
\begin{align} %
\theta_1 \lp e^{2 \pi \mathrm{i} z_{a b}^{( \sfrak )} ( i, j )} \rp
&=
\theta_1 \lp \lp \sqrt{q_1 q_2} Q_b^{( \sfrak + 1 )} \rp^{- 1} e^{2 \pi \mathrm{i} v_{a b}^{( \sfrak )} ( i, j )} \rp, \\ %1
\theta_1 \lp e^{2 \pi \mathrm{i} w_{a b}^{( \sfrak )} ( i, j )} \rp
&=
\theta_1 \lp \lp \sqrt{q_1 q_2} Q_b^{( \sfrak )} \rp^{- 1} e^{2 \pi \mathrm{i} u_{a b}^{( \sfrak )} ( i, j )} \rp, %2
\end{align}
where we used the relation \eqref{deri4} for $z_{a b}^{( \sfrak )} ( i, j )$. With these expressions, the specialization \eqref{gtpre1} of the K\"ahler moduli results in
\begin{align} %
\eqref{deri5} \to 1 \hspace{2em} \text{for } \sfrak < \rfrak - 1,
\end{align}
and
\begin{align} %
\eqref{deri5}
\to
\prod_{( i, j ) \in \mu_{a}^{( \rfrak - 1 )}}
{
\theta_1 \lp q_1^{m} q_2^{n} e^{2 \pi \mathrm{i} v_{a b}^{( \rfrak - 1 )} ( i, j )} \rp
\over
\theta_1 \lp e^{2 \pi \mathrm{i} v_{a b}^{( \rfrak - 1 )} ( i, j )} \rp
}
\hspace{2em} \text{for } \sfrak = \rfrak - 1.
\label{deri6}
\end{align}
Indeed, \eqref{deri6} is the half of the contributions of the Lagrangian brane. This is just what we want because this reduces to $1$ when $m = n = 0$, namely, no Lagrangian brane appear, as required. Actually, this expression matches with the result of \cite{Mori:2016qof}. Moreover, the weights in the sum of the Young diagrams change as
\begin{align} %
\bar{Q}_{f, a}^{( \sfrak )}
\to
\lp q_1 q_2 \rp^{( N - 1 ) - 1 \over 2}
Q_{f, a}^{( \sfrak )}
\prod_{\shortstack{${\scriptstyle c = 1 }$ \\ ${\scriptstyle c \neq b }$ }}^{N}
Q_c^{( \sfrak )},
\label{deri10}
\end{align}
and this is nothing but the ones in the partition function for the web diagram with $( M, N - 1 )$ lines. Our prescription for the refined geometric transition appropriately works for $\sfrak < \rfrak$.

\subsubsection*{For $\sfrak \geq \rfrak$} %%%%%%%%%%%%%%%%%%%%%%%%%
Let us turn to the sector for $\sfrak \geq \rfrak$. In addition to the first step \eqref{gtpre3}, by using \eqref{deri4}, the relation
\begin{align} %
\theta_1 \lp e^{2 \pi \mathrm{i} z_{a b}^{( \sfrak )} ( i, j )} \rp
&=
\theta_1 \lp
\lp \frac{Q_{b}^{( \sfrak + 1 )} Q_\tau}{\sqrt{q_1 q_2}} \rp^{- 1} e^{- 2 \pi \mathrm{i} u_{a b}^{( \sfrak )} ( i, j )} \rp
\end{align}
holds under the restriction \eqref{deri7}. As a result, we have
\begin{align} %
\eqref{deri5}
&\to
\prod_{( i, j ) \in \mu_{a}^{( \sfrak )}}
e^{2 \pi \mathrm{i} ( w_{a b}^{( \sfrak )} ( i, j ) - u_{a b}^{( \sfrak )} ( i, j ) )}
{
\theta_1 \lp \lp \frac{Q_{b}^{( \sfrak + 1 )}}{\sqrt{q_1 q_2}} \rp^{- 1} Q_\tau^{- 1} e^{- 2 \pi \mathrm{i} u_{a b}^{( \sfrak )} ( i, j )} \rp
\theta_1 \lp Q_\tau^{- 1} e^{- 2 \pi \mathrm{i} w_{a b}^{( \sfrak )} ( i, j )} \rp
\over
\theta_1 \lp Q_\tau^{- 1} e^{- 2 \pi \mathrm{i} u_{a b}^{( \sfrak )} ( i, j )} \rp
\theta_1 \lp e^{2 \pi \mathrm{i} v_{a b}^{( \sfrak )} ( i, j )} \rp
} \notag \\ %1
&=
\lp \sqrt{q_1 q_2} Q_{b}^{( \sfrak )} \rp^{- |\mu_{a}^{( \sfrak )}|}
\prod_{( i, j ) \in \mu_{a}^{( \sfrak )}}
{
\theta_1 \lp \lp \frac{Q_{b}^{( \sfrak + 1 )}}{\sqrt{q_1 q_2}} \rp^{- 1} Q_\tau^{- 1} e^{- 2 \pi \mathrm{i} u_{a b}^{( \sfrak )} ( i, j )} \rp
\theta_1 \lp Q_\tau^{- 1} e^{- 2 \pi \mathrm{i} w_{a b}^{( \sfrak )} ( i, j )} \rp
\over
\theta_1 \lp Q_\tau^{- 1} e^{- 2 \pi \mathrm{i} u_{a b}^{( \sfrak )} ( i, j )} \rp
\theta_1 \lp e^{2 \pi \mathrm{i} v_{a b}^{( \sfrak )} ( i, j )} \rp
}, %2
\label{deri1}
\end{align}
Similarly for \eqref{deri10}, the overall factor can be absorbed into the associated weight so that
\begin{align} %
\bar{Q}_{f, a}^{( \sfrak )}
\lp \sqrt{q_1 q_2} Q_{b}^{( \sfrak )} \rp^{- 1}
=
\lp q_1 q_2 \rp^{( N - 1 ) - 1 \over 2}
Q_{f, a}^{( \sfrak )}
\prod_{\shortstack{${\scriptstyle c = 1 }$ \\ ${\scriptstyle c \neq b }$ }}^{N}
Q_c^{( \sfrak )},
\end{align}
which becomes the one for the web diagram with $( M, N - 1 )$ lines. This is the actual reason why our prescription needs the first step \eqref{gtpre3}. Then, the parameter tuning \eqref{gtpre1} as the second step leads to
\begin{align} %
\eqref{deri1} \to 1 \hspace{2em} \text{for } \sfrak > \rfrak,
\end{align}
and
\begin{align} %
\eqref{deri1}
&\to
\prod_{( i, j ) \in \mu_{a}^{( \rfrak )}}
{
\theta_1 \lp Q_\tau^{- 1} e^{- 2 \pi \mathrm{i} ( w_{a b}^{( \rfrak )} ( i, j ) - \e_1 - \e_2 )} \rp
\over
\theta_1 \lp e^{2 \pi \mathrm{i} v_{a b}^{( \rfrak )} ( i, j )} \rp
} \notag \\ %1
&=
\prod_{( i, j ) \in \mu_{a}^{( \rfrak )}}
{
\theta_1 \lp Q_\tau^{- 1} Q_{b a}^{( \rfrak )} q_1^{i + 1/2} q_2^{\mu_{a, i}^{( \rfrak )} - j + 3/2} \rp
\over
\theta_1 \lp Q_\tau^{- 1} Q_{b a}^{( \rfrak )} q_1^{i + 1/2 + m} q_2^{\mu_{a, i}^{( \rfrak )} - j + 3/2 + n} \rp
} %2
\hspace{2em} \text{for } \sfrak = \rfrak,
\label{deri3}
\end{align}
where we implemented the shift \eqref{gtpre2} as the third step of our prescription. Note that $Q_\tau^{- 1} Q_{b a}^{( \sfrak )}$ does not contain $Q_{a}^{( \sfrak )}$ due to \eqref{deri8}. Namely, the shift \eqref{gtpre2} allows the remaining contribution \eqref{deri3} to satisfy the requirement that this becomes trivial when $m = n = 0$.

As the conclusion of this subsection, the refined geometric transition in our scheme correctly produces open string contributions associated to the Lagrangian brane given by \eqref{deri6} and \eqref{deri3}.

\section{$qq$-characters from refined geometric transition}
\label{sec:qq-ch_geom}
In this section, we apply our prescription for the geometric transition to the $qq$-character, which has been recently proposed in the context of the BPS/CFT correspondence~\cite{Nekrasov:2015wsu,Nekrasov:2016qym,Nekrasov:2016ydq}.

We propose that when we consider the geometric transition so that two Lagrange submanifolds emerge, the contributions of two Lagrange submanifolds becomes $\Ysf$-operator, depending on the position of the brane insertion.
%and the positions of the Lagrange submanifolds correspond to the each terms of $\Tsf$-operator.
%Therefore the summation of all possible positions of the Lagrange submanifold becomes $\Tsf$-operator.
Let us examine our prescription with some examples.

%%%%%%%%%%%%%%%%%% section 3-1 %%%%%%%%%%%%%%%%%%%%%%%%%%%%%%%%
\subsection{Seiberg--Witten geometry and $qq$-character}
\label{sec:qq-ch}

Let us briefly remark some background of the $qq$-character in gauge theory.
Nekrasov--Pestun~\cite{Nekrasov:2012xe} pointed out an interesting connection between the quiver gauge theory and the representation theory of the corresponding quiver.
Their statement is that the Seiberg--Witten geometry of the $\Gamma$-quiver gauge theory in 4d is described by the characters of the fundamental representations of $G_\Gamma$-group, where $G_\Gamma$ is the finite Lie group associated with the (ADE) quiver $\Gamma$ under the identification of the quiver with the Dynkin diagram.
In fact, the prescription of Nekrasov--Pestun uses the Weyl reflection to generate the Seiberg--Witten curve, which is generic and applicable to any quiver, even if there is no finite group $G_\Gamma$ corresponding to the quiver $\Gamma$.
Let us check this process with $A_1$ quiver, which is the simplest example.
Since $G_\Gamma = \mathrm{SU}(2)$ for $\Gamma = A_1$, the fundamental character is given by
\begin{align}
 \chi_{\tiny \yng(1)}(\mathrm{SU}(2)) = y + y^{-1}
 \, ,
 \label{eq:ch_A1}
\end{align}
where the first term corresponds to the fundamental weight $y$, and the second term is generated by the Weyl reflection $y \to y^{-1}$.
On the other hand, the Seiberg--Witten curve is an algebraic curve given as a zero locus of the algebraic function
\begin{align}
 \Sigma = \{ (x, y) \mid H(x,y) = 0\}
 \, ,
\end{align}
where $(x,y) \in \Cbb \times \Cbb^*$ for 4d and $(x,y) \in \Cbb^* \times \Cbb^*$ for 5d gauge theory.
For $A_1$ quiver gauge theory with SU($N$) vector multiplet without any matter fields, the function $H(x,y)$ turns out to be
\begin{align}
 H(x,y) = y + y^{-1} - T_N(x)
 \, ,
\end{align}
where $T_N(x)$ is a degree $N$ monic polynomial in $x$, $T_N(x) = x^N + \cdots$.
In other words, the curve is characterized by the polynomial relation%
\footnote{%
There should be the coupling constant dependence on the LHS, but it is be now absorbed by redefinition of the $y$-variable.
}
\begin{align}
 y + y^{-1} = T_N(x)
 \, .
\end{align}
Now it is obvious that the LHS agrees with the SU(2) character \eqref{eq:ch_A1}.
It is possible to derive this polynomial relation from the $\Gamma$-quiver gauge theory partition function with the $\Omega$-deformation~\cite{Nekrasov:2002qd}, and taking the Seiberg--Witten limit $(\epsilon_1,\epsilon_2) \to (0,0)$, which is essentially the same approach as Nekrasov--Okounkov~\cite{Nekrasov:2003rj}.
In particular, the $y$-variable appearing in the algebraic relation is realized as an expectation value of the $\Ysf$-operator, which we focus on in this paper,
\begin{align}
 y(x) = \Big< \mathsf{Y}(x) \Big>
 \, .
\end{align}
The $\Ysf$-operator is a generating function of the chiral ring operators, so that it is realized as a codimension-4 defect operator.
See~\cite{Kim:2016qqs} for its realization as the line operator in 5d gauge theory.
Furthermore the $\Ysf$-operator itself has a cut singularity in the complex plane $x \in \Cbb$ in the Seiberg--Witten limit, and its crossing-cut behavior is indeed described by the Weyl reflection.
%\begin{align}
% y(x+\mathrm{i}0) y(x-\mathrm{i}0) = 1
% \, .
%\end{align}
This is the reason why the Weyl reflection generates the Seiberg--Witten curve.

It was then shown by Nekrasov--Pestun--Shatashvili~\cite{Nekrasov:2013xda} that this representation theoretic structure in gauge theory has a natural $q$-deformation: The Seiberg--Witten curve in the Nekrasov--Shatashvili (NS) limit $(\epsilon_1, \epsilon_2) \to (\hbar, 0)$~\cite{Nekrasov:2009rc} is promoted to the $q$-character, which was originally introduced in the context of the quantum affine algebra~\cite{Frenkel:1998} with emphasis on its connection with the quantum integrable system.
See also \cite{Nakajima:1999,Lusztig,MR1208827} for further developments.
This means that the polynomial relation holds in the NS limit just by replacing the character with $q$-character.
In this case, the Seiberg--Witten curve is not an algebraic curve any longer, but lifted to a quantum curve, which is a difference equation.
For example, for $A_1$ quiver theory, it is given by%
\footnote{
%This expression is based on the 4d gauge theory notation, and thus precisely speaking, this corresponds to the analogous character in (affine) Yangian $\mathscr{Y}(\gfrak_\Gamma)$.
%One can naturally obtain the $q$-character of quantum affine algebra $\mathrm{U}_q(\gfrak_\Gamma)$ from 5d gauge theory.
We use the 5d notation $(q_1,q_2) = (e^{\epsilon_1}, e^{\epsilon_2})$ and define $q = q_1 q_2 = e^{\epsilon_1 + \epsilon_2}$.
The unrefined limit is given by $q_1 = q_2^{-1}$, namely $q = 1$.
}
\begin{align}
 y(x) + \frac{1}{y(q_1^{-1} x)} = T_N(x;\epsilon_1)
 \, .
\end{align}
The RHS is again a degree $N$ monic polynomial in $x$, but the coefficients may depend on the equivariant parameter $\epsilon_1$.
In particular, this difference equation, also called the quantum (deformed) Seiberg--Witten curve~\cite{Poghossian:2010pn,Fucito:2011pn,Fucito:2012xc}, is equivalent to (precisely speaking, the degenerate version of) the TQ-relation of the $G_\Gamma$ XXX/XXZ/XYZ spin chain for 4d/5d/6d gauge theory.
Then, as a corollary, the SUSY vacuum (twisted F-term) condition of the 4d gauge theory in the NS limit is equivalent to the Bethe ansatz equation of the $G_\Gamma$ XXX spin chain.

Recently it has been shown in the context of BPS/CFT correspondence~\cite{Nekrasov:2015wsu,Nekrasov:2016qym,Nekrasov:2016ydq} that a similar polynomial relation actually holds even with generic $\Omega$-background parameters $(\epsilon_1, \epsilon_2)$ by replacing the $q$-character for the NS limit with a further generalized character, called the $qq$-character.
For $A_1$ quiver, in the 5d notation, it is given by%
\footnote{Precisely speaking, $y(q^{-1} x)^{-1}$ means $\displaystyle \Big< \Ysf(q^{-1} x)^{-1} \Big>$ here.}
\begin{align}
 y(x) + \frac{1}{y(q^{-1} x)} = T_N(x;\epsilon_1,\epsilon_2)
 \, .
 \label{eq:qq-ch_A11}
\end{align}
The $qq$-character has a gauge theoretical definition due to the invariance under the deformed Weyl reflection, which is called the iWeyl reflection, reflecting the non-perturbative aspects of the instanton moduli space.
This $qq$-character relation is interpreted as a (non-perturbative version of) Ward identity or Schwinger--Dyson equation since it gives a relation between correlation functions in quiver gauge theory.

The $y$-function, which is the gauge theory average of the $\Ysf$-operator, has pole singularities.
But such singularities are canceled in the combination of $y(x)$ and $y(q^{-1}x)^{-1}$.
In general, the iWeyl reflection shows how to cancel the pole singularity of the $y$-function.
%We will explore how such an algebraic structure appears in refined topological string theory in the following.

\subsubsection{$\Ysf$-operator}

Before discussing the topological string setup, let us mention more about the $\Ysf$-operator to fix our notation.
For generic quiver theory, we define $\Ysf$-operator associated with each gauge node, $\Ysf_i$ for $i \in \Gamma_0$ where $\Gamma_0$ is a set of nodes in the quiver $\Gamma$.
Then, in the 5d (K-theoretic) notation, the contribution of the $\Ysf$-operator for the configuration $\mu$ is~\cite{Nekrasov:2012xe,Nekrasov:2013xda}
\begin{align}
 \Ysf_{i,\mu}(x)
 & =
 \prod_{x' \in \Xcal_i}
 \frac{1 - x'/x}{1 - q_1 x'/x}
 \label{eq:Y-op_def1}
\end{align}
where we put SU($N_i$) gauge group for the $i$-th node, and define
\begin{align}
 \Xcal_i = \{ x_{i,\alpha,k} \}_{\alpha = 1,\ldots, N_i, k = 1,\ldots, \infty}
 \, , \qquad
 x_{i,\alpha,k} = q_2^{\mu_{i,\alpha,k}} q_1^{k-1} Q_{i,\alpha}
 \, , \qquad
 Q_{i,\alpha} = e^{a_{i,\alpha}}
 \, .
\end{align}
The parameter $Q_{i,\alpha}$ is the multiplicative (K-theoretic) Coulomb moduli.
The $\Ysf$-operator has several expressions
\begin{align}
 \Ysf_{i,\mu}(x)
 & =
 \prod_{\alpha=1}^{N_i}
 \left[
 \prod_{(j,k) \in \partial_+ \mu_{i,\alpha}}
 \left( 1 - \frac{q_1^{j-1} q_2^{k-1} Q_{i,\alpha}}{x} \right)
 \prod_{(j,k) \in \partial_- \mu_{i,\alpha}}
 \left( 1 - \frac{q_1^{j} q_2^{k} Q_{i,\alpha}}{x} \right)^{-1}
 \right]
 \nonumber \\
 & =
 \prod_{\alpha=1}^{N_i}
 \left[
 \left(1 - \frac{Q_{i,\alpha}}{x} \right)
 \prod_{(j,k) \in \mu_{i,\alpha}}
 \frac{(1 - q_1^{j} q_2^{k-1} Q_{i,\alpha} / x)
       (1 - q_1^{j-1} q_2^{k} Q_{i,\alpha} / x)}
      {(1 - q_1^{j} q_2^{k} Q_{i,\alpha} / x)
       (1 - q_1^{j-1} q_2^{k-1} Q_{i,\alpha} / x)}
 \right]
 \label{eq:Y-op_def2} 
\end{align}
where $\partial_\pm \mu$ is the outer/inner boundary of the partition $\mu$, where we can add/remove a box, and $q_1^{j-1} q_2^{k-1} Q_{i,\alpha}$ is the $q$-content of the box $(j,k) \in \mu_{i,\alpha}$.
From this expression it is easy to see the asymptotic behavior of the $\Ysf$-operator, which does not depend on the configuration $\mu$,
\begin{align}
 \Ysf_{i,\mu}(x)
 \ \longrightarrow \
 \begin{cases}
  1 & (x \to \infty) \\
  \displaystyle
  %\prod_{\alpha=1}^{N_i} \left( - \frac{Q_{i,\alpha}}{x} \right)
  (-1)^{N_i} Q_i \, x^{-N_i}
  & (x \to 0)
 \end{cases}
\end{align}
where we define the Coulomb moduli product
\begin{align}
 Q_i = \prod_{\alpha=1}^{N_i} Q_{i,\alpha}
 \, .
\end{align}
We remark that $Q_i = 1$ for SU($N_i$) theory, but keep it for latter convenience.

In addition, from the expression \eqref{eq:Y-op_def1} we obtain
\begin{align}
 \Ysf_{i,\mu}(x)
 & =
 \exp
 \left(
  \sum_{n=1}^\infty - \frac{x^{-n}}{n} \Ocal_{i,n}
 \right)
 \, , \qquad
 \Ocal_{i,n} = (1 - q_1^n) \sum_{x \in \Xcal_i} x^n
 \, .
 \label{eq:Y-op_exp}
\end{align}
Here $\Ocal_{i,n}$ is the contribution of the chiral ring operator for the configuration $\mu$, which is given by the single trace operator with respect to the complex adjoint scalar field $\Ocal_{i,n} = \Tr \Phi_i^n$ in 4d, and the loop/surface operator wrapping the compactified $S^1$/$T^2$ in 5d/6d.
Actually, for the gauge theory on $\Rbb^4 \times T^2$, the variable $x$ takes a value in $x \in \check{T}^2$ where $\check{T}^2$ is a dual torus of $T^2$~\cite{Nekrasov:2012xe}.
Thus the $\Ysf$-operator is interpreted as a codimension-4 defect operator, which plays a role as the generating function of the chiral ring operator.

Let us introduce the elliptic $\Ysf$-operator corresponding to 6d gauge theory, which is obtained by replacing the factors in \eqref{eq:Y-op_def1} with the elliptic functions,%
\footnote{The convention of the theta function used here (Dirac) is different from that used in Ref.~\cite{Kimura:2016dys} (Dolbeault).}
\begin{align}
 \Ysf_{i,\mu} (x)
 & =
 \prod_{x' \in \Xcal_i}
 \frac{\theta_{1}(x'/x)}{\theta_{1}(q_1 x'/x)}
 \, .
\end{align}
%\rem{In order to obtain an expression like \eqref{eq:Y-op_exp}, the convention with the theta function $\Theta(x)$ is better.}
This is reduced to the operator in 5d gauge theory \eqref{eq:Y-op_def1} in the limit $\operatorname{Im}\tau\to\infty$.
We also have a similar combinatorial expression to \eqref{eq:Y-op_def2} in the elliptic theory,
\begin{align}
 \Ysf_{i,\mu} (x)
 & =
 \prod_{\alpha=1}^{N_i}
 \left[
 \theta_{1}(Q_{i,\alpha}/x)
 \prod_{(j,k) \in \mu_{i,\alpha}}
 \frac{\theta_{1}(q_1^{j} q_2^{k-1} Q_{i,\alpha} / x)
       \theta_{1}(q_1^{j-1} q_2^{k} Q_{i,\alpha} / x)}
      {\theta_{1}(q_1^{j} q_2^{k} Q_{i,\alpha} / x)
       \theta_{1}(q_1^{j-1} q_2^{k-1} Q_{i,\alpha} / x)}
 \right]
 \, .
 \label{eq:Y-op_def}  
\end{align}
We will use this expression in the following sections.

%%%%%%%%%%%%%%%%%% section 3-2 %%%%%%%%%%%%%%%%%%%%%%%%%%%%%%%%
\subsection{$A_1$ quiver}

Let us consider the $\Ysf$-operator in $A_{1}$ quiver gauge theory. %with a single $\Ysf$-operator. 
The $\Ysf$-operator is a codimension-4 defect operator, and we here try to find its realization using the lower codimension surface defects. Here we give the prescription:
\begin{itemize}
\item[1.]
Consider the geometric transition so that the brane and the anti-brane emerge, and tune the distance between these branes.
\item[2.]
We shift the K\"ahler parameters $Q_{i}^{(s)}\to Q_{i}^{(s)}\sqrt{q_1 q_2}$ in order to make agreement with the Nekrasov partition function.
\item[3.]
Finally, identifying the K\"ahler parameter which corresponds to the position of the branes as the $x$-variable, the summation of all possible configurations of the brane and anti-brane in the Calabi--Yau is regular (invariant under the iWeyl reflection) for arbitrary $x$ with a suitable $\mu$-independent normalization factor.
\end{itemize}
Let us demonstrate this prescription in several examples.

\subsubsection{U(1) theory}
\label{sec:A1_Abelian}

\begin{figure}[htb]
\centering
\includegraphics[width=14cm]{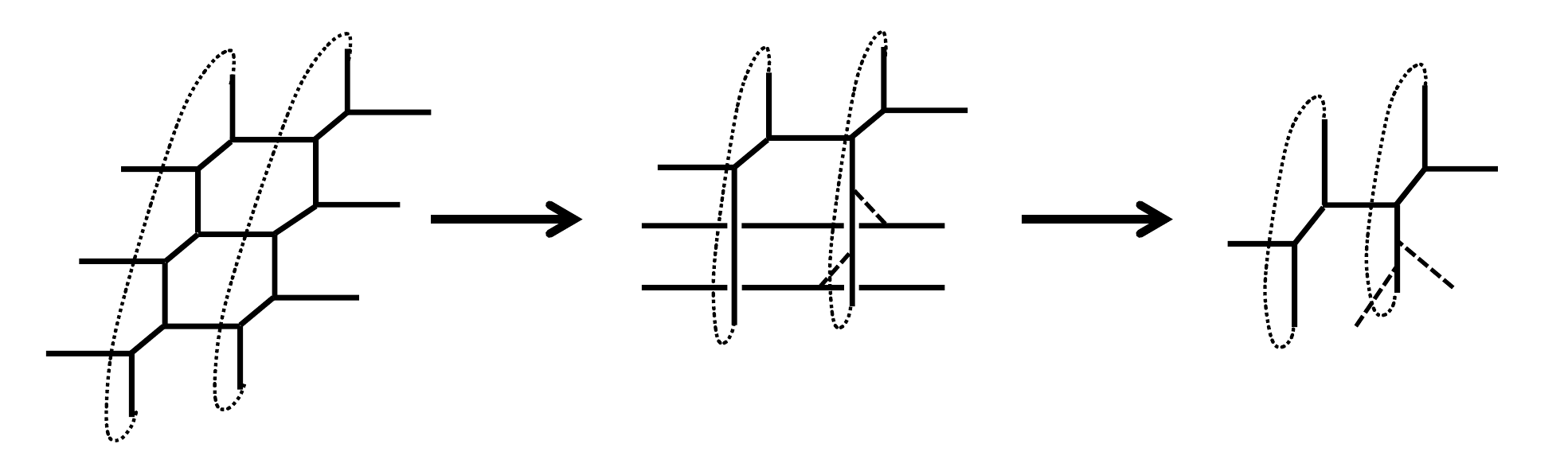}
\caption{The geometric transition.}
\label{GTA_1R}
\end{figure}

For simplicity let us first consider the Abelian gauge theory.
%Note that by comparing with the notation of \cite{Kimura:2015rgi}, this $Q_x$ is $\nu/x$ where $\nu$ is the 5d version of the Coulomb moduli $\nu = e^{a}$ of U(1) gauge theory, and $x$ is the argument of the $\Ysf$-operator. Such a separation will become important when we discuss the non-Abelian case.\rem{In order to be clear in the case of SU($N$), I use the Kimura-san's comment.}
Comparing the $\Ysf$-operator \eqref{eq:Y-op_def} with the contribution of the defect insertion shown in \eqref{deri3}, it turns out to be \textit{a half} of the $\Ysf$-operator.
%\rem{Should be nice to rewrite the surface operator contribution in \eqref{PFU(1)} in terms of the $q$-content shown in \eqref{eq:Y-op_def}: $-\mu_i+j \to j - 1$---I use $q_1=t^{-1}, q_2=q$.}
Thus we can construct the $\Ysf$-operator by merging two surface operators with respect to the $q$-brane and anti-$q$-brane, corresponding to the geometric transition shown in Fig.~\ref{GTA_1R}.
Now the dashed lines on the right and on the left denote the $q$-brane and anti-$q$-brane, respectively.
%\rem{To be mentioned: the anti-$q$-brane would be singular in the sense of \eqref{gtunpre2}.}
We remark that the coupling constant is given by $q^{-1}$ for the anti-$q$-brane instead of $q$, since the sign of the string coupling is opposite to the ordinary one~\cite{Vafa:2001qf}, which also corresponds to applying the negative integer to \eqref{gtunpre1}.
In addition, the most right panel of Fig.~\ref{GTA_1R} shows that two D3-branes are extended to the opposite directions from the centered NS5-brane, and this is consistent with the brane configuration of the supergroup Chern--Simons theory~\cite{Mikhaylov:2014aoa}, which is also similar to the ABJ(M) model~\cite{Aharony:2008ug,Aharony:2008gk}.

%In this case, we consider the following geometric transition that the $q$-brane and anti-$q$-brane emerge.
%\rem{Mention these are $q$- and $t$-branes.---Only $q$-brane emerge.}
The partition function corresponding to Fig.~\ref{GTA_1R} is $\Zcal_{M=2,N=3}$ defined in \eqref{TSpart.fn.}. %$M=2,~N=3$ case in \eqref{TSpart.fn.}.
For this partition function, by setting
\begin{align}
&Q_{2}^{(1)}=Q_{3}^{(1)}=(q_1 q_2)^{-\frac{1}{2}},
\nonumber \\
&
Q_{2}^{(2)}=q_1 (q_1 q_2)^{-\frac{1}{2}},~
Q_{3}^{(2)}=q_1^{-1} (q_1 q_2)^{-\frac{1}{2}},~
Q_{\tau_{2}}^{(2)}=q_1 q_2^{-1},
 \label{eq:Y_parametrize1}
\end{align}
the partition function reduces %\rem{Is this $\Zcal_{2,3}$?---Yes, thanks.}
\begin{align}
\mathcal{Z}_{2,3}
=
\sum_{\mu}(\bar{Q}_{f,1})^{|\mu|}
&\prod_{(i,j)\in\mu}
\frac{
\theta_{1}(Q_{1}^{(2)^{-1}}q_2^{\mu_{i}-j}q_1^{i-1})
\theta_{1}(Q_{1}^{(1)^{-1}}q_2^{-\mu_{i}+j-1}q_1^{-i})
}{
\theta_{1}(q_2^{-\mu_{i}+j}q_1^{\mu_{j}^{t}-i+1})
\theta_{1}(	q_2^{-\mu_{i}+j-1}q_1^{\mu_{j}^{t}-i})
}
\nonumber \\
\times
&\prod_{(i,j)\in\mu}
\frac{
\theta_{1}((q_2^{-1}q_1 Q_{1}^{(2)}Q_{\tau_{3}}^{(2)})^{-1}q_2^{j-1}q_1^{i-1})
\theta_{1}((Q_{1}^{(2)}Q_{\tau_{3}}^{(2)})^{-1}q_2^{j-1}q_1^{i-1})
}{
\theta_{1}((q_2^{-1}Q_{1}^{(2)}Q_{\tau_{3}}^{(2)})^{-1}q_2^{j-1}q_1^{i-1})
\theta_{1}((q_1 Q_{1}^{(2)}Q_{\tau_{3}}^{(2)})^{-1}q_2^{j-1}q_1^{i-1})
},
\end{align}
where we shift $Q_{1}^{(1,2)}\to \sqrt{q_1 q_2}Q_{1}^{(1,2)}$. 
The products in the first line are the contributions of the M-strings without the Lagrange submanifolds. Thus the products in the second line correspond to the contribution of the Lagrange submanifolds.
The latter contributions are consistent with the $\Ysf$-operator defined in \eqref{eq:Y-op_def} for U(1) theory under the identification
\begin{align}
 Q_{x}
 : = \frac{Q_1}{x}
 = (q_1 Q_{1}^{(2)}Q_{\tau_{3}}^{(2)})^{-1}
 \, ,
\end{align}
where $Q_1$ is the multiplicative Coulomb moduli of U(1) theory.
Thus the partition function $\Zcal_{2,3}$ gives rise to the average of the $\Ysf$-operator
\begin{align}
 \Zcal_{2,3}
 \ \stackrel{\eqref{eq:Y_parametrize1}}{\longrightarrow} \
 \Big< \Ysf(x) \Big>
 \, .
\end{align}
This average is defined with respect to the partition function $\Zcal_{2,1}$, which is the 6d U(1) $N_f = 2$ Nekrasov function
\begin{subequations} 
\label{eq:U(1)-weight} 
 \begin{align}
 \Big< \Ocal(x) \Big>
 & =
 \sum_{\mu}
 \Ocal_\mu(x) \,
  \Zcal_\mu^{\mathrm{U}(1)}
  \\
  \Zcal_\mu^{\mathrm{U}(1)}
  & =
 (\bar{Q}_{f,1})^{|\mu|}
 \prod_{(i,j)\in\mu}
 \frac{
 \theta_{1}(Q_{1}^{(2)^{-1}}q_2^{\mu_{i}-j}q_1^{i-1})
 \theta_{1}(Q_{1}^{(1)^{-1}}q_2^{-\mu_{i}+j-1}q_1^{-i})
 }{
 \theta_{1}(q_2^{-\mu_{i}+j}q_1^{\mu_{j}^{t}-i+1})
 \theta_{1}(q_2^{-\mu_{i}+j-1}q_1^{\mu_{j}^{t}-i})
 }
 \end{align}
\end{subequations}
where the parameters $\bar{Q}_{f,1}$, and $Q_1^{(s)}$ correspond to the gauge coupling and the (anti)fundamental mass, respectively.
We remark that we have to multiply the factor $\theta_{1}(Q_x)$ to obtain a precise agreement with the definition of $\Ysf$-operator~\cite{Kimura:2016dys} because the $\mu$-independent factor cannot be fixed in the current formalism.

%Let us denote these contributions as $Y_{\mu}$.
%By defining
%\begin{align}
%\chi_{x}
%=Q_{x}t^{-\mu_{i}+j}q^{i-1},~
%Q_{x}=(q Q_{1}^{(2)}Q_{\tau_{3}}^{(2)})^{-1}
%\end{align}
%$Y_{\mu}$ becomes
%\begin{align}
%Y_{\mu}(Q_x)
%=
%\theta_{1}(Q_{x})
%\prod_{(i,j)\in\mu}
%\frac{
%\theta_{1}(t^{-1} \chi_{x})
%\theta_{1}(q \chi_{x})
%}{
%\theta_{1}(t^{-1}q \chi_{x})
%\theta_{1}(\chi_{x})
%},
%\end{align}
%where we multiply the factor $\theta_{1}(Q_x)$.
%This operator agrees with the Y-operator for 6d gauge theory~\cite{Kimura:2016dys}, which is reduced to the operator in \cite{Kimura:2015rgi,Bourgine:2016vsq} in the limit $\tau\to{\rm i}\infty$. %Therefore we call this as $\Ysf$-operator.
\par
We can also consider the following geometric transition, corresponding to the partition function $\Zcal_{2,3}$ as well.
\begin{figure}[htb]
\centering
\includegraphics[width=14cm]{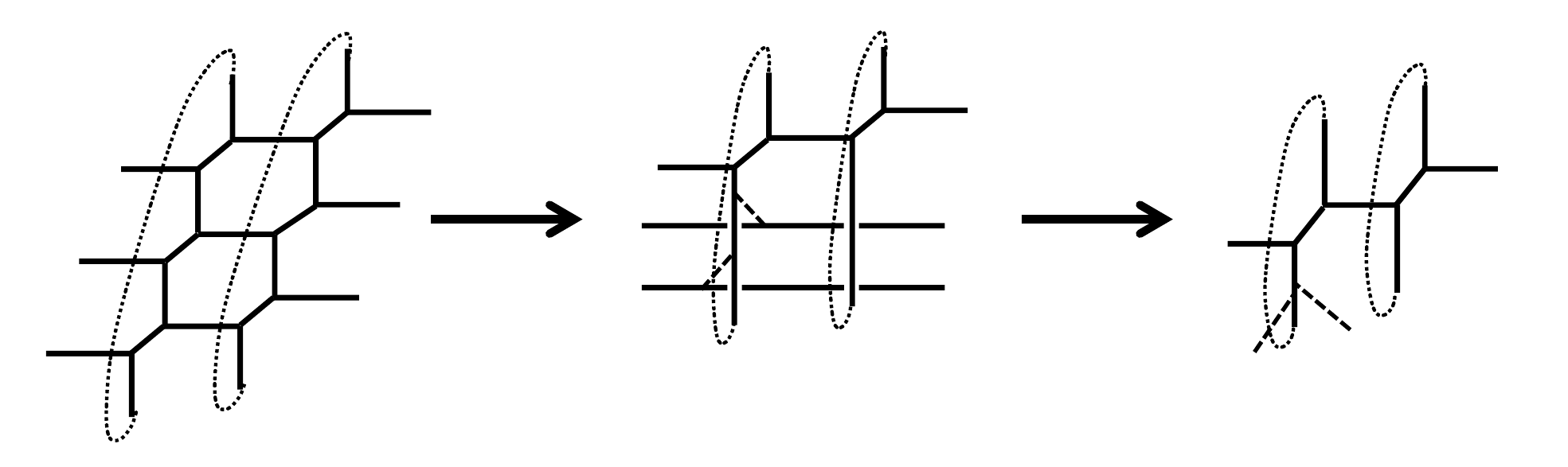}
\caption{The geometric transition.}
\label{GTA_1L}
\end{figure}
This configuration corresponds to the parametrization given by
\begin{align}
&Q_{2}^{(1)}=q_1 (q_1 q_2)^{-\frac{1}{2}},~
Q_{3}^{(1)}=q_1^{-1} (q_1 q_2)^{-\frac{1}{2}},~
Q_{\tau_{2}}^{(1)}=q_{1} q_2^{-1},
\nonumber \\
&Q_{2}^{(2)}=Q_{3}^{(2)}=(q_1 q_2)^{\frac{1}{2}},
 \label{eq:Y_parametrize2}
\end{align}
and define
\begin{align}
 Q_{x} = \frac{Q_1}{x} = (q_1 Q_{\tau_{3}}^{(1)})^{-1}
 \, .
\end{align}
In this case the contribution of the Lagrange submanifolds reads
%\rem{The overall factor is $\theta_{1}(Q_x)^{-1}$?---Yes, thanks.}
\begin{align}
%Y'_{\mu}(Q_x)
%=
%\theta_{1}(Q_{x})^{-1}
%\prod_{(i,j)\in\mu}
%\frac{
%\theta_{1}(t^{-1}q \chi_{x})
%\theta_{1}( \chi_{x})
%}{
%\theta_{1}(t^{-1}q\times t^{-1} \chi_{x})
%\theta_{1}(t^{-1}q\times q\chi_{x})
% }.
 \prod_{(i,j)\in\mu}
 \frac{\theta_{1}(q_2^{i} q_1^j Q_x)
       \theta_{1}(q_2^{i-1} q_1^{j-1} Q_x)}
      {\theta_{1}(q_2^{i+1} q_1^{j} Q_x)
       \theta_{1}(q_2^{i} q_1^{j+1} Q_x)}
 \, .
\end{align}
However this naive expression does not work.
We have to shift the argument in the numerator as discussed in Sec.~\ref{Deri}, to obtain a consistent result,
\begin{align}
%\theta_{1}(t^{-1}q \chi_{x})
%\theta_{1}( \chi_{x})
%\to
%\theta_{1}(t^{-1}q\times t^{-1}q \chi_{x}),
%\theta_{1}(t^{-1}q \chi_{x})
 \theta_{1}(q_2^{i} q_1^j Q_x)
 \theta_{1}(q_2^{i-1} q_1^{j-1} Q_x)
 \ \longrightarrow \
 \theta_{1}(q_2^{i+1} q_1^{j+1} Q_x)
 \theta_{1}(q_2^{i} q_1^{j} Q_x)
 \, .
\end{align}
Under the identification $Q_x = Q_1/x$, this configuration gives rise to the $\Ysf$-operator inverse by multiplying a factor $\theta_{1}(q Q_x)^{-1}$,
\begin{align}
 \frac{1}{\Ysf_\mu(q^{-1} x)}
 & =
 \theta_{1}(q Q_x)^{-1}
 \prod_{(i,j)\in\mu}
 \frac{\theta_{1}(q_2^{i} q_1^{j} (q Q_x))
       \theta_{1}(q_2^{i-1} q_1^{j-1} (q Q_x))}
      {\theta_{1}(q_2^{i} q_1^{j-1} (q Q_x))
       \theta_{1}(q_2^{i-1} q_1^{j} (q Q_x))}
 \, .
\end{align}
%we find
%\begin{align}
%Y'_{\mu}(Q_x)=\frac{1}{Y_{\mu}(t^{-1}qQ_x)}.
%\end{align}
Thus the partition function $\Zcal_{2,3}$ under the parametrization \eqref{eq:Y_parametrize2} leads to the average of the $\Ysf$-operator inverse
\begin{align}
 \Zcal_{2,3}
 \ \stackrel{\eqref{eq:Y_parametrize2}}{\longrightarrow} \
 \left< \frac{1}{\Ysf(q^{-1} x)}\right>
 \, .
\end{align}

Although the $\Ysf$-operator and its inverse themselves have pole singularities, we can construct a regular function using these two operators, as discussed in Sec.~\ref{sec:qq-ch}.
In this case, the fundamental $qq$-character of $A_1$ quiver, which has no singularity, is given by the average of the $\Tsf$-operator defined
\begin{align}
 \chi_{\tiny \yng(1)}(A_1;q_1, q_2)
 = \Big< \Tsf(x) \Big>
 & :=
 \Big< \Ysf(x) \Big>
 + \qfrak \, \Psf(x) \Big< \Ysf(q^{-1} x)^{-1} \Big>
 \label{eq:qq-ch_A1}
\end{align}
with the gauge coupling $\qfrak = \bar{Q}_{f,1}$ and the matter factor
\begin{align}
% \Tsf_{\mu}(x)
% & = 
% \Ysf_{\mu}(x)
% + \qfrak \, \frac{\Psf(x)}{\Ysf_{\mu}(q^{-1}x)}
% \, ,
% \\
 \Psf(x)
 & =
% \theta_{1}(Q_{m}Q^{-1}_{x})\theta_{1}(Q_{m}^{-1}Q^{-1}_{x}tq^{-1})
 \theta_{1}(Q_{1}^{(1)}Q_1^{-1} x)\theta_{1}(Q_{1}^{(2)^{-1}} Q_1^{-1} q^{-1} x)
 \, .
 \label{eq:matter-fac_U(1)}
\end{align}
The average is taken with respect to the 6d U(1) Nekrasov function \eqref{eq:U(1)-weight} as before.
This shows that the $\Tsf$-operator average is given by the $qq$-character discussed in Sec.~\ref{sec:qq-ch}, and its regularity is proven using the iWeyl reflection
\begin{align}
 \Ysf(x)
 \ \longrightarrow \
 \qfrak \, \frac{\Psf(x)}{\Ysf(q^{-1} x)}
 \, .
 \label{eq:iWeyl_A1}
\end{align}
We provide a proof of the regularity of this $qq$-character in Appendix~\ref{secB}.
We remark that, comparing with \eqref{eq:qq-ch_A11}, we have additional factors $\qfrak$ and $\Psf(x)$ in this case.
The former one can be absorbed by redefinition of the $\Ysf$-operator $\Ysf \to \qfrak^{\frac{1}{2}} \Ysf$, and the latter is due to the (anti)fundamental matters, which is necessary for gauge/modular anomaly cancellation in 6d gauge theory.

%Let us discuss a pictorial interpretation of the $qq$-character.
The $\Ysf$-operator and its inverse $\Ysf^{-1}$ correspond to the brane insertion to the right and left NS5-branes, respectively, as shown in Figs.~\ref{GTA_1R} and~\ref{GTA_1L}.
These are all the possibilities for the brane insertion because there are only two NS5-branes for $A_1$ quiver theory where the right and left branes are connected by a suspended D5-brane.
On the other hand, as mentioned in Sec.~\ref{sec:qq-ch}, the $qq$-character is generated by the iWeyl reflection \eqref{eq:iWeyl_A1} converting the $\Ysf$-operator to its inverse, $\Ysf(x) \to \Ysf(q^{-1}x)^{-1}$.
The iWeyl reflection is a consequence of creation/annihilation of instantons~\cite{Nekrasov:2015wsu}, which is a fluctuation on the suspended brane.
Since the fluctuation affects the branes on the both sides, the brane insertion on the right is transferred to the left through the iWeyl reflection.

%and the partition function of $A_1$ quiver $\mathcal{Z}_{2,1}$ as
%\begin{align}
%&\mathcal{Z}_{2,1}
%=
%\sum_{\mu}\bar{Q}_{f,1}^{|\mu|}\mathcal{Z}_{\mu}^{A_{1}},
%\\
%&\mathcal{Z}_{\mu}^{A_{1}}
%=
%\prod_{(i,j)\in\mu}
%\frac{
%\theta_{1}(Q_{m}^{-1}t^{-\mu_{i}+j}q^{i-1})
%\theta_{1}(Q_{m}^{-1}t^{\mu_{i}-j+1}q^{-i})
%}{
%\theta_{1}(t^{\mu_{i}-j}q^{\mu_{j}^{t}-i+1})
%\theta_{1}(t^{\mu_{i}-j+1}q^{\mu_{j}^{t}-i})
%},
%\end{align}
%the expectation value of $\Tsf$-operator which is defined as
%\begin{align}
%\langle T(Q_{x})\rangle
%&= 
%\sum_{\mu}
%\bar{Q}_{f,1}^{|\mu|}\mathcal{Z}_{\mu}^{A_1}T_{\mu}(Q_{x}),
%\end{align}
%is regular for arbitrary $Q_{x}$.
%\rem{Putting the gauge coupling $\bar{Q}_{f,1}$ in front of the second term of the $\Tsf$-operator, namely, $T = Y + \bar{Q}_{f,1} P / Y$, one can use the partition function $\Zcal_{M=N=1}$ to obtain a regular average.---OK, I modified.}
%\rem{For SU($N$) theory, regarding the asymptotic behavior, the average of the $\Tsf$-operator turns out a product of $N$ theta functions. Thus for U(1) theory it should be a single theta function. We can check this by expansion with respect to the gauge coupling. In particular, the $O(\qfrak^0)$ term will give $\theta_{1}(Q_x)$, and there will be no higher-order correction.}

\subsubsection{SU($N$) theory}

\begin{figure}[htb]
\centering
\includegraphics[width=14cm]{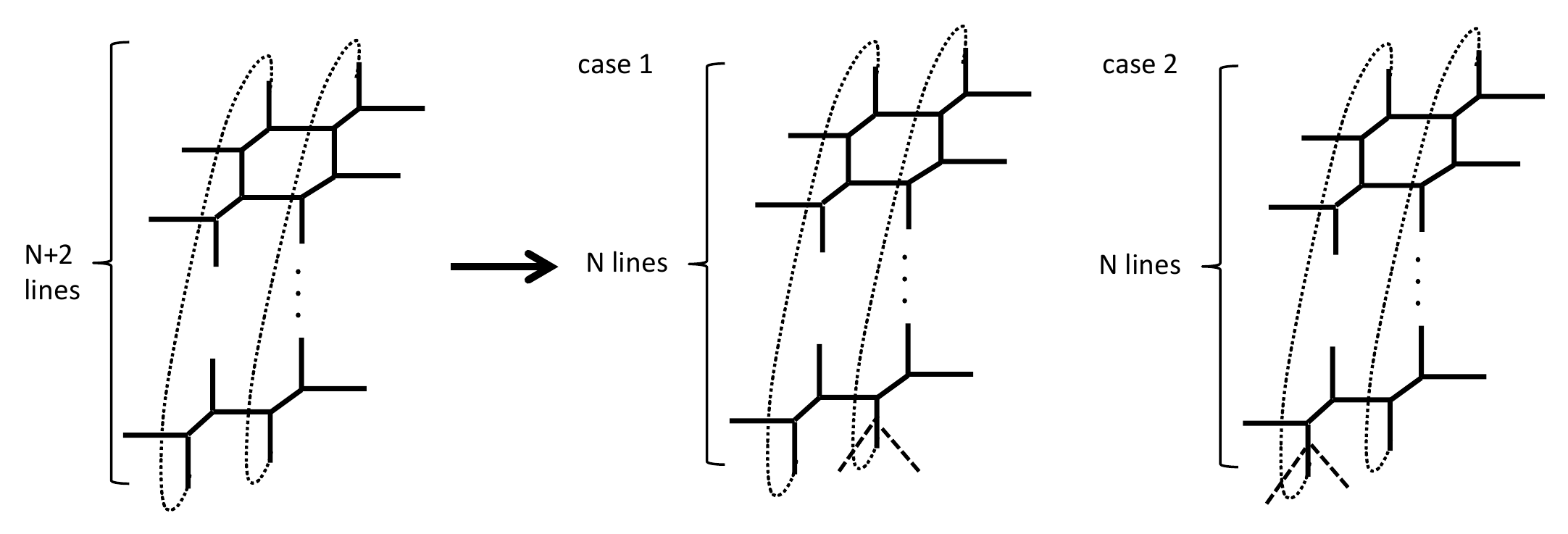}
\caption{The geometric transition.}
\label{GTA_1NA}
\end{figure}

One can easily generalize this result to the non-Abelian case. Let us consider the following geometric transition corresponding to SU($N$) theory with the insertion (Fig.~\ref{GTA_1NA}).
In this case we have two possible brane insertion to the right and left NS5-branes, which is actually the same as U(1) theory discussed in Sec.~\ref{sec:A1_Abelian}.
For the case 1, where the defect brane is inserted to the right NS5-brane, we obtain the $\Ysf$-operator
\begin{align}
 \Ysf_{\vec{\mu}} (x)
 & =
 \prod_{a=1}^{N}
 \left[
 \theta_{1}(Q_{a}/x)
 \prod_{(j,k) \in \mu_{a}}
 \frac{\theta_{1}(q_2^{j} q_1^{k-1} Q_{a} / x)
       \theta_{1}(q_2^{j-1} q_1^{k} Q_{a} / x)}
      {\theta_{1}(q_2^{j} q_1^{k} Q_{a} / x)
       \theta_{1}(q_2^{j-1} q_1^{k-1} Q_{a} / x)}
 \right]
\end{align}
under the parametrization
\begin{subequations}
 \label{eq:Y_parametrize3}
  \begin{align}
&Q_{N+1}^{(1)}=Q^{(1)}_{N+2}=(q_1 q_2)^{-\frac{1}{2}},~
\nonumber \\
&Q_{N+1}^{(2)}=q_1 (q_1 q_2)^{-\frac{1}{2}} ,~
Q^{(2)}_{N+2}=q_1^{-1} (q_1 q_2)^{-\frac{1}{2}},~Q_{\tau_{N+2}}^{(2)}=q_1 q_2^{-1} ,
\\
&
% \chi_{x_{a}}=Q_{x_{a}}t^{-\mu_{a,i}+j}q^{i-1},~
% Q_{x_{a}}=
\begin{cases}
 (Q_{1}^{(2)})^{-1}\tilde{Q}^{(2)^{-1}}_{N+2} =: %Q_{1}Q^{-1}_{x_0}
 Q_1 / x & (a=1)\\
 (Q_{a}^{(2)}\prod_{i=1}^{a-1}Q_{\tau_{i}}^{(2)})^{-1}\tilde{Q}^{(2)^{-1}}_{N+2} =: % Q_{a}Q^{-1}_{x_0}
 Q_a / x & (a=2,\ldots,N)
\end{cases}
   \label{eq:Coulomb_mod}
  \end{align}   
\end{subequations}
%\rem{For non-Abelian case, we have to divide $Q_{x_a}$ into the ``center of mass'' part and the Coulomb moduli part, like $Q_{x_a} \to Q_a/Q_{x_0}$. Then we discuss regularity of the $\Tsf$-operator with respect to a single parameter $Q_{x_0}$. Please compare the expression \eqref{eq:Y-op_inf_prod} with (3.28) of \cite{Kimura:2016dys}.---OK, I modified. How about this notation ?}
%
%we find the contribution of the $\Ysf$-operator $Y_{\vec{\mu}}(Q_{x})$ is
%\begin{align}
%Y_{\vec{\mu}}(Q_{x_{a}})=\prod_{a=1}^{N}
%\biggl[
%\theta_{1}( Q_{x_{a}})
%\prod_{(i,j)\in\mu_{a}}
%\frac{
%\theta_{1}( t^{-1} \chi_{x_{a}})
%\theta_{1}( q \chi_{x_{a}})
%}{
%\theta_{1}( t^{-1}q \chi_{x_{a}})
%\theta_{1}( \chi_{x_{a}})
%}
%\biggr],
%\end{align}
%where we define $\vec{\mu}=(\mu_1,\mu_2,...,\mu_N)$.
%
where we define $N$-tuple partition $\vec{\mu}=(\mu_1,\mu_2,\ldots,\mu_N)$, and the $\mu$-independent factor $\displaystyle \prod_{a=1}^N \theta_1(Q_a/x)$ is multiplied by hand.
Thus the partition function $\Zcal_{2,N+2}$ tuned with the parameters \eqref{eq:Y_parametrize3} gives rise to the average of the $\Ysf$-operator
\begin{align}
 \Zcal_{2,N+2}
 \ \stackrel{\eqref{eq:Y_parametrize3}}{\longrightarrow} \
 \Big< \Ysf(x) \Big>
 \, .
\end{align}
%We remark that we have to multiply the $\vec{\mu}$-independent factor $\displaystyle \prod_{a=1}^N \theta_{1}(Q_a/x)$ to obtain the $\Ysf$-operator.
The operator average is now taken with respect to 6d SU($N$) $N_f = 2N$ Nekrasov function
\begin{subequations}
\begin{align}
 \Big< \Ocal(x) \Big>
 & =
 \sum_{\vec{\mu}}
 \Ocal_{\vec{\mu}}(x) \, \Zcal_{\vec{\mu}}^{\mathrm{SU}(N)}
 \\
 \Zcal_{\vec{\mu}}^{\mathrm{SU}(N)}
 & =
 Q_f^{|\vec{\mu}|}
\prod_{a=1}^{N}
\prod_{(i,j)\in\mu_{a}}
\prod_{b=1}^{N}
\frac{
\theta_{1}(Q_{ab}^{(2)^{-1}}q_2^{\mu_{a,i}-j}q_1^{i-1})
\theta_{1}(Q_{ba}^{(1)^{-1}}q_2^{-\mu_{a,i}+j-1}q_1^{-i})
}{
\theta_{1}(\hat{Q}_{ba}^{(1)^{-1}}q_2^{-\mu_{a,i}+j}q_1^{\mu_{b,j}^{t}-i+1})
\theta_{1}(\hat{Q}_{ab}^{(1)^{-1}}q_2^{\mu_{a,i}-j+1}q_1^{-\mu_{b,j}^{t}+i})
}  
\end{align}
\end{subequations}
where we define the total instanton number $\displaystyle |\vec{\mu}| = \sum_{a=1}^N |\mu_a|$.
Imposing the condition $Q_{\tau_i}^{(1)}=Q_{\tau_i}^{(2)}Q_{i}^{(2)^{-1}}Q_{i+1}^{(2)}$, the Coulomb moduli parameter in this SU($N$) Nekrasov function is related to that defined in \eqref{eq:Coulomb_mod} as
\begin{align}
 \hat{Q}_{ab}^{(1)} & =
 \begin{cases}
  Q_b / Q_a & (a>b) \\
  Q_\tau Q_b / Q_a & (a<b)
 \end{cases}
 \, .
\end{align}

Similarly we obtain the $\Ysf$-operator inverse $\Ysf^{-1}$ from the case 2 with the defect brane inserted to the left.
The $\Ysf$-operator and its inverse have pole singularities as before, but we can use essentially the same combination as \eqref{eq:qq-ch_A1} to obtain a regular function, which is the $qq$-character
\begin{align}
 \chi_{\tiny \yng(1)}(A_1;q_1, q_2)
 =
 \Big< \Tsf(x) \Big>
 & =
 \Big< \Ysf(x) \Big>
 + \qfrak \, \Psf(x) \Big< \Ysf(q^{-1} x)^{-1} \Big>
 \label{eq:qq-ch_A1_SU(N)}
\end{align}
where the coupling constant and the (anti)fundamental contribution are now given by $\qfrak = Q_f$, and
\begin{align}
  \Psf(x) :=
 \prod_{a=1}^{N}
 \theta_{1}( Q^{(1)}_{a} Q_a^{-1} x)
 \theta_{1}( Q_{a}^{(2)^{-1}} Q_a^{-1} q^{-1} x)
 \, .
\end{align}
One can show the regularity of the $qq$-character (the $\Tsf$-operator average) in a similar way to U(1) theory, using the iWeyl reflection \eqref{eq:iWeyl_A1}.
We remark that the expression of the $qq$-character for SU($N$) theory \eqref{eq:qq-ch_A1_SU(N)} coincides with that for U(1) theory \eqref{eq:qq-ch_A11} apart from the matter factor $\Psf(x)$.
The $qq$-character provides a universal relation, which does not depend on the gauge group rank, but does only on the quiver structure.

\subsubsection{Higher $qq$-character}
\label{sec:higher_ch}

%%%%%%%%%%%%%%%%%% section 3-4 %%%%%%%%%%%%%%%%%%%%%%%%%%%%%%%%
%\subsection{$A_1$-quiver with two $\Ysf$-operators}
%Finally we consider the case that there are two $\Ysf$-operators. For simplicity let us consider the $A_1$-quiver Abelian case.

The Seiberg--Witten curve and its quantizations for $\Gamma$-quiver theory are described using the fundamental ($q$- and $qq$-)characters of $G_\Gamma$-group.
In addition, we can consider the higher-representation $qq$-character, which plays a role to determine the OPE of the generating currents of quiver W-algebras~\cite{Kimura:2015rgi}.
In this case, we have to consider several $\Ysf$-operators at the same time, and construct a regular function which is invariant under the iWeyl reflection.
Let us demonstrate how to treat multiple $\Ysf$-operators in U(1) theory for simplicity.

\begin{figure}[htb]
\centering
\includegraphics[width=13cm]{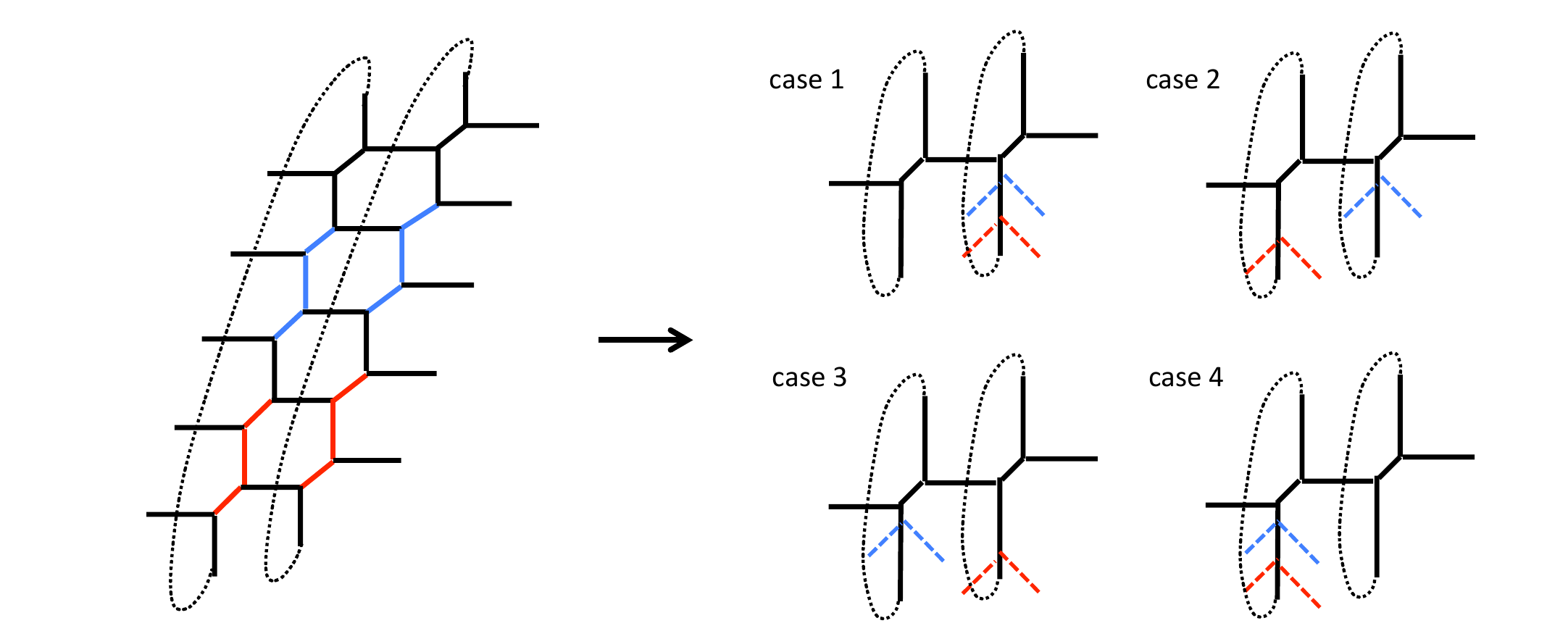}
\caption{In this geometric transition we obtain the $\Tsf$-operator which consists of two $\Ysf$-operators for $A_1$ quiver. We set the K\"ahler parameters in the blue and red parts.}
\label{GTA_1Y2}
\end{figure}

We start with the web diagram shown in Fig.~\ref{GTA_1Y2}.
In this case we tune the following parameters to obtain two $\Ysf$-operators,
\begin{subequations}
\begin{align}
Q_{2}^{(1)},~Q_{3}^{(1)},~Q_{2}^{(2)},~Q_{3}^{(2)},
\label{Blue}
\\
Q_{4}^{(1)},~Q_{5}^{(1)},~Q_{4}^{(2)},~Q_{5}^{(2)}.
\label{Red}
\end{align}
\end{subequations}
The parameters \eqref{Blue} and \eqref{Red} correspond to the blue brane and the red brane in Fig.~\ref{GTA_1Y2}, respectively. We show how to set the parameter in order to realize the brane configuration in each case:
\begin{subequations}
\begin{align}
\text{\underline{Case 1}}: \ & \Ysf(x_1) \Ysf(x_2)\nonumber \\
 &\begin{cases}
  Q_{2}^{(1)}=(q_1 q_2)^{-\frac{1}{2}},~Q_{3}^{(1)}=(q_1 q_2)^{-\frac{1}{2}},~
  Q_{4}^{(1)}=(q_1 q_2)^{-\frac{1}{2}},~Q_{5}^{(1)}=(q_1 q_2)^{-\frac{1}{2}},
  \\
  Q_{2}^{(2)}=q_1 (q_1 q_2)^{-\frac{1}{2}},~Q_{3}^{(2)}=q_1^{-1}(q_1 q_2)^{-\frac{1}{2}},~
  Q_{4}^{(2)}=q_1 (q_1 q_2)^{-\frac{1}{2}},~Q_{5}^{(2)}=q_1^{-1}(q_1 q_2)^{-\frac{1}{2}}
 \end{cases}
 \\
\text{\underline{Case 2}}: \ &\Ysf(x_1)/\Ysf(q^{-1} x_2)\nonumber \\
 &\begin{cases}
  Q_{2}^{(1)}=(q_1 q_2)^{-\frac{1}{2}},~Q_{3}^{(1)}=(q_1 q_2)^{-\frac{1}{2}},~
  Q_{4}^{(1)}=q_1 (q_1 q_2)^{-\frac{1}{2}},~Q_{5}^{(1)}=q_1^{-1}(q_1 q_2)^{-\frac{1}{2}},
  \\
  Q_{2}^{(2)}=q_1 (q_1 q_2)^{-\frac{1}{2}},~Q_{3}^{(2)}=q_1^{-1}(q_1 q_2)^{-\frac{1}{2}},~
  Q_{4}^{(2)}= (q_1 q_2)^{\frac{1}{2}},~Q_{5}^{(2)}=(q_1 q_2)^{\frac{1}{2}}
 \end{cases}
 \\
\text{\underline{Case 3}}: \ &\Ysf(x_2)/\Ysf(q^{-1} x_1)\nonumber \\
 &\begin{cases}
  Q_{2}^{(1)}=q_1 (q_1 q_2)^{-\frac{1}{2}},~Q_{3}^{(1)}= q_1^{-1}(q_1 q_2)^{-\frac{1}{2}},~
  Q_{4}^{(1)}=(q_1 q_2)^{-\frac{1}{2}},~Q_{5}^{(1)}=(q_1 q_2)^{-\frac{1}{2}},
  \\
  Q_{2}^{(2)}= (q_1 q_2)^{\frac{1}{2}},~Q_{3}^{(2)}=(q_1 q_2)^{\frac{1}{2}},~
  Q_{4}^{(2)}=q_1 (q_1 q_2)^{-\frac{1}{2}},~Q_{5}^{(2)}=q_1^{-1}(q_1 q_2)^{-\frac{1}{2}}
 \end{cases}
 \\
\text{\underline{Case 4}}: \ &\left(\Ysf(q^{-1} x_1) \Ysf(q^{-1} x_2)\right)^{-1}\nonumber \\
 &\begin{cases}
  Q_{2}^{(1)}=q_1(q_1 q_2)^{-\frac{1}{2}},~Q_{3}^{(1)}=q_1^{-1}(q_1 q_2)^{-\frac{1}{2}},~
  Q_{4}^{(1)}=q_1(q_1 q_2)^{-\frac{1}{2}},~Q_{5}^{(1)}=q_1^{-1}(q_1 q_2)^{-\frac{1}{2}},
  \\
  Q_{2}^{(2)}=(q_1 q_2)^{\frac{1}{2}},~Q_{3}^{(2)}=(q_1 q_2)^{\frac{1}{2}},~
  Q_{4}^{(2)}=(q_1 q_2)^{\frac{1}{2}},~Q_{5}^{(2)}=(q_1 q_2)^{\frac{1}{2}}
 \end{cases}
\end{align}
\end{subequations}
In the cases 2, 3, 4, we have to perform the $q_1q_2$-shift as before, where we define
\begin{subequations}
\begin{align}
&Q_{x_1}:=\frac{Q_{1}}{x_1}=(q_1 Q_{1}^{(2)}\prod_{i=3}^{5}Q^{(2)}_{\tau_{i}})^{-1}=(q_1 \prod_{i=3}^{5}Q^{(1)}_{\tau_{i}}),\\
&Q_{x_2}:=\frac{Q_{1}}{x_2}=(q_1 Q_{1}^{(2)}Q^{(2)}_{\tau_{5}})^{-1}=(q_1 Q^{(1)}_{\tau_{5}})
\end{align}
\end{subequations}
Then the partition function $\Zcal_{2,5}$ gives rise to the two-point function of the $\Ysf$-operator, by multiplying the $\mu$-independent factor,
\begin{align}
 \Zcal_{2,5}
 \ \longrightarrow \
 \begin{cases}
  \Big< \Ysf(x_1) \Ysf(x_2) \Big> & \text{(case 1)} \\[1em]
  \displaystyle \left< \frac{\Ysf(x_1)}{\Ysf(q^{-1} x_2)} \right>
  & \text{(case 2)} \\[1em]
  \displaystyle \left< \frac{\Ysf(x_2)}{\Ysf(q^{-1} x_1)} \right>
  & \text{(case 3)} \\[1em]
  \displaystyle \left< \left(\Ysf(q^{-1} x_1)\Ysf(q^{-1} x_2)\right)^{-1} \right>
  & \text{(case 4)}
 \end{cases}
\end{align}
where the average is taken with respect to the U(1) Nekrasov function~\eqref{eq:U(1)-weight}.
Then the average of the $\Tsf$-operator defined 
\begin{align}
 \chi_{\tiny \yng(2)}(A_1;q_1, q_2)
 & = \Big< \Tsf^{[2]}(x_1, x_2) \Big>
 \nonumber \\
 & :=
 \Big< \Ysf(x_1) \Ysf(x_2) \Big>
 + \qfrak \, \Psf(x_1) \Ssf\left(\frac{x_2}{x_1}\right)
 \left< \frac{\Ysf(x_2)}{\Ysf(q^{-1} x_1)} \right>
 \nonumber \\
 & \hspace{3em}
 + \qfrak \, \Psf(x_2) \Ssf\left(\frac{x_1}{x_2}\right)
 \left< \frac{\Ysf(x_1)}{\Ysf(q^{-1} x_2)} \right>
 + \qfrak^2 \,
 \frac{\Psf(x_1) \Psf(x_2)}{\Ysf(q^{-1} x_1) \Ysf(q^{-1} x_2)}
\end{align}
yields the $qq$-character of the degree-2 symmetric representation for $A_1$ quiver, and its regularity is again shown using the iWeyl reflection \eqref{eq:iWeyl_A1}.
Now the $\Ssf$-factor is defined~\cite{Kimura:2016dys}
\begin{align}
 \Ssf(x) & =
 \frac{\theta_{1}(q_1 x) \theta_{1}(q_2 x)}{\theta_{1}(q x) \theta_{1}(x)}
\end{align}
and the matter factor $\Psf(x)$ is the same as \eqref{eq:matter-fac_U(1)}.
This $qq$-character is regular even in the collision limit $x_2 \to x_1$, involving a derivative term, which is a specific feature to the $qq$-character~\cite{Nekrasov:2015wsu}.
In this limit, the cycle between the blue and red ones shrinks in Fig.~\ref{GTA_1Y2}.
We show the proof of the regularity in Appendix \ref{secB}.
We remark that we put the $\mu$-independent factors $\Ssf(x)$ and $\Psf(x)$ to define the $\Tsf$-operator because it's a matter of the normalization of the partition function.

In general, the $n$-point function of the $\Ysf$-operator for SU($N$) theory is obtained from the partition function $\Zcal_{2,N+2n}$ with $2^n$ possible brane insertions,
\begin{align}
 \Zcal_{2,N+2n}
 \ \longrightarrow \
 \Big< \Ysf(x_1) \cdots \Ysf(x_n) \Big>
 \, , \
 \left<  \frac{\Ysf(x_2) \cdots \Ysf(x_n)}{\Ysf(q^{-1} x_1)} \right>
 \, , \
 \left<  \frac{\Ysf(x_3) \cdots \Ysf(x_n)}
              {\Ysf(q^{-1} x_1) \Ysf(q^{-1} x_2)} \right> 
 \, , \ldots
\end{align}
We can construct the $qq$-character of the degree-$n$ representation $R_n= \underbrace{\square \cdots \square}_{n}$ for $A_1$ quiver by summing up all the possible $n$-point functions of the $\Ysf$-operator~\cite{Nekrasov:2015wsu,Kimura:2015rgi,Kimura:2016dys}, with a suitable $\Ssf$-factor inserted,
\begin{align}
 \chi_{R_n} (A_1;q_1, q_2)
 = \Big< \Tsf^{[n]}(x_1,\ldots,x_n) \Big>
 & :=
 \Big< \Ysf(x_1) \cdots \Ysf(x_n) \Big>
 + \cdots
 \, .
\end{align}

\if0

Then, by defining the variables $Q_{x_1}, Q_{x_2}$ appropriately in each case and summing all cases \rem{We had better show the defs of $Q_{x_{1,2}}$}, we finally obtain the $\Tsf$-operator,
\begin{align}
T(Q_{x_1},Q_{x_2})=&
Y_{\mu}(Q_{x_1})Y_{\mu}(Q_{x_2})+\bar{Q}_{f,1}P(Q_{x_2})\mathcal{S}(Q_{x_1}Q^{-1}_{x_2})\frac{Y_{\mu}(Q_{x_1})}{Y_{\mu}(t^{-1}q Q_{x_2})}
\nonumber \\
&+\bar{Q}_{f,1}P(Q_{x_{1}})\mathcal{S}(Q_{x_2}Q^{-1}_{x_1})\frac{Y_{\mu}(Q_{x_2})}{Y_{\mu}(t^{-1}q Q_{x_1})}
+\bar{Q}_{f,1}^2\frac{P(Q_{x_1})P(Q_{x_2})}{Y_{\mu}(t^{-1}q Q_{x_1})Y_{\mu}(t^{-1}qQ_{x_2})},
\end{align}
where
\begin{align}
\mathcal{S}(Q_x)=
\frac{
\theta_{1}( t Q_{x})\theta_{1}( q^{-1} Q_{x})
}{
\theta_{1}( tq^{-1} Q_{x})\theta_{1}(  Q_{x})
},
\end{align}
and $P(Q_{x})$ is the same as in the case of $A_{1}$ quiver theory. The expectation value of this operator
\begin{align}
\langle T(Q_{x_1},Q_{x_2}) \rangle = \sum_{\mu}\bar{Q}_{f,1}^{|\mu|}\mathcal{Z}^{A_{1}}_{\mu}T(Q_{x_1},Q_{x_2})
\end{align}
is regular for arbitrary $Q_{x_1}$ and $Q_{x_2}$.
%\rem{Does it mean $Q_{x_1}/Q_{x_2} = e^{2\pi \mathrm{i} \tau \mathbb{Z}}$? Even in this case, it's regular, but we have to carefully expand the $qq$-character with some derivative terms.---I write this matter in appendix B}
We show the regularity in the appendix \ref{secB}. Note that in order to ensure the regularity of the $\Tsf$-operator, we multiple the factor $\mathcal{S}(Q_{x})$.

\fi

%%%%%%%%%%%%%%%%%% section 3-3 %%%%%%%%%%%%%%%%%%%%%%%%%%%%%%%%
\subsection{$A_2$ quiver} % with single $\Ysf$-operator}

\begin{figure}[htb]
\centering
\includegraphics[width=14cm]{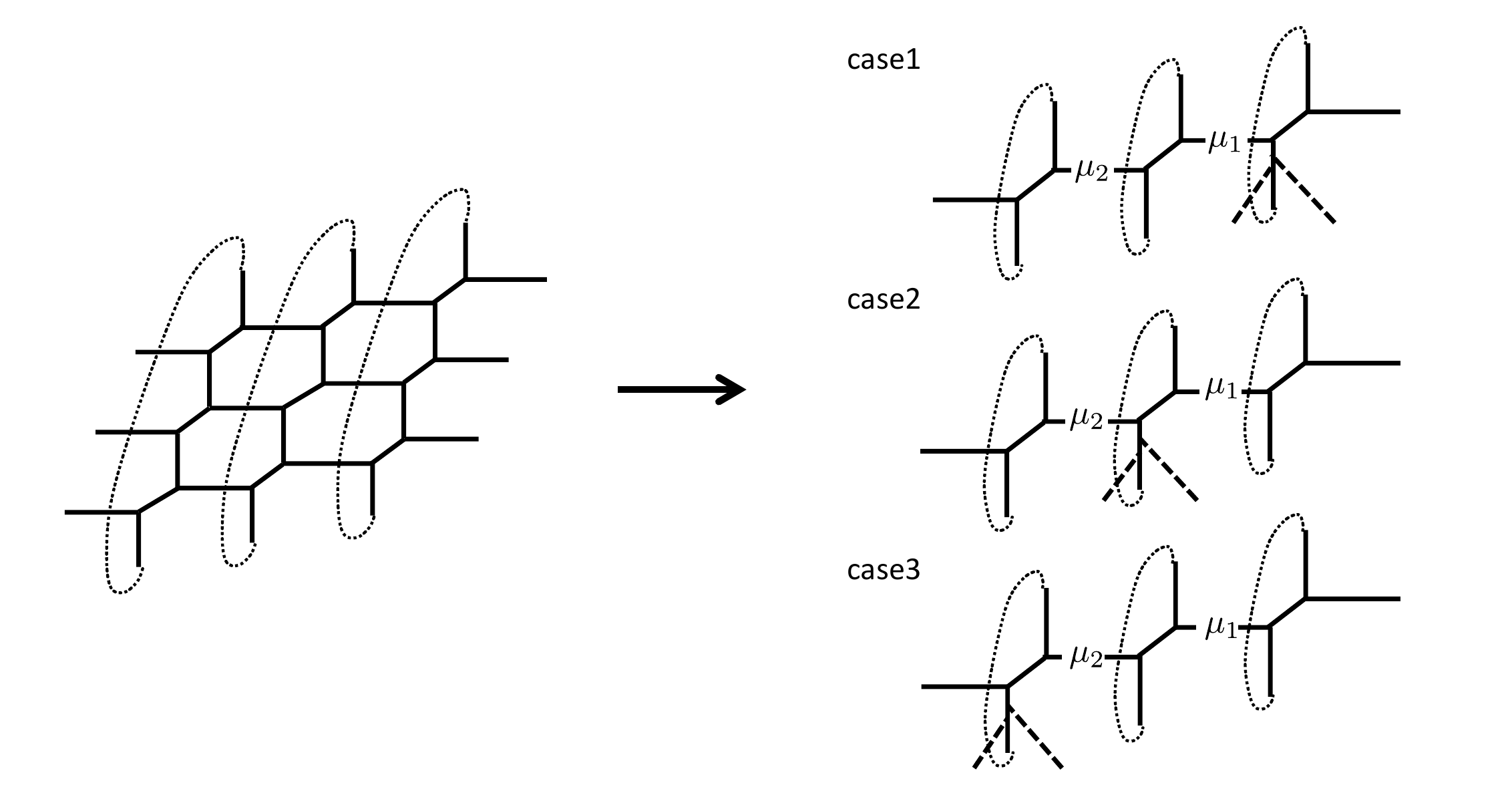}
\caption{In this geometric transition we obtain the $\Tsf$-operator for $A_2$ quiver.}
\label{GTA_2}
\end{figure}

Next we consider the $A_2$ quiver gauge theory to examine the $qq$-character using the refined geometric transition.
As mentioned in Sec.~\ref{sec:qq-ch}, the Seiberg--Witten curve and its quantization are associated with the fundamental representation character of $G_\Gamma$-group for $\Gamma$-quiver gauge theory.
Thus in this case it is deeply related to the representation theory of SU(3) group.
Since the $qq$-character generated by the iWeyl reflection does not depend on the gauge group rank, let us focus on the Abelian $A_2$ quiver theory, $\mathrm{U}(1) \times \mathrm{U}(1)$, for simplicity.
We have three possible ways to insert the defect brane as shown in Fig.~\ref{GTA_2}.
%Next we consider the $A_2$-quiver Abelian gauge theory with single $\Ysf$-operator. The corresponding web diagram is as Fig.~\ref{GTA_2}.

\subsubsection*{Case 1}

We consider the defect brane inserted to the right-most NS5-brane.
In this case, the calculation is essentially the same as that for $A_1$ quiver shown in Fig.~\ref{GTA_1R}.
We apply the following configuration
\begin{align}
&Q_{2}^{(1)}=Q_{3}^{(1)}=(q_1 q_2)^{-\frac{1}{2}},~Q_{2}^{(2)}=Q_{3}^{(2)}=(q_1 q_2)^{-\frac{1}{2}},~
\nonumber \\
&
Q_{2}^{(3)}=q_1(q_1 q_2)^{-\frac{1}{2}},~
Q_{3}^{(3)}=q_1^{-1}(q_1 q_2)^{-\frac{1}{2}},~
Q_{\tau_{2}}^{(3)}=q_1 q_2^{-1},
 \label{eq:Y_parametrize_case1}
\end{align}
with the Coulomb moduli parameter
\begin{align}
 %Q_{x}=
 \frac{Q_{1,1}}{x} = (q_1 Q_{1}^{(3)}Q_{\tau_{3}}^{(3)})^{-1}
 \, .
\end{align}
Comparing with the $\Ysf$-operator definition \eqref{eq:Y-op_def}, the contribution of the defect brane leads to $\Ysf_{1,\mu_2}(x)$ by multiplying the factor $\theta_{1}(Q_{1,1}/x)$.
Thus the partition function $\Zcal_{3,3}$ gives rise to the average of $\Ysf_1(x)$ under the parametrization \eqref{eq:Y_parametrize_case1}:
\begin{align}
 \Zcal_{3,3}
 \ \stackrel{\eqref{eq:Y_parametrize_case1}}{\longrightarrow} \
 \Big< \Ysf_1(x) \Big>
\end{align}
where the operator average is taken with respect to 6d $\mathrm{U}(1) \times \mathrm{U}(1)$ Nekrasov function
\begin{subequations} 
 \label{eq:A2-weight} 
\begin{align}
 \Big< \Ocal(x) \Big>
 & =
 \sum_{\mu_1, \mu_2} \Ocal_{\mu_1,\mu_2}(x) \,
 \mathcal{Z}_{\mu_1 ,\mu_2}^{\mathrm{U}(1) \times \mathrm{U}(1)}
 \\
 \mathcal{Z}_{\mu_1 ,\mu_2}^{\mathrm{U}(1) \times \mathrm{U}(1)}
 & =
 \bar{Q}_{f,1}^{|\mu_1|} \bar{Q}_{f,2}^{|\mu_2|}
 \prod_{(i,j)\in\mu_1}
 \frac{
 \theta_{1}( Q_{1}^{(3)^{-1}}q_2^{\mu_{1,i}-j}q_1^{i-1})
 \theta_{1}( Q_{1}^{(2)^{-1}}q_2^{-\mu_{1,i}+j-1}q_1^{\mu_{2,j}^{t}-i})
 }{
 \theta_{1}( q_2^{-\mu_{1,i}+j}q_1^{\mu_{1,j}^{t}-i+1})
 \theta_{1}( q_2^{-\mu_{1,i}+j-1}q_1^{\mu_{1,j}^{t}-i})
 }
 \nonumber \\
 & \qquad \times
 \prod_{(i,j)\in\mu_2}
 \frac{
 \theta_{1}( Q_{1}^{(2)^{-1}}q_2^{\mu_{2,i}-j}q_1^{-\mu_{1,j}^{t}+i-1})
 \theta_{1}( Q_{1}^{(1)^{-1}}q_2^{-\mu_{2,i}+j-1}q_1^{-i})
 }{
 \theta_{1}( q_2^{-\mu_{2,i}+j}q_1^{\mu_{2,j}^{t}-i+1})
 \theta_{1}( q_2^{-\mu_{2,i}+j-1}q_1^{\mu_{2,j}^{t}-i})
 } ,
\end{align}
\end{subequations}
%with the mass parameter defined $Q_m = Q_1^{(2)}$.

%\begin{align}
%Y^{\text{case 1}}
%&=
%\theta_{1}(Q_{x})
%\prod_{(i,j)\in\mu_{2}}
%\frac{
%\theta_{1}(t^{-1} \chi_{x})
%\theta_{1}(q \chi_{x})
%}{
%\theta_{1}(t^{-1}q \chi_{x})
%\theta_{1}(\chi_{x})
%}
%\nonumber \\
%&=
%Y_{\mu_2}(Q_x).
%\end{align}
where we define the gauge couplings $\bar{Q}_{f,1,2}$ and the Young diagrams $\mu_{1,2}$ as follows,
\begin{align}
&
\bar{Q}_{f,1}=\bar{Q}_{f,1}^{(2)},~
\bar{Q}_{f,2}=\bar{Q}_{f,1}^{(1)},
\\
&
\mu_1 =\mu_1^{(2)},~
\mu_2 =\mu_1^{(1)}.
\end{align}

\subsubsection*{Case 2}

In this case, the defect brane is inserted to the middle brane.
This configuration corresponds to the following parametrization
%In this case, by setting
\begin{align}
&Q_{2}^{(1)}=Q_{3}^{(1)}=(q_1 q_2)^{-\frac{1}{2}},~Q_{2}^{(3)}=Q_{3}^{(3)}=(q_1 q_2)^{\frac{1}{2}},~
\nonumber \\
&
Q_{2}^{(2)}=q_1(q_1 q_2)^{-\frac{1}{2}},~
Q_{3}^{(2)}=q_1^{-1}(q_1 q_2)^{-\frac{1}{2}},~
Q_{\tau_{2}}^{(3)}=q_1 q_2^{-1},
 \label{eq:Y_parametrize_case2}
\end{align}
and %$Q_{x}=(qQ_{\tau_{3}}^{(2)})^{-1}$,
two Coulomb moduli parameters defined
\begin{align}
 \frac{Q_{1,1}}{x} = (q_1 Q_{\tau_{3}}^{(2)})^{-1}
 \, , \qquad
 \frac{Q_{2,1}}{x} = (q_1 Q_1^{(2)} Q_{\tau_{3}}^{(2)})^{-1}
 \, .
 \label{GTA_2x}
\end{align}
We remark that the difference between $Q_{1,1}$ and $Q_{1,2}$ is given by the factor $Q_1^{(2)} =: Q_m$, which is interpreted as the bifundamental mass parameter, because such a bifundamental mass can be absorbed by the shift of U(1) Coulomb moduli~\cite{Nekrasov:2013xda}.
In this paper we do not explicitly write the bifundamental mass parameter.

In this case, the contribution of the Lagrange submanifolds reads %$Y^{\text{case 2}}$,
\begin{align}
\prod_{(i,j)\in\mu_1}
\frac{
\theta_{1}( q_2^{i} q_1^{j-1} Q_{2,1} / x)
\theta_{1}( q_2^{i-1} q_1^{j} Q_{2,1} / x)
}{
 \theta_{1}( q_2^{i} q_1^{j} Q_{2,1} / x)
 \theta_{1}( q_2^{i-1} q_1^{j-1} Q_{2,1} / x) 
}
\prod_{(i,j)\in\mu_2}
\frac{
 \theta_{1}( q_2^{i-1} q_1^{j-1} Q_{1,1} / x)
 \theta_{1}( q_2^{i} q_1^{j} Q_{1,1} / x)  
}{
 \theta_{1}( q_2^{i+1} q_1^{j} Q_{1,1} / x)
 \theta_{1}( q_2^{i} q_1^{j+1} Q_{1,1} / x)
 }
 \, .
\end{align}
%\begin{align}
%Y^{\text{case 2}}
%=
%\prod_{(i,j)\in\mu_1}
%\frac{
%\theta_{1}( t^{-1} Q_{1}^{(2)^{-1}}\chi_{x})
%\theta_{1}( q Q_{1}^{(2)^{-1}}\chi_{x})
%}{
%\theta_{1}( t^{-1}q Q_{1}^{(2)^{-1}}\chi_{x})
%\theta_{1}( Q_{1}^{(2)^{-1}}\chi_{x})
%}
%\prod_{(i,j)\in\mu_2}
%\frac{
%\theta_{1}( \chi_{x})
%\theta_{1}( t^{-1}q\chi_{x})
%}{
%\theta_{1}( t^{-1}q\times t^{-1} \chi_{x})
%\theta_{1}( t^{-1}q\times q \chi_{x})
%}.
%\end{align}
In order to obtain a consistent result, we have to shift the parameters of the numerator in the second factor, as discussed in Sec.~\ref{Deri},
\begin{align}
 \theta_{1}( q_2^{i-1} q_1^{j-1} Q_{1,1} / x)
 \theta_{1}( q_2^{i} q_1^{j} Q_{1,1} / x)
 \ \longrightarrow \
 \theta_{1}( q_2^{i} q_1^{j} Q_{1,1} / x)
 \theta_{1}( q_2^{i+1} q_1^{j+1} Q_{1,1} / x)
 \, .
\end{align}
Multiplying the $\mu$-independent factors, $\theta_{1}(Q_{2,1}/x)$ and $\theta_{1}(q Q_{1,1} / x)^{-1}$, the $\mu_1$- and $\mu_2$-contributions are written as $\Ysf_2(x)$ and $\Ysf_1^{-1}(q^{-1} x)$, respectively.
Thus the partition function $\Zcal_{3,3}$ becomes the average of the $\Ysf$-operator ratio, by tuning the parameters as \eqref{eq:Y_parametrize_case2},
\begin{align}
 \Zcal_{3,3}
 \ \stackrel{\eqref{eq:Y_parametrize_case2}}{\longrightarrow} \
 \left<
  \frac{\Ysf_2 (x)}{\Ysf_1 (q^{-1} x)}
 \right>
 \, .
\end{align}
The average is again taken with respect to the U(1) $\times$ U(1) Nekrasov function \eqref{eq:A2-weight}.

%we obtain
%\begin{align}
%Y^{\text{case 2}}
%=
%\frac{Y_{\mu_1}(Q^{-1}_{m}Q_{x})}{Y_{\mu_2}(t^{-1}qQ_{x})}.
%\end{align}

\subsubsection*{Case 3}

The remaining situation is that the defect brane is inserted to the left-most brane.
In this case, the calculation is essentially the same as Fig.~\ref{GTA_1L} for $A_1$ quiver theory.
Applying the parametrization
\begin{align}
&Q_{2}^{(2)}=Q_{3}^{(2)}=(q_1 q_2)^{\frac{1}{2}},~Q_{2}^{(3)}=Q_{3}^{(3)}=(q_1 q_2)^{\frac{1}{2}},~
\nonumber \\
&
Q_{2}^{(1)}=q_1(q_1 q_2)^{-\frac{1}{2}},~
Q_{3}^{(1)}=q_1^{-1}(q_1 q_2)^{-\frac{1}{2}},~
Q_{\tau_{2}}^{(3)}=q_1 q_2^{-1},
\nonumber \\
 & %Q_{x}:= (qQ_{\tau_{3}}^{(1)})^{-1}
 \frac{Q_{2,1}}{x} = (q_1 Q_1^{(2)} Q_{\tau_{3}}^{(1)})^{-1} 
 \label{eq:Y_parametrize_case3}
\end{align}
with a suitable $q_1q_2$-shift of the arguments to be consistent with the geometric transition, the partition function $\Zcal_{3,3}$ yields
\begin{align}
 \Zcal_{3,3}
 \ \stackrel{\eqref{eq:Y_parametrize_case3}}{\longrightarrow} \
 \left< \frac{1}{\Ysf_2(q^{-1} x)} \right>
 \, .
\end{align}

\subsubsection*{$qq$-characters}

Now we can construct the $qq$-character using all the possible brane insertions.
The $qq$-character of the fundamental representation for $A_2$ quiver theory, denoted by \textbf{3}, is given by the $\Tsf$-operator average,
\begin{align}
 \chi_{\textbf{3}}(A_2;q_1, q_2)
 & = \Big< \Tsf_1(x) \Big>
 \nonumber \\
 & :=
 \Big< \Ysf_1(x) \Big>
 + \qfrak_1 \, \Psf_1(x)
 \left< \frac{\Ysf_2 (x)}{\Ysf_1 (q^{-1} x)} \right>
 + \qfrak_1 \qfrak_2 \, \Psf_1(x) \Psf_2(x)
 \left< \frac{1}{\Ysf_2 (q^{-1} x)} \right> 
 \label{eq:T1-op}
\end{align}
where the coupling constants are given by $\qfrak_1 = \bar{Q}_{f,1}$ and $\qfrak_2 = \bar{Q}_{f,2}$, and the matter factors are defined
\begin{align}
 \Psf_1(x) = \theta_{1}(q^{-1} Q_1^{(3)^{-1}} Q_{1,1}^{-1} x)
 \, , \qquad
 \Psf_2(x) = \theta_{1}(Q_1^{(1)^{-1}}Q_1^{(2)^{-1}} Q_{1,1} / x)
 \, .
\end{align}
Although each factor in \eqref{eq:T1-op} has pole singularities as before, the $qq$-character itself is a regular entire function in $x$, as shown in Appendix \ref{secB}.
The local pole cancellation is performed by the iWeyl reflection
\begin{align}
 \Ysf_1(x)
 \ \longrightarrow \
 \qfrak_1 \, \Psf_1(x) \frac{\Ysf_2(x)}{\Ysf_1(q^{-1} x)}
 \, , \quad
 \Ysf_2(x)
 \ \longrightarrow \
 \qfrak_2 \, \Psf_2(x) \frac{\Ysf_1(q^{-1} x)}{\Ysf_2(q^{-1} x)}
 \, .
 \label{eq:iWeyl_A2}
\end{align}

\begin{figure}[htb]
\centering
\includegraphics[width=14cm]{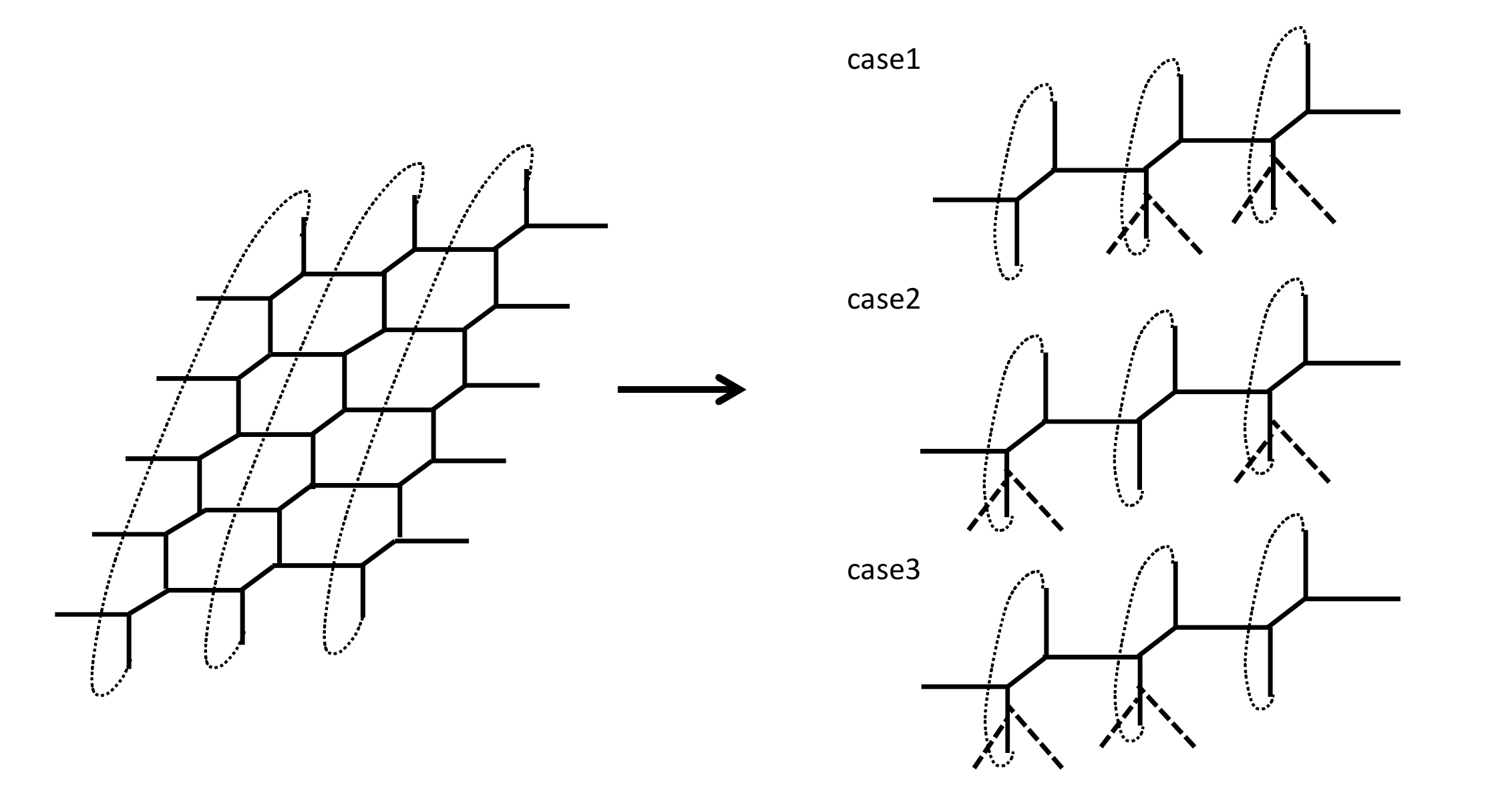}
\caption{The geometric transition which emerge the two defect branes. The summation of them corresponds to the $qq$-character of $\bar{\bf{3}}$.}
\label{GTA_2another}
\end{figure}

For $A_2$ quiver, we have another representation, which is the anti-fundamental representation denoted by $\bar{\textbf{3}}$.
The corresponding $qq$-character is generated by applying the iWeyl reflection \eqref{eq:iWeyl_A2} to the highest weight $\Ysf_2(x)$,
\begin{align}
 \chi_{\bar{\textbf{3}}}(A_2;q_1, q_2)
 & = \Big< \Tsf_2(x) \Big>
 \nonumber \\
 & :=
 \Big< \Ysf_2(x) \Big>
 + \qfrak_2 \, \Psf_2(x)
 \left< \frac{\Ysf_1 (q^{-1} x)}{\Ysf_2 (q^{-1} x)} \right>
 + \qfrak_1 \qfrak_2 \, \Psf_1(q^{-1} x) \Psf_2(x)
 \left< \frac{1}{\Ysf_1 (q^{-2} x)} \right> 
 \, .
 \label{eq:T2-op}
\end{align}
We remark that the operator $\Ysf_2(x)$ itself cannot be constructed by a single insertion of the defect brane, but is realized as a composite operator:
\begin{align}
 \Ysf_2(x) = \Ysf_1(q^{-1} x) \times \frac{\Ysf_2(x)}{\Ysf_1(q^{-1} x)} 
 \, .
\end{align}
In other words, the operator $\Ysf_2(x)$ is obtained by two insertions of the defect branes to the right-most and the middle branes (see the case 1 in Fig.~\ref{GTA_2another}).
Similarly the remaining terms in \eqref{eq:T2-op} are obtained as
\begin{align}
 \frac{\Ysf_1(q^{-1} x)}{\Ysf_2(q^{-1} x)}
 = \Ysf_1(q^{-1} x) \times \frac{1}{\Ysf_2(q^{-1} x)}
 \quad (\text{case 2})
 \\[.5em]
 \frac{1}{\Ysf_1(q^{-2} x)} =
 \frac{\Ysf_2(q^{-1} x)}{\Ysf_1(q^{-2} x)} \times \frac{1}{\Ysf_2(q^{-1} x)}
 \quad (\text{case 3})
\end{align}
Thus the $qq$-character of $\bar{\textbf{3}}$ for $A_2$ quiver is given by summing all the possible configurations with two defect branes shown in Fig.~\ref{GTA_2another}.

\subsection{Generic quiver}

The argument discussed above is extended to generic (simply-laced) quiver gauge theory.

\subsubsection{$A_r$ quiver} % with single $\Ysf$-operator}
\label{sec:Ar-quiver}

For $A_r$ quiver, there exist $r$ weights, associated with the gauge nodes, and the fundamental representation is obtained from each (highest) weight, which is the antisymmetric representation of SU($r+1$).
The $qq$-character of the degree $n$ antisymmetric representation $R_n'$ ($n = 1, \ldots, r$) is given by~\cite{Nekrasov:2015wsu}
\begin{align}
 \chi_{R'_n}(A_r; q_1,q_2)
 & =
 \Big< \Tsf_n(x) \Big>
 \nonumber \\
 &
 :=
 \left( \prod_{k=1}^{n} \mathfrak{Q}_k \right)^{-1}
 \Psf_1(q^{-n} x)
 \sum_{1 \le i_1 < \cdots < i_n \le r+1}
 \left< \prod_{k=1}^n \Lambda_{i_k} (q^{-n+k} x) \right>
 \nonumber \\
 & =
 \Big< \Ysf_n(x) \Big>
 + \qfrak_n
 \left<
 \frac{\Ysf_{n-1}(q^{-1} x) \Ysf_{n+1}(x)}{\Ysf_n(q^{-1} x)}
 \right>
 + \cdots
 \label{eq:qq-ch_Ar}
\end{align}
where $\qfrak_n$ is the gauge coupling of the $n$-th gauge node, and we define
\begin{align}
 \Lambda_i(x) & = \mathfrak{Q}_i \,
 \frac{\Ysf_i(x)}{\Ysf_{i-1}(q^{-1} x)}
 \label{eq:Lambda_Ar}
\end{align}
with $\Ysf_0(x) = \Psf_1(x)$,  $\Ysf_{r+1}(x) = \Psf_r(x)$ and
\begin{align}
 \mathfrak{Q}_i = \prod_{k=1}^{i-1} \qfrak_k
 \, .
\end{align}
We can see that the $qq$-character is generated by the iWeyl reflection
\begin{align}
 \Ysf_n(x)
 \ \longrightarrow \
 \qfrak_n \,
 \frac{\Ysf_{n-1}(q^{-1} x) \Ysf_{n+1}(x)}{\Ysf_n(q^{-1} x)}
 \, .
\end{align}
In this case there are $r+1$ NS5-branes, so that $r+1$ possibilities for the brane insertion.
Indeed the factor $\Lambda_i(x)$ defined as \eqref{eq:Lambda_Ar} corresponds to the insertion of single defect brane.
Thus the $qq$-character of $R_n'$ is realized as the summation of all the possible configurations with $n$ brane insertions, since it involves a product of $n$ $\Lambda$-factors as shown in \eqref{eq:qq-ch_Ar}.

\subsubsection{$DE$ quiver}

Let us then discuss $DE$ quiver theory.
In this case, it is not straightforwardly possible to obtain the toric Calabi--Yau threefold reproducing $DE$ quiver gauge theory, due to the trivalent node in the quiver.
Recently it has been proposed that $DE$-type gauge theory can be constructed from the (non-toric) Calabi--Yau geometry~\cite{Hayashi:2017jze}, and thus it is expected that we can discuss the $qq$-character by inserting the defect brane to such a $DE$-type configuration.

The simplest non-trivial $DE$-type theory is $D_4$ quiver.
In this case there are four fundamental representations corresponding to the nodes in $D_4$ quiver, three 8-dimensional and one 28-dimensional representations.
The three \textbf{8}-representations are essentially equivalent to each other, which is so-called the SO(8) triality.
In particular, for the \textbf{28}-representation, the corresponding $qq$-character involves a derivative term, due to the collision limit of the $\Ysf$-operators~\cite{Nekrasov:2015wsu,Kimura:2016ebq}, corresponding to the vanishing cycle as discussed in Sec.~\ref{sec:higher_ch}, and it would be interesting to study the geometric meaning of the collision limit.

\subsubsection{Beyond $ADE$ quiver}

For $ADE$ quiver, all the fundamental representations are finite dimensional, and thus the ($qq$-)character is given by a finite (elementary symmetric) polynomial of $\{\Lambda_i\}$, which is a ratio of the $\Ysf$-operator~\eqref{eq:Lambda_Ar}.
In general, we can consider the quiver, which does not correspond to the finite $ADE$-type Dynkin diagram, namely affine and hyperbolic quivers.
Although, in such a case, the fundamental representations become infinite dimensional, we can discuss the $qq$-character generated by the iWeyl reflection.
For example, the affine quiver $\hat{A}_r$ is realized using the infinitely-long linear quiver $A_\infty$ by imposing periodicity.
Thus there are infinitely many possibilities for the brane insertion.
This is a geometric interpretation of the infinite sum in the affine $qq$-character.
For the simplest case $\hat{A}_0$ corresponding to 4d $\Ncal=2^*$ (5d $\Ncal = 1^*$) theory, the $qq$-character is described as a summation over the partition~\cite{Nekrasov:2015wsu,Kimura:2015rgi}.

%%%%%%%%%%%%%%%%%% section 4 %%%%%%%%%%%%%%%%%%%%%%%%%%%%%%%%%
\section{Summary and Discussion}
\label{sec:summary}
%In this paper, we have proposed the prescription of the geometric transition in refined topological string especially with respect to the inner brane.
%In order to obtain a proper contribution of the brane insertion, we have to shift the parameter, which becomes trivial in the unrefined limit.
%We then have applied this prescription to the codimension-4 defect operator, called the $\Ysf$-operator.
%The pole singularity of the $\Ysf$-operator is cancelled out in a proper combination of the $\Ysf$-operators, which is given by the $qq$-character.
%We have examined the pole cancellation in the $qq$-character as a nontrivial check of our prescription of the refined geometric transition.
In this paper, we have proposed the prescription of the geometric transition in the refined topological string enforced along the preferred direction. In order to obtain a proper contribution of the brane insertion, in addition to the specialization of the K\"ahler moduli, we have to shift the variable by hand to satisfy consistency, which becomes trivial in the unrefined limit. We then have applied this prescription to the codimension-4 defect operator, called the $\Ysf$-operator as its stringy realization. The pole singularity of the $\Ysf$-operator is cancelled out in a proper combination of the $\Ysf$-operators, which is given by the $qq$-character. We have examined the pole cancellation in the $qq$-character as a nontrivial check of our prescription of the refined geometric transition.

Let us finally provide several open questions which we would like to resolve. As commented, the refined large $N$ duality between the resolved and deformed conifold has been clarified in terms of the refined Chern--Simons theory \cite{Kameyama:2017ryw}. Nevertheless, the corresponding brane configuration is not clear from their argument, and as the first issue, we would pursue that our geometric transition may give a actual brane picture compatible with their result. Second, it may be possible that our prescription in Section \ref{New} is generalized so as to incorporate the labels $( p, q )$ of the fivebrane charges, as mentioned there. The third thing is concerned with the exact definition of the refined version of the open topological vertex formalism. As far as we know, it is not yet established, and thus, the direct computation of the open string amplitude respecting the Lagrangian brane on the inner brane is still a nontrivial problem. In the unrefined case, the Schur function is suitable to capture the holonomy of D-branes corresponding to the insertion of the Lagrangian brane. It is expected from the results of \cite{Kameyama:2017ryw} that the Schur function would be replaced with the Macdonald function in the refined case as done for the refined topological vertex in \cite{Awata:2005fa}. Combining the expression obtained via the refined geometric transition, we hope that the successful direct approach would be reported in the near future.

We also hold some technical and qualitative issues on the $\Ysf$-operator.
In the topological string approach, there is an ambiguity of the normalization.
Actually the $\Ysf$-operator and the $qq$-character have factors independent of the partition $\mu$, and we need to add such a factor by hand to obtain a proper result.
It would be interesting to clarify a systematic way to discuss the $\mu$-independent factor in the framework of refined topological string.

The brane configuration of the $\Ysf$-operator proposed in this paper is due to the comparison with the gauge theory definition.
The current construction of the codimension-4 $\Ysf$-operator uses the codimension-2 surface defects with the $q$-brane and anti-$q$-brane.
Such a relation between defect operators with different codimensions is not yet obvious.
One possible interpretation is the tachyon condensation, which could be related to the (refined) supergroup Chern--Simons theory~\cite{Vafa:2001qf}.
For example, it is interesting to compare the Y-operator contribution with the partition function of the refined U(1$|$1) Chern--Simons theory~\cite{Kimura:2015dfa}.
More detailed analysis is necessary for understanding its geometric meaning in refined theory.

%To be discussed:
%\begin{itemize}
% \item It would be interesting to derive the factor $\mathcal{S}(Q_{x})$ from the refined topological string theory.
% \item It would be interesting to explain why the $\Tsf$-operator is obtained by summing the all possible brane configuration.
% \item Compare the Y-operator contribution with the partition function of the refined U(1$|$1) Chern--Simons theory~\cite{Kimura:2015dfa}
% \item Brane configuration of the supergroup Chern--Simons theory~\cite{Mikhaylov:2014aoa}
%\end{itemize}

\subsection*{Acknowledgments}
We would like to thank Shamil Shakirov, Masato Taki, Satoshi Yamaguchi, and Yegor Zenkevich for giving helpful comments.
The work of T. K. was supported in part by Keio Gijuku Academic Development Funds, JSPS Grant-in-Aid for Scientific Research (No. JP17K18090), the MEXT-Supported Program for the Strategic Research Foundation at Private Universities ``Topological Science'' (No. S1511006), and JSPS Grant-in-Aid for Scientific Research on Innovative Areas ``Topological Materials Science'' (No. JP15H05855).
The work of H. M. and Y. S. was supported in part by the JSPS Research Fellowship for Young Scientists.

\appendix
%%%%%%%%%%%%%%%%%% section A %%%%%%%%%%%%%%%%%%%%%%%%%%%
\section{Definitions and notations} \label{secA}

%%%%%%%%%%%%%%%%%% section A.1 %%%%%%%%%%%%%%%%%%%%%%%%%%
\subsection{Mathematical preliminaries} \label{Mpre}

\subsection*{Young diagram} %%%%%%%%%%%%%%%%%%%%%%%%%%%%%%%%%%%%
\begin{figure}[t] % Young diagram
\begin{center}
\begin{tabular}{ccc}
	\begin{minipage}[b]{.3\hsize}
		\begin{center}
		\subfigure[Positions of boxes]{\label{yd1}\includegraphics[width=4cm,bb=250 110 600 500,clip]{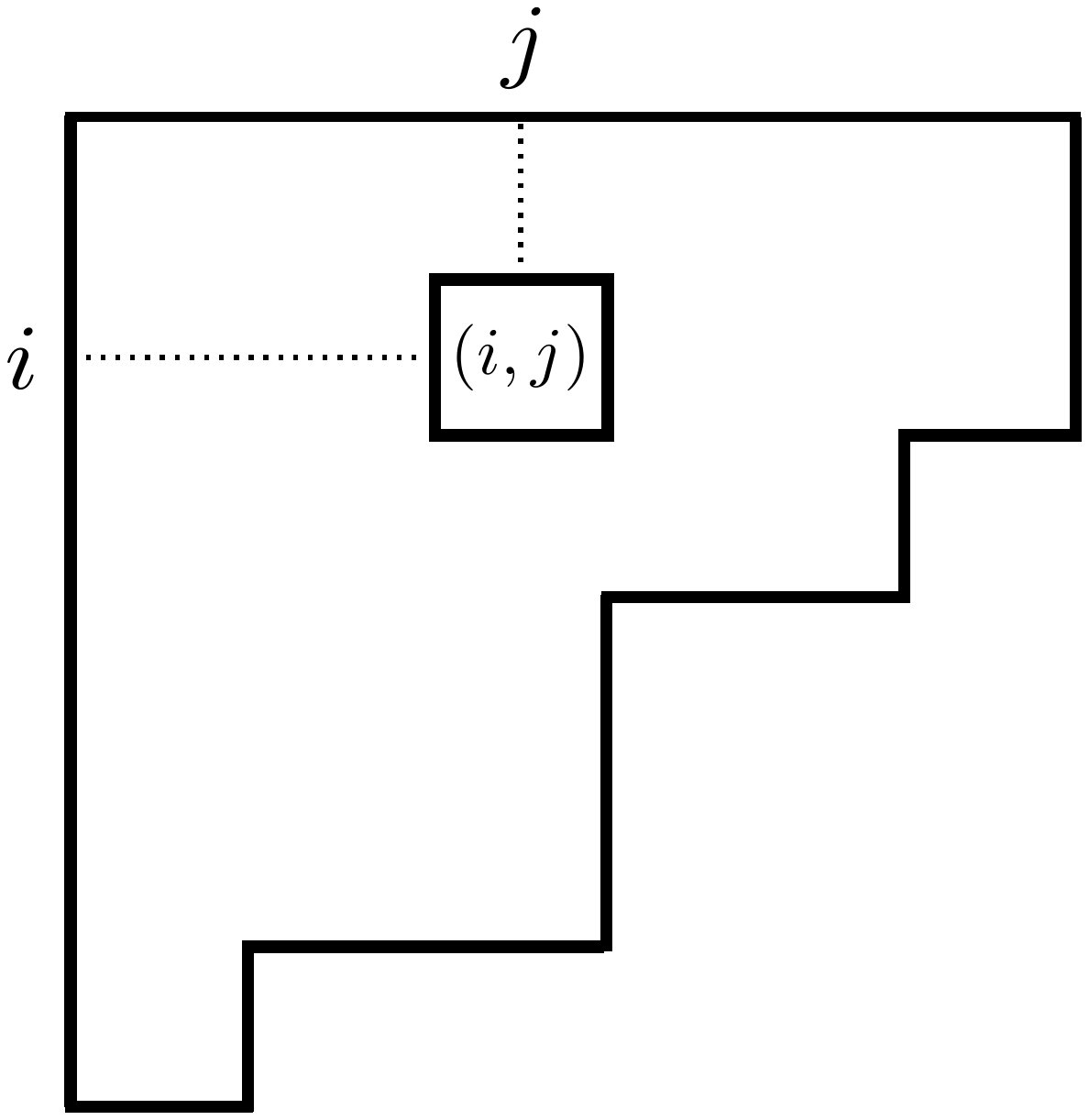}}
		\end{center}
	\end{minipage}
	&
	\hspace{.25em}
	\begin{minipage}[b]{.3\hsize}
		\begin{center}
		\subfigure[$l$-tuple diagram]{\label{yd2}\includegraphics[width=4cm,bb=250 110 600 500,clip]{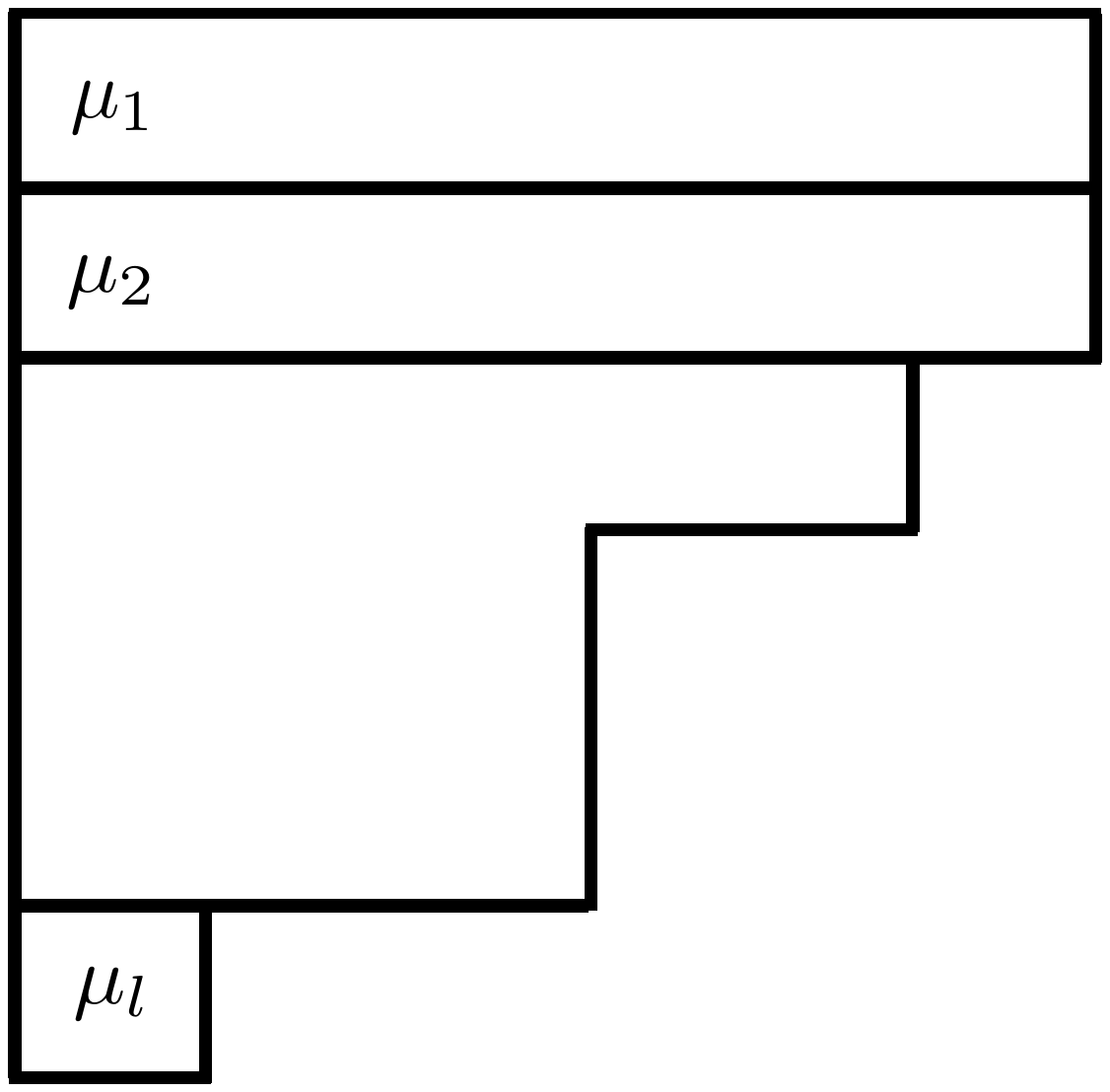}}
		\end{center}
	\end{minipage}
	&
	\hspace{.25em}
	\begin{minipage}[b]{.3\hsize}
		\begin{center}
		\subfigure[Transpose]{\label{yd3}\includegraphics[width=4cm,bb=250 110 600 500,clip]{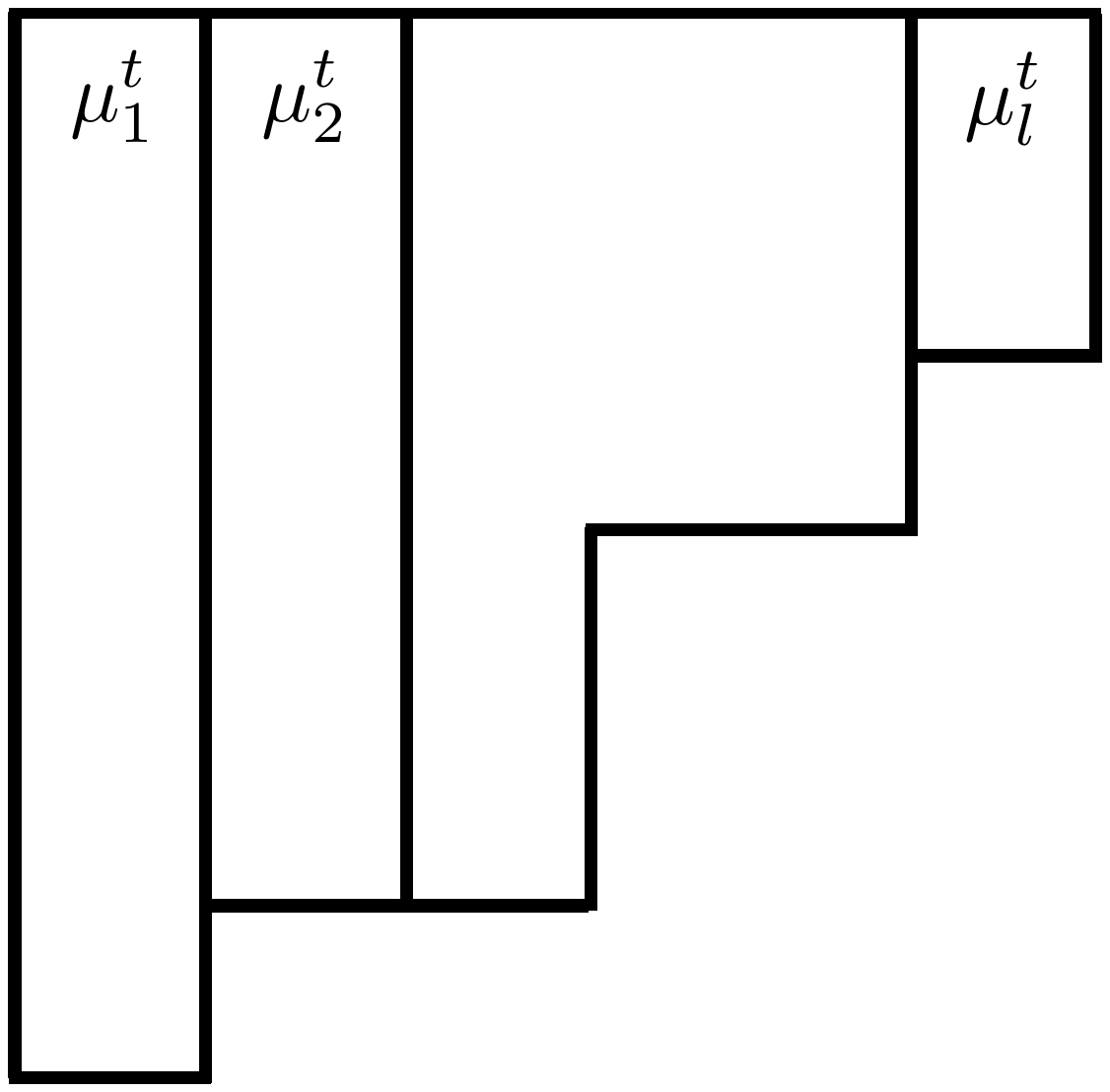}}
		\end{center}
	\end{minipage}
\end{tabular}
\caption{\label{yd}The Young diagram and its parameters.}
\end{center}
\end{figure}

To define the Young diagram, we take the decreasing sequence of nonnegative integers that is regularly used for the instanton counting problem. Let $( i, j )$ be positions of boxes in the diagram (shown in Fig.\,\ref{yd1}), then we denote as $\mu$ a Young diagram of the following set of $l$-tuple diagrams (Fig.\,\ref{yd2}):
\begin{xalignat}{2} %
\mu &= \lc \mu_{i} \in \Zbb_{\geq 0} | \mu_{1} \geq \mu_{2} \geq \cdots \geq \mu_{l} \rc, &
\mu^{t} &= \lc \mu_{j}^{t} \in \Zbb_{\geq 0} | \mu_{j}^{t} = \# \{ i | \mu_{i} \geq j \} \rc,
\label{ydef1}
\end{xalignat}
where the transpose of $\mu$ is indicated by the superscript $t$ (Fig.\,\ref{yd3}). For a given Young diagram $\mu$, we use the following simplified symbols:
\begin{xalignat}{3} %
|\mu| &= \sum_{i = 1}^{l} \mu_{i},
&
||\mu||^2 &= \sum_{i = 1}^{l} \mu_{i}^{2},
&
\prod_{( i, j ) \in \mu} f ( i, j ) &= \prod_{i = 1}^{l} \prod_{j = 1}^{\mu_i} f ( i, j ).
%= \prod_{j = 1}^{\check{d} ( Y )} \prod_{i = 1}^{Y_i^{\text{T}}} f ( i, j ),
\label{ydef2}
\end{xalignat}
The first one in \eqref{ydef2} is the total number of boxes of $\mu$. The partitions $\{ \mu_i \}$ and $\{ \mu_j^t \}$ concretely characterize the instanton partition function, which can be removed by using
\begin{align} %
    \begin{aligned}
    \sum_{j = 1}^{\mu_i} \lp \mu_i - j \rp &= \sum_{j = 1}^{\mu_i} \lp j - 1 \rp && \text{for fixed $i$}, \\ %1
    \sum_{i = 1}^{\mu_j^t} \lp \mu_j^t - i \rp &= \sum_{i = 1}^{\mu_j^t} \lp i - 1 \rp &&  \text{for fixed $j$}. %2
    \end{aligned}
\end{align}
In the paper, these are implicitly applied as expressing the $\Ysf$-operator in a convenient fashion from the general form obtained via the refined geometric transition in Section \ref{New}.

%We define the Young diagram $\mu$ as the following figure:
%\begin{figure}[htbp]
%\centering
%    \includegraphics[width=15cm]{figure/young.pdf}
%    \caption{The Young diagram.  We define $\mu_{i}$ as (a), $\mu^{t}_{j}$ as (b), and the coordinates $(i,j)$ as (c). $\mu_{i}$ is the number of boxes in the $i$-th horizontal line. $\mu_{j}^{t}$ is the number of boxes in the $j$-th vertical line.}
%    \label{Young}
%\end{figure}

\subsection*{Theta function} %%%%%%%%%%%%%%%%%%%%%%%%%%%%%%%%%%%%
The topological string amplitude for the compactified web diagram of our interest is nicely expressed in terms of the theta function,
\begin{align} %
\theta_{1} ( z| \tau )
=
- i e^{\pi \mathrm{i} \tau \over 4} e^{\pi \mathrm{i} z}
\prod_{k = 1}^{\infty}
\lp 1 - e^{2 \pi \mathrm{i} k \tau} \rp
\lp 1 - e^{2 \pi \mathrm{i} k \tau} e^{2 \pi i z} \rp
\lp 1 - e^{2 \pi \mathrm{i} ( k - 1 ) \tau} e^{- 2 \pi \mathrm{i} z} \rp,
\label{theta1a}
\end{align}
where a variable is $z \in \Cbb$, and $\tau \in \Cbb$ is a constant with $\text{Im} ( \tau ) > 0$. Equivalently, the theta function is frequently used in the multiplicative form,
\begin{align} %
\theta_{1} ( x; q )
=
- \mathrm{i} q^{1 \over 8} x^{1 \over 2} ( q, q x, x^{- 1}; q )_{\infty},
\label{theta1m}
\end{align}
where $x := e^{2 \pi \mathrm{i} z}$, $q := e^{2 \pi \mathrm{i} \tau}$, and the $q$-Pochhammer symbol ($q$-shifted factorial) is defined by
\begin{align} % q-shifted factorial
( x ; q )_{n} = \left\{
	\begin{aligned}
	& 1 && \mbox{for } n = 0, \\[.75em] %1
	& \prod_{k = 0}^{n - 1} ( 1- x q^{k} ) && \mbox{for } n \geq 1, \\ %2
	& \prod_{k = 1}^{- n} ( 1- x q^{- k} )^{- 1} && \mbox{for } n \leq - 1. %3
	\end{aligned} \right.
\label{qshifted}
\end{align}
In addition, $( x ; q )_\infty := \lim_{n \to \infty} ( x ; q )_{n}$ with $| q | < 1$ and we use the shorthand notation
\begin{align}
( x_{1}, x_{2}, \cdots, x_{r} ; q )_{n} := ( x_{1} ; q )_{n} ( x_{2} ; q )_{n} \cdots ( x_{r}; q )_{n}.
\end{align}
Note that \eqref{theta1a} and \eqref{theta1m} are nothing but the Jacobi's triple product identity. This theta function actually has the simple inversion property and satisfies the $q$-difference equation,
\begin{align} %
\theta_{1} ( x^{- 1}; q )
&=
- \theta_{1} ( x; q ), \label{inversiont1} \\ %1
\theta_{1} ( x q^n; q )
&=
( - x )^{- n}
q^{- \frac{n^2}{2}}
\theta_{1} ( x; q ) \hspace{1em} \text{for $n \in \Zbb$}. \label{difft1} %2
\end{align}

We further give another type of the theta function defined by
\begin{align}
\theta ( x; q )
=
\frac{1}{( q; q )_\infty} \sum_{n \in \Zbb} ( - 1 )^{n} q^{\frac{1}{2} n ( n - 1 )} x^{n}
=
( x, q x^{- 1}; q )_\infty.
\end{align}
This theta function is simply translated into $\theta_{1} ( x; q )$ via the Jacobi's triple product identity,
\begin{align} %
\theta_{1} ( x; q )
=
\mathrm{i} q^{1 \over 8} x^{- {1 \over 2}} ( q; q )_{\infty}\, \theta ( x; q ).
\label{connection}
\end{align}
We can immediately verify that this theta function actually satisfies the $q$-difference equations,
\begin{align}
\theta ( x^{- 1}; q )
&= - x^{- 1} \theta ( x; q )
= \theta ( x q; q ), \label{inversion} \\ %1
\theta ( x q^{n}; q )
&=
( - x )^{- n}
q^{- \frac{n ( n - 1 )}{2}}
\theta ( x; q ), \label{ethetav1} \\ %2
\theta ( x q^{n}; q; p )_{m}
&=
( - x )^{- n m}
q^{- \frac{n m ( n - 1 )}{2}}
p^{- \frac{n m ( m - 1 )}{2}}
\theta ( x; q; p )_{m}, \label{ethetav2} %3
\end{align}
where we define
\begin{align} %
\theta ( x; q; p )_{m}
:=
\prod_{s = 0}^{m - 1} \theta ( x p^{s}; q ).
\end{align}
We remark that the $q \to 0$ limit of the theta function becomes simply
\begin{align} %
\lim_{q \to 0} \theta ( x; q ) = 1 - x.
\end{align}
It will be turned out that this limiting formula is actually the operation of the dimensional reduction from 6d to 5d at the level of the partition function.

%\rem{We define the two kinds of the theta function as follows:
%\begin{align}
%\theta(x)&=
%-\mathrm{i} e^{\frac{\mathrm{i} \pi \tau}{4}}x^{\frac{1}{2}}
%\prod_{n=0}^{\infty}
%\Bigl\{
%(1-e^{2 \pi \mathrm{i}(n+1) \tau})
%(1-e^{2 \pi \mathrm{i}(n+1) \tau}x)
%(1-e^{2 \pi \mathrm{i} n \tau}x^{-1})
%\Bigr\}
%,
%\\
%\Theta(x)&=
% \prod_{n=0}^\infty
% (1-e^{2 \pi \mathrm{i}(n+1) \tau}x)
% (1-e^{2 \pi \mathrm{i} n \tau}x^{-1}) 
% %\mathrm{i}e^{-\frac{i\pi\tau}{4}}x^{-\frac{1}{2}}(Q_{\tau};Q_{\tau})_{\infty}^{-1}\theta(x)
%\end{align}
%This theta function satisfies the following relations:
%\begin{align}
%&\theta(Q_{\tau}x)=-x^{-1}Q_{\tau}^{-\frac{1}{2}}\theta(\tau ; x),
%\\
%&\theta(x)=-\theta(x^{-1}).
%\end{align}
%}

\subsection*{Elliptic gamma function} %%%%%%%%%%%%%%%%%%%%%%%%%%%%%%%%
The elliptic gamma function is defined by
\begin{align} %
\G_{e} ( x )
:=
\G ( x; p, q )
=
\prod_{n, m \geq 0}
\frac{1 - x^{- 1} p^{n + 1} q^{m + 1}}
{1 - x p^{n} q^{m}},
\end{align}
with $|p|, |q| < 1$, and $x \in \Cbb^{\ast}$. For specific values of $x$, the elliptic gamma function get simplified as
\begin{align} %
\G_{e} ( p )
= \frac{( q; q )_\infty}{( p; p )_\infty}, &&
\G_{e} ( q )
= \frac{( p; p )_\infty}{( q; q )_\infty}, &&
\G_{e} ( - 1 )
= \frac{1}{2 ( - p; p )_\infty ( - q; q )_\infty}.
\end{align}
The certain combinations of elliptic gamma function are related to the theta function defined above as follows:
\begin{align} %
\G_{e} ( x )
\G_{e} ( x^{- 1} )
=
\frac{1}{\theta ( x; p ) \theta ( x^{- 1}; q )}
=
\frac{1}{\theta ( x; q ) \theta ( x^{- 1}; p )}
\end{align}
because $p$ and $q$ are encoded symmetrically into the elliptic gamma function, in addition, we find the difference equations involving the theta function,
\begin{align} %
\G_{e} ( x p )
&=
\theta ( x; q )
\G_{e} ( x ), \hspace{5.75em}
\G_{e} ( x q )
=
\theta ( x; p )
\G_{e} ( x ), \\ %1
\G_{e} ( x p^{n} )
&=
%\prod_{r = 0}^{n - 1} \theta ( x p^{r}; q )
%=
\theta ( x; q; p )_{n}
\G_{e} ( x ), \hspace{3.5em}
\G_{e} ( x q^{m} )
=
%\prod_{s = 0}^{m - 1} \theta ( x q^{s}; p )
%=
\theta ( x; p; q )_{m}
\G_{e} ( x ), \\ %2
\G_{e} ( x p^{n} q^{m} )
&=
( - x )^{- m n}
p^{- \frac{1}{2} n m ( n - 1 )}
q^{- \frac{1}{2} n m ( m - 1 )}
%\prod_{r = 0}^{n - 1} \theta ( x p^{r}; q )
\theta ( x; q; p )_{n}
%\prod_{s = 0}^{m - 1} \theta ( x q^{s}; p )
\theta ( x; p; q )_{m}
\G_{e} ( x ). %3
\end{align}
for $n, m \in \Zbb$. Note that the first line represents the finite difference equations of the first order \cite{Spiridonov:2009za} that can lead to the second line, in other words, the last relation can be derived in the recursive manner from the first one. Furthermore, there are the limiting relations \cite{Spiridonov:2009za},
\begin{align} % the limiting relations
\lim_{p \to 0}
\G_{e} ( x )
&=
\frac{1}{( x; q )_\infty}, \\ %1
\lim_{x \to 1}
( 1 - x )
\G_{e} ( x )
&=
\frac{1}{( p; p )_\infty ( q; q )_\infty}. %2
\end{align}
Moreover, we have the reflection identity,
\begin{align}
\G_{e} \lp ( p q )^{\frac{a}{2}} x^{b} \rp
\G_{e} \lp ( p q )^{\frac{2 - a}{2}} x^{- b} \rp
=
1.
\label{egamma2}
\end{align}
The usage of the elliptic gamma function is underlying a nontrivial property linking its specific ratio to the theta function involving Young diagrams \cite{Nieri:2015dts} (see also \cite{Iqbal:2015fvd}),
\begin{align} %
\prod_{( i, j ) \in \mu} \theta ( Q p^{\mu_{i} - j} t^{\nu_{j}^{t} - i + 1}; q )
\prod_{( i, j ) \in \nu} \theta ( Q p^{- \nu_{i} + j - 1} t^{- \mu_{j}^{t} + i}; q )
=
\prod_{i, j \geq 1}
\frac{
\Gamma_{e} ( Q t^{j - i + 1}; p, q )
\Gamma_{e} ( Q p^{\mu_{i} - \nu_{j}} t^{j - i}; p, q )
}{
\Gamma_{e} ( Q t^{j - i}; p, q )
\Gamma_{e} ( Q p^{\mu_{i} - \nu_{j}} t^{j - i + 1}; p, q )
}.
\end{align}
Note that it has been reported in \cite{Nekrasov:2003rj} that there exists a similar formula involving the gamma function for the Nekrasov function for the 4d theory. Further, the 5d Nekrasov function is similarly written in terms of the $q$-gamma function.

%\begin{align} %
%\prod_{( i, j ) \in \mu} \theta ( Q q_1^{-\mu_{i} + j} q_2^{\nu_{j}^{t} - i + 1} )
%\prod_{( i, j ) \in \nu} \theta ( Q q_1^{ \nu_{i} - j + 1} q_2^{- \mu_{j}^{t} + i} )
%=
%\prod_{i, j \geq 1}
%\frac{
%\Gamma_{e} ( Q q_2^{j - i + 1}; q_1^{-1}, Q_{\tau} )
%\Gamma_{e} ( Q q_1^{-\mu_{i} + \nu_{j}} q_2^{j - i};  q_1^{-1}, Q_{\tau} )
%}{
%\Gamma_{e} ( Q q_2^{j - i}; q_1^{-1}, Q_{\tau} )
%\Gamma_{e} ( Q q_1^{-\mu_{i} + \nu_{j}} q_2^{j - i + 1}; q_1^{-1}, Q_{\tau} )
%}.
%\end{align}
%\rem{A similar formula for the Nekrasov function using the $\Gamma$-function has been already mentioned in \cite{Nekrasov:2003rj} for 4d theory.}

%We define the elliptic gamma function as follows:
%\begin{align}
%\Gamma_{e}(x;p,q)=\frac{(pqx^{-1};p,q)_{\infty}}{(x;p,q)_{\infty}},~(x;p,q)_{\infty}=\prod_{j,k\geq 0}(1-p^j q^k x).
%\end{align}
%This function satisfies
%\begin{align}
%\prod_{(i,j)\in\mu}\theta(Q t^{\mu_{i}-j}q^{\nu_{j}^{t}-i+1};\tau)
%\prod_{(i,j)\in\nu}\theta(Q t^{-\nu_{i}+j-1}q^{-\mu_{j}^{t}+i};\tau)
%=
%\prod_{i,j\geq1}
%\frac{
%\Gamma_{e}(Qq^{j-i+1};t,Q_{\tau})\Gamma_{e}(Qt^{\mu_{i}-\nu_{j}}q^{j-i};t,Q_{\tau})
%}{
%\Gamma_{e}(Qq^{j-i};t,Q_{\tau})\Gamma_{e}(Qt^{\mu_{i}-\nu_{j}}q^{j-i+1};t,Q_{\tau})
%}
%\end{align}
%and
%\begin{align}
%\theta(x)
%=\frac{\Gamma_{e}(qx;q,Q_{\tau})}{\Gamma_{e}(x;q,Q_{\tau})}.
%\end{align}

%%%%%%%%%%%%%%%%%% section A.2 %%%%%%%%%%%%%%%%%%%%%%%%%%
\subsection{Refined topological vertex} \label{Rtv}
In this paper, we rely on the Iqbal--Koz\c{c}az--Vafa formalism \cite{Iqbal:2007ii} for the refined topological vertex $C_{\lambda \mu \nu} ( t, q )$ given by
\begin{align} %
C_{\lambda \mu \nu} ( t, q )
=
t^{- \frac{|| \mu^{t} ||^{2}}{2}} q^{\frac{|| \mu ||^2 + ||\nu||^{2}}{2}}
\tilde{Z}_{\nu} ( t, q )
\sum_{\eta}
\lp \frac{q}{t} \rp^{\frac{|\eta| + |\lambda| - |\mu|}{2}}
s_{\lambda^{t}/\eta} ( t^{- \rho} q^{- \nu} ) s_{\mu/\eta} ( t^{- \nu^{t}} q^{- \rho} ),
\end{align}
\label{rtv}
where $s_{\la/\mu} ( x )$ is the skew Schur function and
\begin{xalignat}{2} %
\tilde{Z}_{\nu} ( t, q ) &= \prod_{( i, j ) \in \nu} \frac{1}{1 - q^{\nu_{i} - j} t^{\nu_{j}^{t} - i + 1}}, &
\rho &= \lc - {1 \over 2},\, - {3 \over 2},\, - {5 \over 2}, \cdots \rc.
\end{xalignat}
The function $\tilde{Z}_{\nu} ( t, q )$ is essentially the Macdonald function $P_{\nu} ( x; q, t )$ \cite{macdonald1998symmetric}
\begin{align} %
\widetilde{Z}_{\nu} ( t, q )
=
t^{- {|| \nu^{\text{T}} ||^2 \over 2}}
P_{\nu} ( t^{- \rho}; q, t ).
\end{align}
We do not go further details of the refined topological vertex and trace back the calculation of the partition function \eqref{TSpart.fn.} that has been accomplished in \cite{Haghighat:2013tka}. Note that the parameters $( q, t^{- 1} )$ are replaced in the main context of the paper with $( q_1, q_2 )$, respectively. We would like to comment on the fact that this partition function is absolutely reproduced by using the Awata--Kanno formalism for $C_{\lambda \mu \nu} ( t, q )$ \cite{Awata:2005fa, Awata:2008ed}.

%%%%%%%%%%%%%%%%%% section B %%%%%%%%%%%%%%%%%%%%%%%%%%%
\section{Regularity} \label{secB}
In this appendix we show the regularity of the $qq$-character in the case of $A_{1}$, $A_{2}$ quiver with the single $\Ysf$-operator, and $A_{1}$ quiver with two $\Ysf$-operators. The strategy is as follows:
\begin{itemize}
\item[1.] We write the partition function and the $\Ysf$-operator to the infinite product form.
\item[2.] We calculate the ratio $\mathcal{Z}_{\mu}^{{\rm U}(1)}/\mathcal{Z}_{\mu+1}^{{\rm U}(1)}$ and the product $\Ysf_{\mu}\Ysf_{\mu+1}$, where $\mu+1$ denotes the Young diagram that we add the one box to some row $\mu_{I}$, namely $\mu_I \to \mu_I + 1$.
\item[3.] Then, we find that the ratio of the partition functions relates to the product of the $\Ysf$-operators. 
\end{itemize}
We will demonstrate these steps. Note that we consider the regularity for the variable $Q_{x}$ instead of the $x$-variable while we focus on U(1) theory.

%%%%%%%%%%%%%%%%%% section B-1 %%%%%%%%%%%%%%%%%%%%%%%%%%%%%%%

%\subsection{$A_1$ quiver with the single $\Ysf$-operator}
\subsection{$A_1$ quiver}

\subsubsection{${\rm U}(1)$ gauge theory with single $\Ysf$-operator}\label{secB1}
To begin with, let us consider the simplest case. By using the formula in Appendix \ref{secA}, we write the partition function and the $\Ysf$-operator to the infinite product form as follows,
\begin{align}
\mathcal{Z}_{\mu}^{{\rm U}(1)}
&=
(-1)^{|\mu|}(Q_{1}^{(1)}Q_{1}^{(2)^{-1}})^{\frac{|\mu|}{2}}q_2^{\frac{|\mu|}{2}}q_2^{\frac{||\mu||^2}{2}}q_1^{\sum_{(i,j)\in\mu}i}
\nonumber \\
&\quad\times
\prod_{i,j\geq1}
\frac{
\Gamma_{e}(Q_{1}^{(2)^{-1}}q_2^{-1}q_1^{j-i} )\Gamma_{e}(Q_{1}^{(2)^{-1}}q_2^{\mu_{j}-1}q_1^{j-i-1} )
\Gamma_{e}(Q_{1}^{(1)}q_1^{j-i+1} )\Gamma_{e}(Q_{1}^{(1)}q_2^{\mu_{j}}q_1^{j-i} )
}{
\Gamma_{e}(Q_{1}^{(2)^{-1}}q_2^{-1}q_1^{j-i-1})\Gamma_{e}(Q_{1}^{(2)^{-1}}q_2^{\mu_{j}-1}q_1^{j-i} )
\Gamma_{e}(Q_{1}^{(1)}q_1^{j-i} )\Gamma_{e}(Q_{1}^{(1)}q_2^{\mu_{j}}q_1^{j-i+1} )
}
\nonumber \\
&\quad\quad\quad\times
\frac{
\Gamma_{e}(q_2^{-1}q_1^{j-i-1})\Gamma_{e}(q_2^{-\mu_{i}+\mu_{j}-1}q_1^{j-i} )
}{
\Gamma_{e}(q_2^{-1}q_1^{j-i} )\Gamma_{e}(q_2^{-\mu_{i}+\mu_{j}+1}q_1^{j-i-1} )
},
\\
\Ysf_{\mu}(x)
&=
-{\rm i}e^{\frac{{\rm i}\pi\tau}{4}}Q_{x}^{\frac{1}{2}}
\prod_{i\geq1}
\frac{
\theta(Q_{x}q_2^{\mu_{i}}q_1^{i-1})
}{
\theta(Q_{x}q_2^{\mu_{i}}q_1^{i})
}
 ,
 \label{eq:Y-op_inf_prod}
\end{align}
%\rem{Convention to be checked: $q_1 \leftrightarrow q_2$}
where we denote the elliptic gamma function $\Gamma_{e}(x;q_2^{-1},Q_{\tau})=:\Gamma_{e}(x)$ for simplicity, and $Q_x = Q_1 / x$.
Note that the $\mu$-independent factors are interpreted as the one-loop contribution, and the remaining ones are the full partition function.
By using the reflection of the theta function $\theta_{1}(x)=-\theta_{1}(x^{-1})$, the $\Ysf$-operator can also be written as
\begin{align}
\Ysf_{\mu}(x)=
{\rm i}e^{\frac{{\rm i}\pi\tau}{4}}Q_{x}^{-\frac{1}{2}}
\prod_{i\geq1}
\frac{
\theta(Q_{x}^{-1}q_2^{-\mu_{i}}q_1^{-i+1})
}{
\theta(Q_{x}^{-1}q_2^{-\mu_{i}}q_1^{-i})
}.
\end{align}
This coincides with the definition in \cite{Kimura:2016dys}, up to a trivial factor.
Let us consider the ratio $\mathcal{Z}_{\mu}^{{\rm U}(1)}/\mathcal{Z}_{\mu+1}^{{\rm U}(1)}$ and the product $\Ysf_{\mu}(q^{-1}x)\Ysf_{\mu+1}(x)$. After some calculations, we have
\begin{align}
&\frac{
\mathcal{Z}_{\mu}^{{\rm U}(1)}
}{
\mathcal{Z}_{\mu+1}^{{\rm U}(1)}
}
=
\frac{-q_2^{-\mu_{I}-1}q_1^{-I}(Q_{1}^{(1)}Q_{1}^{(2)^{-1}})^{-\frac{1}{2}}}
{\theta(Q_{m}^{-1}q_2^{\mu_{I}}q_1^{I-1})\theta(Q_{m}q_2^{\mu_{I}+1}q_1^{I})}
\prod_{i\geq1,i\neq I}
\frac{
\theta(q_2^{-\mu_{I}+\mu_{i}-1}q_1^{i-I-1})\theta(q_2^{-\mu_{i}+\mu_{I}}q_1^{I-i})
}{
\theta(q_2^{-\mu_{I}+\mu_{i}-1}q_1^{i-I})\theta(q_2^{-\mu_{i}+\mu_{I}}q_1^{I-i-1})
}
,\\
&\Ysf_{\mu}(q^{-1}x)\Ysf_{\mu+1}(x)
=
q_2^{-\frac{1}{2}}q_1^{-\frac{1}{2}}
e^{\frac{{\rm i}\pi\tau}{2}}
\prod_{i\geq1}
\frac{
\theta(Q_{x}^{-1}q_2^{-\mu_{i}-1}q_1^{-i})\theta(Q_{x}q_2^{\mu'_{i}}q_1^{i-1})
}{
\theta(Q_{x}^{-1}q_2^{-\mu_{i}-1}q_1^{-i-1})\theta(Q_{x}q_2^{\mu'_{i}}q_1^{i})
}
\label{prodY},
\end{align}
where $\mu+1=:\mu'$ denotes the Young diagram that we add the one box to some row $\mu_{I}$, namely $\mu_I \to \mu_I + 1$, as we defined in the beginning of this section. Then by using the relation
\begin{align}
\Psf(x)
=
Q^{-1}_{x}(Q_{1}^{(1)}Q_{1}^{(2)^{-1}})^{\frac{1}{2}}q^{-\frac{1}{2}}e^{\frac{{\rm i}\pi\tau}{2}}
\theta(Q_{1}^{(1)}Q_{x}^{-1})\theta(Q_{1}^{(2)^{-1}}q^{-1}Q_{x}^{-1}),
\end{align}
we find 
\begin{align}
\frac{
\mathcal{Z}_{\mu}^{\text{U(1)}}
}{
\mathcal{Z}_{\mu+1}^{\text{U(1)}}
}
=
 - \qfrak^{-1}
 \frac{\Ysf_{\mu}(q^{-1}x)\Ysf_{\mu+1}(x)}{\Psf(x)}\Biggr|_{Q_{x}=q_2^{-\mu_{I}-1}q_1^{-I}}
 \, ,
\end{align}
%\rem{We need the gauge coupling $\qfrak$ due to the factor counting the instanton number $\qfrak^{|\mu|}$.}
which implies
\begin{align}
 \Res{ Q_{x} = q_2^{-\mu_{I}-1} q_1^{-I} }
 \left[
 \Ysf_{\mu+1}(x) \Zcal_{\mu+1}^\text{U(1)}
 +
 \qfrak \, \frac{\Psf(x)}{\Ysf_\mu(q^{-1} x)}
 \Zcal_\mu^\text{U(1)}
 \right]
 = 0
\end{align}
%\rem{In the expressions of $\Ysf_\mu(Q_x)$ and $\Psf(Q_x)$, there are $Q_x$ factors themselves. Thus we should mention that they seem not modular invariant in the sense of $\theta$-function, but they are indeed modular in the sense of $\theta_{1}$-function.}
This means that the $\Ysf$-operators $\Ysf_{\mu}(x)$ and $\Ysf_{\mu}(q^{-1}x)^{-1}$ have the poles, but the summation is regular since these poles cancelled with each other. Therefore we obtain the $\Tsf$-operator average for U(1) theory~\eqref{eq:qq-ch_A1}, which is regular for arbitrary $Q_{x}$, by the summation over the partition $\mu$.

%%%%%%%%%%%%%%%%%% section B-2 %%%%%%%%%%%%%%%%%%%%%%%%%%%%%%%

\subsubsection{U(1) gauge theory with two $\Ysf$-operators}
In this subsection we show the regularity for the ${\rm U}(1)$ theory with the two $\Ysf$-operators. The calculation is almost done in the previous subsection. In this case we have to rewrite the factor $\Ssf(x)$ in terms of the $\Ysf$-operator. This factor can be written as
\begin{align}
\Ssf\biggl(\frac{x_1}{x_2}\biggr)
 =
\Ssf\biggl(q^{-1}\frac{x_2}{x_1}\biggr)
= 
\frac{
\theta( q_{1}^{-1} Q_{x_{1}}Q_{x_{2}}^{-1})\theta( q_2^{-1} Q_{x_{1}}Q_{x_{2}}^{-1})
}{
\theta( q^{-1} Q_{x_{1}}Q_{x_{2}}^{-1})\theta(  Q_{x_{1}}Q_{x_{2}}^{-1})
}
.
\end{align}
We remark $Q_{x_1} = Q_1 / x_1$ and $Q_{x_2} = Q_1 / x_2$.
Also we show the ration of the $\Ysf$-operator,
\begin{align}
&\frac{
\Ysf_{\mu+1}(x)
}{
\Ysf_{\mu}(x)
}
=
\frac{
\theta(Q_{x}q_2^{\mu_{I}+1}q_1^{I-1})\theta(Q_{x}q_2^{\mu_{I}}q_1^{I})
}{
\theta(Q_{x}q_2^{\mu_{I}+1}q_1^{I})\theta(Q_{x}q_2^{\mu_{I}}q_1^{I-1})
}
,
\\
&\frac{
\Ysf_{\mu+1}(q^{-1}x)
}{
\Ysf_{\mu}(q^{-1}x)
}
=
\frac{
\theta(Q_{x}^{-1}q_2^{-\mu_{I}-2}q_1^{-I})\theta(Q_{x}^{-1}q_2^{-\mu_{I}-1}q_1^{-I-1})
}{
\theta(Q_{x}^{-1}q_2^{-\mu_{I}-2}q_1^{-I-1})\theta(Q_{x}^{-1}q_2^{-\mu_{I}-1}q_1^{-I})
}.
\end{align}
These two expressions are related each other,
\begin{align}
\frac{
\Ysf_{\mu+1}(x_1)
}{
\Ysf_{\mu}(x_1)
}
=
\Ssf\biggl(\frac{x_1}{x_2}\biggr)\biggl|_{Q_{x_2}=q_2^{-\mu_{I}-1}q_1^{-I}}
,~
\frac{
\Ysf_{\mu+1}(q^{-1}x_1)
}{
\Ysf_{\mu}(q^{-1}x_1)
}
=
\Ssf\biggl(\frac{x_1}{x_2}\biggr)\biggl|_{Q_{x_2}=q_2^{-\mu_{I}-1}q_1^{-I}}.
\end{align}
One can obtain the similar equations for $Q_{x_{2}}$. Then, according to the discussion in the appendix \ref{secB1}, we can show the regularity for the arbitrary $Q_{x_1}$ and $Q_{x_2}$.
\par
However, when we take the collision limit $Q_{x_1}=Q_{x_2}$, the $\Ssf$-factor might have the pole.
In order to consider this matter, let us consider the following case,
\begin{align}
Q_{x_1}=Q_x ,~Q_{x_2}=w Q_x
\end{align}
and take the limit $w\to1$. Then, by using the following formula
\begin{align}
 \theta(x;p)
 =
 (x;p)_\infty (px^{-1};p)_\infty
 \ \stackrel{x \to 1}{\longrightarrow} \
 (1 - x) (p;p)_\infty^2 ,
\end{align}
and
\begin{align}
 \Ocal(w x)
 & = \Ocal(e^{\log x + \log w})
 \nonumber \\
 & =
 \Ocal(e^{\log x})
 + \log w \frac{\partial}{\partial \log x} \Ocal(e^{\log x})
 + O((\log w)^2)
 \nonumber \\
 & = \Ocal(x)
 - (1 - w) \frac{\partial}{\partial \log x} \Ocal(x)
 + O((1-w)^2),
 \\
&\qquad \Bigl(\log w = \log (1 - (1 - w)) = - (1 - w) + O((1-w)^2)\Bigr)
 \nonumber
\end{align}
%from the following fact,
%\begin{align}
%\Ssf(w)=
%\frac{
%\theta( q_{1}^{-1} w)\theta( q_2^{-1} w)
%}{
%\theta( q^{-1} w)\theta(  w)
%}
%\xrightarrow{w\sim1}
%\frac{1}{(1-w^{-1})}
%\frac{
%\theta( q_{1}^{-1} )\theta( q_2^{-1})
%}{
%\theta( q^{-1})(Q_{\tau},Q_{\tau})_{\infty}^{2}
%},
%\end{align}
we have 
\begin{align}
&\Psf(wx)\Ssf(w^{-1})
\frac{\Ysf_{\mu}(x)}{\Ysf_{\mu}(q wx)}
+\Psf(x)\Ssf(w)\frac{\Ysf_{\mu}(wx)}{\Ysf_{\mu}(qx)}
\nonumber\\
&=
\Ssf(w^{-1})\Ysf_{\mu}(x)
\nonumber \\
&\quad\times\biggl(\Psf(x)-(1-w)\partial_{{\rm log}x}\Psf(x) +O((1-w)^2)\biggr)
\biggl(\frac{1}{\Ysf_{\mu}(q x)}+(1-w)\frac{\partial_{{\rm log}x}\Ysf_{\mu}(q x)}{\Ysf^{2}_{\mu}(q x)} + O((1-w)^2)\biggr)
\nonumber \\
&\quad+\Psf(x)\Ssf(w)\frac{1}{\Ysf_{\mu}(q x)}
\biggl(\Ysf_{\mu}(x) -(1-w) \partial_{{\rm log}x}\Ysf_{\mu}(x)+O((1-w)^2)\biggr)
\nonumber\\
&\xrightarrow{w\to1}
\Psf(x)\frac{\Ysf_{\mu}(x)}{\Ysf_{\mu}(q x)}
\biggl(
\mathfrak{c}(q_1,q_2)
-
\frac{
\theta( q_{1} )\theta( q_2)
}{
\theta( q)(Q_{\tau},Q_{\tau})_{\infty}^{2}
}
\partial_{\log x}{\rm log} \biggl[\frac{\Ysf_{\mu}(x)\Ysf_{\mu}(qx)}{\Psf(x)}\biggr]
\biggr),
\end{align}
where
\begin{align}
\mathfrak{c}(q_1 ,q_2)
&=\lim_{w\to1}(\Ssf(w)+\Ssf(w^{-1}))
\nonumber \\
&=\lim_{w\to1}
\Biggl[
\frac{w-1}{\theta( w)}
\partial_{w}\biggl[\frac{\theta( q_{1}^{-1} w)\theta( q_2^{-1} w)}{\theta( q^{-1} w)}\biggr]
+
\frac{w^{-1}-1}{\theta(w^{-1})}
\partial_{w^{-1}}\biggl[\frac{\theta( q_{1}^{-1} w^{-1})\theta( q_2^{-1} w^{-1})}{\theta( q^{-1} w^{-1})}\biggr]
\Biggr].
\end{align}
One can show that this coefficient $\mathfrak{c}(q_1 ,q_2 )$ is regular. Therefore, even if $Q_{x_1}=Q_{x_2}$, the expectation value of the $\Tsf$-operator is regular.
%\rem{To be checked: we use the formul\ae
%\begin{align}
% \theta(x;p)
% =
% (x;p)_\infty (px^{-1};p)_\infty
% \ \stackrel{x \to 1}{\longrightarrow} \
% (1 - x) (p;p)_\infty^2
%\end{align}
%\begin{align}
% \Ocal(w x)
% & = \Ocal(e^{\log x + \log w})
% \nonumber \\
% & =
% \Ocal(e^{\log x})
% + \log w \frac{\partial}{\partial \log x} \Ocal(e^{\log x})
% + O((\log w)^2)
% \nonumber \\
% & = \Ocal(x)
% - (1 - w) \frac{\partial}{\partial \log x} \Ocal(x)
% + O((1-w)^2)
%\end{align}
%where $\log w = \log (1 - (1 - w)) = - (1 - w) + O((1-w)^2).$
%}

%%%%%%%%%%%%%%%%%% section B-3 %%%%%%%%%%%%%%%%%%%%%%%%%%%%%%%

%\subsection{$A_{2}$ quiver with the single $\Ysf$-operator}
\subsection{$A_{2}$ quiver}

Let us consider the regularity for the $\Tsf$-operator average in $A_{2}$ quiver theory. Again by using some formulas in Appendix \ref{secA}, we obtain
\begin{align}
\mathcal{Z}_{\mu_1 , \mu_2}^{{\rm U}(1)\times{\rm U}(1)}
=&
q_2^{\frac{||\mu_{1}||^2}{2}+\frac{|\mu_{1}|}{2}}q_1^{\sum_{(i,j)\in\mu_{1}}i}
q_2^{-\frac{||\mu_{2}||^2}{2}+\frac{|\mu_{2}|}{2}}q_1^{|\mu_{2}|-\sum_{(i,j)\in\mu_{2}}i}
\nonumber \\
&\times\prod_{i,j\geq1}
\frac{
\Gamma_{e}(Q_{1}^{(2)}q_1^{j-i+1} )\Gamma_{e}(Q_{1}^{(2)}q_2^{-\mu_{2,i}+\mu_{1,j}}q_1^{j-i} )
}{
\Gamma_{e}(Q_{1}^{(2)}q_1^{j-i} )\Gamma_{e}(Q_{1}^{(2)}q_2^{-\mu_{2,i}+\mu_{1,j}}q_1^{j-i+1} )
}
\nonumber \\
&\times\prod_{i,j\geq1}
\frac{
\Gamma_{e}(q_2^{-1}q_1^{j-i-1} )\Gamma_{e}(q_2^{-\mu_{1,i}+\mu_{1,j}+1}q_1^{j-i} )
}{
\Gamma_{e}(q_2^{-1}q_1^{j-i} )\Gamma_{e}(q_2^{-\mu_{1,i}+\mu_{1,j}-1}q_1^{j-i-1} )
}
\frac{
\Gamma_{e}(q_2^{-1}q_1^{j-i-1})\Gamma_{e}(q_2^{-\mu_{2,i}+\mu_{2,j}-1}q_1^{j-i} )
}{
\Gamma_{e}(q_2^{-1}q_1^{j-i} )\Gamma_{e}(q_2^{-\mu_{2,i}+\mu_{2,j}-1}q_1^{j-i-1} )
}
\nonumber \\
&\times\prod_{i,j\geq1}
\frac{
\Gamma_{e}(Q_{1}^{(1)^{-1}}q_{2}^{-1}q_1^{j-i})\Gamma_{e}(Q_{1}^{(1)^{-1}}q_2^{\mu_{1,j}-1}q_1^{j-i-1} )
\Gamma_{e}(Q_{1}^{(3)^{-1}}q_{2}^{-1}q_1^{j-i})\Gamma_{e}(Q_{1}^{(3)^{-1}}q_2^{-\mu_{2,i}-1}q_1^{j-i-1} )
}{
\Gamma_{e}(Q_{1}^{(1)^{-1}}q_{2}^{-1}q_1^{j-i-1} )\Gamma_{e}(Q_{1}^{(1)^{-1}}q_2^{\mu_{1,j}-1}q_1^{j-i} )
\Gamma_{e}(Q_{1}^{(3)^{-1}}q_{2}^{-1}q_1^{j-i-1} )\Gamma_{e}(Q_{1}^{(3)^{-1}}q_2^{-\mu_{2,i}-1}q_1^{j-i} )
}.
\end{align}
%Also we can write the partition function by using the reflection $\theta_{1}(x)=-\theta_{1}(x^{-1})$,
%\begin{align}
%\mathcal{Z}_{\mu_1 , \mu_2}^{{\rm U}(1)\times{\rm U}(1)}
%=
%&
%q_1^{-\frac{||\mu_{2}||^2}{2}+\frac{|\mu_{2}|}{2}}q_2^{-|\mu_{2}|-\sum_{(i,j)\in\mu_{2}}i}
%q_1^{\frac{||\mu_{1}||^2}{2}+\frac{|\mu_{1}|}{2}}q_2^{\sum_{(i,j)\in\mu_{1}}i}
%\nonumber \\
%&\times\prod_{i,j\geq1}
%\frac{
%\Gamma_{e}(Q_{1}^{(2)^{-1}}q_1^{-1}q_2^{j-i} )\Gamma_{e}(Q_{1}^{(2)^{-1}}q_1^{-\mu_{2,i}+\mu_{1,j}-1}q_2^{j-%i-1} )
%}{
%\Gamma_{e}(Q_{1}^{(2)^{-1}}q_1^{-1}q_2^{j-i-1})\Gamma_{e}(Q_{1}^{(2)^{-1}}q_1^{-\mu_{2,i}+\mu_{1,j}-1}q_2^{j-%i} )
%}
%\nonumber \\
%&\times\prod_{i,j\geq1}
%\frac{
%\Gamma_{e}(q_1^{-1}q_2^{j-i-1} )\Gamma_{e}(q_1^{-\mu_{1,i}+\mu_{1,j}-1}q_2^{j-i} )
%}{
%\Gamma_{e}(q_1^{-1}q_2^{j-i} )\Gamma_{e}(q_1^{-\mu_{1,i}+\mu_{1,j}-1}q_2^{j-i-1} )
%}
%\frac{
%\Gamma_{e}(q_1^{-1}q_2^{j-i-1})\Gamma_{e}(q_1^{-\mu_{2,i}+\mu_{2,j}-1}q_2^{j-i} )
%}{
%\Gamma_{e}(q_1^{-1}q_2^{j-i} )\Gamma_{e}(q_1^{-\mu_{2,i}+\mu_{2,j}-1}q_2^{j-i-1} )
%}
%\nonumber \\
%&\times\prod_{i,j\geq1}
%\frac{
%\Gamma_{e}(Q_{1}^{(3)}q_2^{j-i+1} )\Gamma_{e}(Q_{1}^{(3)}q_1^{-\mu_{2,i}}q_2^{j-i} )
%\Gamma_{e}(Q_{1}^{(1)}q_2^{j-i+1} )\Gamma_{e}(Q_{1}^{(1)}q_1^{\mu_{1,i}}q_2^{j-i} )
%}{
%\Gamma_{e}(Q_{1}^{(3)}q_2^{j-i} )\Gamma_{e}(Q_{1}^{(3)}q_1^{-\mu_{2,i}}q_2^{j-i+1} )
%\Gamma_{e}(Q_{1}^{(1)}q_2^{j-i} )\Gamma_{e}(Q_{1}^{(1)}q_1^{\mu_{1,i}}q_2^{j-i+1} )
%}.
%\end{align}
Then, we have
\begin{align}
\frac{\mathcal{Z}_{\mu_1 , \mu_2}^{{\rm U}(1)\times{\rm U}(1)}}
{\mathcal{Z}_{\mu_1+1 , \mu_2}^{{\rm U}(1)\times{\rm U}(1)}}
=
&\qfrak_1^{-1}\frac{
q_2^{-\mu_{1,I}-1}q_1^{-I}
}{
\theta(Q_{1}^{(3)^{-1}}q_1^{\mu_{1,I}}q_2^{I-1} )
}
\prod_{i\geq1}
\frac{
\theta(Q_{1}^{(2)}q_2^{\mu_{1,I}-\mu_{2,i}+1}q_1^{I-i} )
}{
\theta(Q_{1}^{(2)}q_2^{\mu_{1,I}-\mu_{2,i}+1}q_1^{I-i+1} )
}
\nonumber \\
&\qquad\qquad\qquad\qquad
\times\prod_{i\geq1,i\neq I}
\frac{
\theta(q_2^{-\mu_{1,I}+\mu_{1,i}-1}q_1^{i-I-1} )
\theta(q_2^{-\mu_{1,i}+\mu_{1,I}}q_1^{I-i} )
}{
\theta(q_2^{-\mu_{1,I}+\mu_{1,i}-1}q_1^{i-I} )
\theta(q_2^{-\mu_{1,i}+\mu_{1,I}}q_1^{I-i-1} )
}
,
\\
\frac{\mathcal{Z}_{\mu_1 , \mu_2}^{{\rm U}(1)\times{\rm U}(1)}}
{\mathcal{Z}_{\mu_1 , \mu_2 +1}^{{\rm U}(1)\times{\rm U}(1)}}
=&
\qfrak_2^{-1}\frac{
q_2^{\mu_{2,I}}q_1^{I-1}
}{
\theta(Q_{1}^{(1)^{-1}}q_2^{-\mu_{2,I}-1}q_1^{-I} )
}
\prod_{i\geq1}
\frac{
\theta(Q_{1}^{(2)}q_2^{-\mu_{2,I}+\mu_{1,i}}q_1^{i-I+1} )
}{
\theta(Q_{1}^{(2)}q_2^{-\mu_{2,I}+\mu_{1,i}}q_1^{i-I} )
}
\nonumber \\
&\qquad\qquad\qquad\qquad
\times\prod_{i\geq1,i\neq I}
\frac{
\theta(q_2^{-\mu_{2,I}+\mu_{2,i}-1}q_1^{i-I-1} )
\theta(q_2^{-\mu_{2,i}+\mu_{2,I}}q_1^{I-i} )
}{
\theta(q_2^{-\mu_{2,I}+\mu_{2,i}-1}q_1^{i-I} )
\theta(q_2^{-\mu_{2,i}+\mu_{2,I}}q_1^{I-i-1} )
}
.
\end{align}
The product of $\Ysf$-operators is given by \eqref{prodY}.
Then, we find that
\begin{align}
&\frac{\mathcal{Z}_{\mu_1 , \mu_2}^{{\rm U}(1)\times{\rm U}(1)}}
{\mathcal{Z}_{\mu_1 +1, \mu_2}^{{\rm U}(1)\times{\rm U}(1)}}
=
-\qfrak_1^{-1}\frac{
\Ysf_{\mu_{1}}(q^{-1}x)\Ysf_{\mu_{1}+1}(x)
}{
\Psf_{1}(x)\Ysf_{\mu_{2}}(x)
}\Biggl|_{Q_{x}=q_2^{-\mu_{1,I}-1}q_1^{-I}},~
\\
&\frac{\mathcal{Z}_{\mu_1 , \mu_2}^{{\rm U}(1)\times{\rm U}(1)}}
{\mathcal{Z}_{\mu_1 ,\mu_2 +1}^{{\rm U}(1)\times{\rm U}(1)}}
=
-\qfrak_2^{-1}\frac{
\Ysf_{\mu_{2}}(x)\Ysf_{\mu_{2}+1}(q^{-1}x)
}{
\Psf_{2}(x)\Ysf_{\mu_{1}}(q^{-1}x)
}\Biggl|_{Q_{x}=Q_{1}^{(2)^{-1}}q_2^{-\mu_{2,I}-1}q_1^{-I}}.
\end{align}
%\rem{Gauge couplings to be added $(\qfrak_1,\qfrak_2)$.}
Note that the variable $x$ is given by \eqref{GTA_2x}. Therefore the average $\Big< \Tsf_1(x) \Big>$ is regular for the arbitrary $x$.

%\clearpage
%%%%%%%%%%%%%%%%%% References %%%%%%%%%%%%%%%%%%%%%%%%%%
%\bibliographystyle{utphys}
%\bibliography{ref_top}

\providecommand{\href}[2]{#2}\begingroup\raggedright\endgroup

\end{document}